\documentclass[longauth]{aa}  
\usepackage{graphicx}
\usepackage{txfonts}
\usepackage[breaklinks=true]{hyperref}
\usepackage{xspace}
\usepackage[dvipsnames]{xcolor}
\usepackage[normalem]{ulem}

\usepackage{listings}

\definecolor{dkgreen}{rgb}{0,0.6,0}
\definecolor{gray}{rgb}{0.5,0.5,0.5}
\definecolor{mauve}{rgb}{0.58,0,0.82}
\newcommand{\orcit}[1]{\protect\href{https://orcid.org/#1}{\protect\includegraphics[width=8pt]{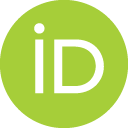}}}
\newcommand{\instref}[1]{\inst{\ref{inst:#1}}}

\newcommand{\gaia}{\textit{Gaia}\xspace}
\newcommand{\gdr}[1]{\gaia~DR#1\xspace}
\newcommand{\gedr}{\gaia~EDR3\xspace}

\newcommand{\xp}{{BP/RP}\xspace}
\newcommand{\gband}{\ensuremath{G}-band\xspace}
\newcommand{\bprp}{\ensuremath{G_{\rm BP}-G_{\rm RP}}\xspace}
\newcommand{\gbp}{\ensuremath{G_{\rm BP}}\xspace}
\newcommand{\grp}{\ensuremath{G_{\rm RP}}\xspace}
\newcommand{\cstar}{\ensuremath{c_*}\xspace}

\newcommand{\obmtrev}{OBMT-Rev\xspace}

\newcommand{\secname}{Sect.}
\newcommand{\equref}[1]{Eq.~\ref{eq:#1}}
\newcommand{\secref}[1]{\secname~\ref{sec:#1}}

\newcommand{\figref}[1]{Fig.~\ref{fig:#1}}

\authorrunning{F.~De~Angeli~et~al.}

\begin{document} 

\title{\gaia Data Release 3: Processing and validation of BP/RP low-resolution spectral data}

\author{
F.~De~Angeli\orcit{0000-0003-1879-0488}\instref{ioa}\fnmsep\thanks{Corresponding author: F.~De~Angeli\newline
e-mail: \href{mailto:fda@ast.cam.ac.uk}{\tt fda@ast.cam.ac.uk}}
\and
M.~Weiler\instref{ub}
\and
P.~Montegriffo\orcit{0000-0001-5013-5948}\instref{oabo}
\and
D.~W.~Evans \orcit{0000-0002-6685-5998}\instref{ioa}
\and
M.~Riello \orcit{0000-0002-3134-0935}\instref{ioa}
\and
R.~Andrae\orcit{0000-0001-8006-6365}\instref{mpia}
\and
J.~M.~Carrasco \orcit{0000-0002-3029-5853}\instref{ub}
\and
G.~Busso \orcit{0000-0003-0937-9849}\instref{ioa}
\and
P.~W.~Burgess\instref{ioa}
\and
C.~Cacciari \orcit{0000-0001-5174-3179}\instref{oabo}
\and
M.~Davidson\instref{ifa}
\and
D.~L.~Harrison \orcit{0000-0001-8687-6588}\inst{\ref{inst:ioa},\ref{inst:kavli}}
\and
S.~T.~Hodgkin \orcit{0000-0002-5470-3962}\instref{ioa}
\and
C.~Jordi \orcit{0000-0001-5495-9602}\instref{ub}
\and
P.~J.~Osborne\instref{ioa}
\and
E.~Pancino~\orcit{0000-0003-0788-5879}\inst{\ref{inst:oafi},\ref{inst:asi}}
\and
G.~Altavilla~\orcit{0000-0002-9934-1352}\inst{\ref{inst:oaroma},\ref{inst:asi}}
\and
M.~A.~Barstow \orcit{0000-0002-7116-3259}\instref{lei}
\and
C.~A.~L.~Bailer-Jones\instref{mpia}
\and
M.~Bellazzini \orcit{0000-0001-8200-810X}\instref{oabo}
\and
A.~G.~A.~Brown \orcit{0000-0002-7419-9679}\instref{leiden}
\and
M.~Castellani \orcit{0000-0002-7650-7428}\instref{oaroma}
\and
S.~Cowell\instref{ioa}
\and
L.~Delchambre\instref{liege}
\and
F.~De~Luise~\orcit{0000-0002-6570-8208}\instref{oate}
\and
C.~Diener\instref{ioa}
\and
C.~Fabricius \orcit{0000-0003-2639-1372}\instref{ub}
\and
M.~Fouesneau\instref{mpia} 
\and
Y.~Fremat\instref{brussels}
\and
G.~Gilmore~\orcit{0000-0003-4632-0213}\instref{ioa}
\and
G.~Giuffrida\instref{oaroma}
\and
N.~C.~Hambly \orcit{0000-0002-9901-9064}\instref{ifa}
\and
S.~Hidalgo \orcit{0000-0002-0002-9298}\instref{iac}
\and
G.~Holland\instref{ioa}
\and
Z.~Kostrzewa-Rutkowska\inst{\ref{inst:leiden},\ref{inst:sron}}
\and
F.~van~Leeuwen\instref{ioa}
\and 
A.~Lobel\instref{brussels}
\and
S.~Marinoni \orcit{0000-0001-7990-6849}\inst{\ref{inst:oaroma},\ref{inst:asi}}
\and
N.~Miller\instref{ioa}
\and
C.~Pagani\instref{lei}
\and
L.~Palaversa\inst{\ref{inst:zag},\ref{inst:ioa}}
\and
A.~M.~Piersimoni \orcit{0000-0002-8019-3708}\instref{oate}
\and
L.~Pulone \orcit{0000-0002-5285-998X}\instref{oaroma}
\and
S.~Ragaini\instref{oabo}
\and
M.~Rainer \orcit{0000-0002-8786-2572}\inst{\ref{inst:oafi},\ref{inst:brera}}
\and
P.~J.~Richards\instref{stfc}
\and
G.~T.~Rixon\inst{\ref{inst:ioa}} 
\and
D.~Ruz-Mieres \orcit{0000-0002-9455-157X}\instref{ioa}
\and
N.~Sanna \orcit{0000-0001-9275-9492}\instref{oafi}
\and 
L.~M.~Sarro\instref{uned}
\and
N.~Rowell\instref{ifa}
\and
R.~Sordo\instref{oapd}
\and
N.~A.~Walton \orcit{0000-0003-3983-8778}\instref{ioa}
\and
A.~Yoldas\instref{ioa}
}


\institute{
Institute of Astronomy, University of Cambridge, Madingley Road, Cambridge CB3 0HA, UK\label{inst:ioa}
\and
Institut de Ci\`encies del Cosmos (ICC), Universitat de Barcelona (IEEC-UB), c/ Mart\'{\i} i Franqu\`es, 1, 08028 Barcelona, Spain
\label{inst:ub}
\and
INAF -- Osservatorio di Astrofisica e Scienza dello Spazio di Bologna, Via Gobetti 93/3, 40129 Bologna, Italy
\label{inst:oabo}
\and
Max Planck Institute for Astronomy, K\"{o}nigstuhl 17, 69117 Heidelberg, Germany\label{inst:mpia}
\and
Institute for Astronomy, School of Physics and Astronomy, University of Edinburgh, Royal Observatory, Blackford Hill, Edinburgh, EH9~3HJ, UK
\label{inst:ifa}
\and
Kavli Institute for Cosmology, Institute of Astronomy, Madingley Road, Cambridge, CB3 0HA, UK\label{inst:kavli}
\and
INAF -- Osservatorio Astrofisico di Arcetri, Largo E. Fermi, 5, 50125 Firenze, Italy\label{inst:oafi}
\and
INAF -- Osservatorio Astronomico di Roma, via Frascati 33, 00078 Monte Porzio Catone (Roma), Italy\label{inst:oaroma}
\and
Space Science Data Center - ASI, Via del Politecnico SNC, 00133 Roma, Italy\label{inst:asi}
\and
School of Physics \& Astronomy, University of Leicester, Leicester LE9 1UP, UK\label{inst:lei}
\and
Leiden Observatory, Leiden University, Niels Bohrweg 2, 2333 CA Leiden, The Netherlands\label{inst:leiden}
\and
Institut d’Astrophysique et de G\'eophysique, Universit\'e de Li\`ege, 19c, All\'ee du 6 Ao\^ut, B-4000 Li\`ege, Belgium\label{inst:liege}
\and
INAF - Osservatorio Astronomico d'Abruzzo, Via Mentore Maggini, 64100 Teramo, Italy\label{inst:oate}
\and
Royal Observatory of Belgium, Ringlaan 3, 1180 Brussels, Belgium\label{inst:brussels}
\and
IAC - Instituto de Astrofisica de Canarias, Via L\'{a}ctea s/n, 38200 La Laguna S.C., Tenerife, Spain\label{inst:iac}
\and
SRON Netherlands Institute for Space Research, Niels Bohrweg 4, 2333 CA Leiden, The Netherlands\label{inst:sron}
\and
Ruđer Bo\v{s}kovi\'c Institute, Bijeni\v{c}ka cesta 54, Zagreb, Croatia\label{inst:zag}
\and 
INAF - Osservatorio Astronomico di Brera, Via E. Bianchi, 46, 23807 Merate (LC), Italy\label{inst:brera}
\and
STFC, Rutherford Appleton Laboratory, Harwell, Didcot, OX11 0QX, United Kingdom\label{inst:stfc}
\and
Dpto. de Inteligencia Artificial, UNED, c/ Juan del Rosal 16, 28040 Madrid, Spain\label{inst:uned}
\and
INAF - Osservatorio Astronomico di Padova, Vicolo Osservatorio 5, 35122 Padova, Italy\label{inst:oapd}
}


\date{Received Month Day, Year; accepted Month Day, Year}

\abstract
{Blue (BP) and Red (RP) Photometer low-resolution spectral data is one of the exciting new products in \textit{Gaia} Data Release 3 (\textit{Gaia} DR3). These data have also been used to derive astrometry and integrated photometry in \textit{Gaia} Early Data Release 3 and astrophysical parameters and Solar System Object reflectance spectra in \textit{Gaia} DR3.}
{In this paper we give an overview of the processing techniques that allow converting satellite raw data of multiple transits per source into combined spectra calibrated onto an internal reference system, resulting in low-resolution BP and RP mean spectra. We describe how we overcome challenges due to the complexity of the on-board instruments and to the various observation strategies. Furthermore, we show highlights from the scientific validation of the results.
This work covers the internal calibration of BP/RP spectra onto a self-consistent mean instrument, while the calibration of the BP/RP spectra to the absolute reference system of physical flux and wavelength is covered in \cite{Montegriffo2022}. This should be seen as an essential companion to this paper.}
{We calibrate about 65 billion individual transit spectra onto the same mean BP/RP instrument through a series of calibration steps, including background subtraction, calibration of the CCD geometry and an iterative procedure for the calibration of CCD efficiency as well as variations of the line-spread function and dispersion across the focal plane and in time. The calibrated transit spectra are then combined for each source in terms of an expansion into continuous basis functions. We discuss the configuration of these basis functions.}
{Time-averaged mean spectra covering the optical to near-infrared wavelength range $[330,1050]$ nm are published for approximately $220$ million objects. Most of these are brighter than $G=17.65$ but some BP/RP spectra are published for sources down to $G=21.43$. Their signal-to-noise ratio varies significantly over the wavelength range covered and with magnitude and colour of the observed objects, with sources around $G=15$ having S/N above $100$ in some wavelength ranges. The top-quality BP/RP spectra are achieved for sources with magnitudes $9<G<12$, having S/N reaching $1000$ in the central part of the RP wavelength range. Scientific validation suggests that the internal calibration was generally successful. However, there is some evidence for imperfect calibrations at the bright end $G<11$, where calibrated BP/RP spectra can exhibit systematic flux variations that exceed their estimated flux uncertainties. We also report that due to long-range noise correlations, BP/RP spectra can exhibit wiggles when sampled in pseudo-wavelength.}
{The \textit{Gaia} DR3 data products are the expansion coefficients and corresponding covariance matrices for BP and RP separately. Users are encouraged to work with the data in this format, with full covariance information showing that correlations between coefficients are typically very low. Documentation and instructions on how to access and use BP/RP spectral data from the archive are also provided.}

\maketitle

\section{Introduction}\label{sec:introduction}

The European Space Agency (ESA) mission \gaia \citep{Prusti2016} has already released to the astronomical community three catalogues of increasing richness in terms of content, precision and accuracy. Researchers from many branches of astrophysics have shown great interest in the published data leading to the publication of more than 6000 refereed papers based on \gaia data to date
\footnote{See the list of refereed papers since launch available at \href{https://ui.adsabs.harvard.edu/public-libraries/fWFE_JYLRZG2jwgwKetH8w}{https://ui.adsabs.harvard.edu/public-libraries/fWFE\_JYLRZG2jwgwKetH8w}.}.

With respect to the previous \gaia Early Data Release 3,
\gaia Data Release 3 \citep[\gdr3][]{DR3-top-level} introduces a number of new data products based on the same source catalogue, including a total of 1.8 billion objects and based on a period of 34 months of satellite operations. A large fraction of the objects in the catalogue has astrophysical parameters determined from the medium (Radial Velocity Spectrometer, RVS) and low-resolution (Blue and Red Photometers, BP and RP) spectral data as well as from the photometric data \citep{GSPPhot,DR3-DPACP-157}. For many of these objects the actual
RVS and/or \xp data itself are part of the release; RVS spectra are released for about 1 million sources, while mean low-resolution 
\xp spectra are available for about 220 million objects, selected to have a reasonable number of observations and to be sufficiently bright to ensure good signal to noise ratio at this stage in the mission. New estimates of mean radial velocities, variable-star classification and epoch photometry are released for a subset of sources. A large set of Solar System objects, including new discoveries, with preliminary orbital solutions and individual epoch observations are available in the \gdr3 release. A selection of these have also their reflectance spectra estimated from the epoch \xp spectral data \citep{DR3-DPACP-89}. The release includes also results for non-single stars, Quasars and Extended Objects. Finally, an additional data set is also released, called the \gaia Andromeda Photometric Survey (GAPS), consisting of the photometric time series for all sources located in a 5.5 degree radius field centred on the Andromeda galaxy \citep{DR3-DPACP-142}.
A number of papers have been prepared by the Data Processing and Analysis Consortium (DPAC) describing all aspects of the data processing and the results of the performance verification activities. In this paragraph we have only included specific citations to papers that have made use of the \xp spectral data. A full list is available at \href{https://www.cosmos.esa.int/web/gaia/dr3-papers}{https://www.cosmos.esa.int/web/gaia/dr3-papers}.

This paper focuses on the \xp low-resolution spectral data and on the processing that led to the generation of the \xp spectra included in \gdr3. Some aspects of the \xp processing have already been introduced in recent papers which should be considered essential companions to this one. In particular, calibrations that were also required for the generation of the \xp integrated photometry have been detailed in \cite{Riello2021} and will be described only very briefly in this paper. The algorithm adopted for the internal calibration of the \xp spectral data is presented in the dedicated paper \cite{Carrasco2021}. We refer to \cite{Carrasco2021} for the detailed justification of the model definition and complement that work providing information on the actual model configuration adopted to generate the \gdr{3} BP/RP spectra. The focus of this paper is the processing leading to the generation of a homogeneous catalogue of source spectra from the raw \gaia BP/RP observations. While \gdr{3} does not provide access to individual observations, knowing the complexities related to the instruments, observing strategies, and processing is important to understand the final product. This paper also contains useful information about the representation of the spectra and the strategies adopted to optimise such representation and minimise the noise in the final spectra. The validation shown in this paper focuses on these aspects.
The calibration of the \xp spectral data to the absolute reference system (both in terms of flux and wavelength) is detailed in \cite{Montegriffo2022}, also accompanying the \gdr3 release. Users interested in systematic effects present in the final BP/RP products should refer to that paper presenting the results of the validation of the externally calibrated data with respect to external absolute spectra.
Finally, \cite{DPACP-127} presents the over all results of the independent DPAC validation process, with useful insights into the limitations and recommendations for \xp spectral data.

The paper outline is the following: in Sect. \ref{sec:inputs} we describe the general concept of low-resolution spectroscopic data and the specific aspects of the \gaia\ \xp data that are relevant for this paper; Sect. \ref{sec:processing} is dedicated to the data processing, with considerations on the processing strategies, algorithms and results; a description of the composition of the \xp spectral catalogue in \gdr{3} is provided in Sect. \ref{sec:outputs}; highlights from the internal validation activities are given in Sect. \ref{sec:validation} and Sect. \ref{sec:recommendations} offers some recommendations for the users. 

\section{Input data}\label{sec:inputs}

During its operations, the \gaia satellite scans the entire sky every 6 months while spinning around its principal axis and precessing around the Earth-Sun direction. The light from two Fields of View (FoVs) is focused on the same focal plane. Images of sources crossing the focal plane move over an array of Charge-Coupled Devices (CCDs) operating in Time Delayed Integration (TDI) mode, such that the charges generated by a point-like astronomical source are clocked through the CCD at the same speed of the apparent motion of the source due to the satellite scanning motion. In the following, we will use \textit{transit} to refer to a full focal plane crossing of a source and \textit{CCD transit} when referring to the crossing of a single CCD, generating one observation.

Throughout this paper, time will be expressed in On-Board-Mission-Time (OBMT) in units of satellite revolutions ($1$ \obmtrev $= 21\,600$ s). A formula to convert OBMT to barycentric coordinate time (TCB) is provided by Eq. (3) in \cite{Prusti2016}.

In the focal plane array (see Fig. 4 in \citealp{Prusti2016}, or Fig. 2 in \citealp{Carrasco2021}), the CCDs are arranged in rows (in the along scan direction, AL) and strips (in the across scan direction, AC).
The largest section of the focal plane array (including 62 Astrometric Field, AF, CCDs, arranged in 7 rows of 9 CCD each, except for one row where there are only 8) is dedicated to the collection of the observations in the broad G-band which are used for the astrometric measurements and for the photometry.
Following these, two strips of 7 CCDs each are dedicated to the BP and RP instruments.
Finally, 4 rows and 3 strips of CCDs collect the RVS observations. Obviously, not all sources crossing the focal plane will also cross the RVS CCDs.

Colour information for all sources is essential to achieve the high-accuracy that characterises the \gaia astrometry. An initial design where the flux of sources in a variety of medium bands would be measured on different CCD strips to fulfill this requirement \citep{Jordi2006}, was abandoned in favour of low-resolution aperture prism spectroscopy. This observational technique is frequently used to obtain a large number of spectra with a single exposure in large-scale astronomic surveys, starting from the Draper catalogue in the early 20th century \citep{pickering} all the way to future applications such as in Euclid \citep{euclid} and NGRST \citep[formerly known as WFIRST][]{wfirst}. The \xp instruments were added to the satellite payload to collect this data covering the wavelength ranges [$330$, $680$] nm and [$640$, $1050$] nm respectively with varying resolution depending on the position in the spectrum and on the CCD \citep[the resolution covers the range $100$ to $30$ for BP and $100$ to $70$
for RP in $\lambda/\Delta\lambda$, see Fig. 3 in ][]{Carrasco2021}.

In normal operation mode, observations transmitted to the ground from the satellite are cut-outs of a small area surrounding the position where each source was detected on board. In the case of \xp observations, because of the need to cover the full range of the dispersed light, these cut-outs (\textit{windows} in \gaia terminology) need to be much longer in the direction in which the light is dispersed, which is aligned with the AL direction. This is why the size of the \xp windows is 60 pixels in AL (as opposed to a maximum of 18 pixels for the AF windows assigned to the brightest objects) by 12 pixels in AC direction, corresponding to an area in the sky of approximately $3.5$ arcsec by $2.1$ arcsec\footnote{The angular dimensions of each pixel are approximately $58.9$ and $176.8$ mas in the AL and AC direction respectively.}. This affects the possibility to assign different windows to nearby sources in crowded regions. As a consequence of this, not all detections  result in a \xp observation and the average number of \xp observations is lower than the average number of transits per source on the focal plane.
Partly overlapping windows can in some cases be allocated by the on-board software. When this happens, the window of the brightest sources is transmitted fully to the ground, while only the non-overlapping section of the other window is transmitted to the ground. These \textit{truncated} windows are not included in the data leading to \gdr3 as they are normally rather disturbed by the nearby brighter source and require special treatment which will only be implemented for future data releases.

Observations on board can be taken in different configurations depending on the on-board magnitude estimate of the source. The activation of a given configuration can also affect simultaneous observations nearby.

\begin{figure}[!h]
    \centering
    \includegraphics[width=0.48\textwidth]{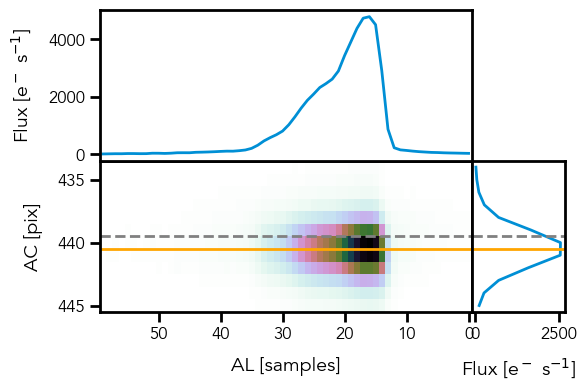}
    \caption{An example of a 2D spectrum is shown in the central panel. The dashed and continuous horizontal lines show the AC centre of the window and the AC predicted position based on the source astrometry, the satellite attitude and the BP CCD geometry. The top and right panels show the result of binning in the AC and AL direction respectively. The AL coordinate is given in units of samples.}
    \label{fig:2dVs1d}
\end{figure}

Different configuration aspects include the AC resolution within a window which is only achieved for sources brighter than $11.5$~mag in the \gband, while windows assigned to fainter sources are binned in the AC direction on board before transmission thus resulting in a spectrum with 60 AL samples, where each sample contains the overall flux measurement from 12 pixels. 
Figure \ref{fig:2dVs1d} shows the case of a 2D spectrum. The top panel shows the 1D spectrum resulting from the binning in the AC direction.
The different window strategy for sources fainter than $11.5$~mag has the purpose of limiting the volume of the data that needs downloading from the satellite and reducing the readout noise. In the following we will refer to these as different window classes (or WCs) and in particular to 2D (where the AC resolution is preserved) vs 1D (where binning AC occurs) spectra respectively. It should be noted that all \xp spectra available in \gdr{3} will be 1D (i.e. flux values corresponding to positions in the AL coordinate or wavelength when the external calibration is applied). Spectra acquired with a 2D configuration on-board are flattened to 1D during the calibration process: a simple sum of the samples in the same AC column is adopted for consistency with the on-board AC binning algorithm.

\begin{figure}[!h]
    \centering
    \includegraphics[width=0.48\textwidth]{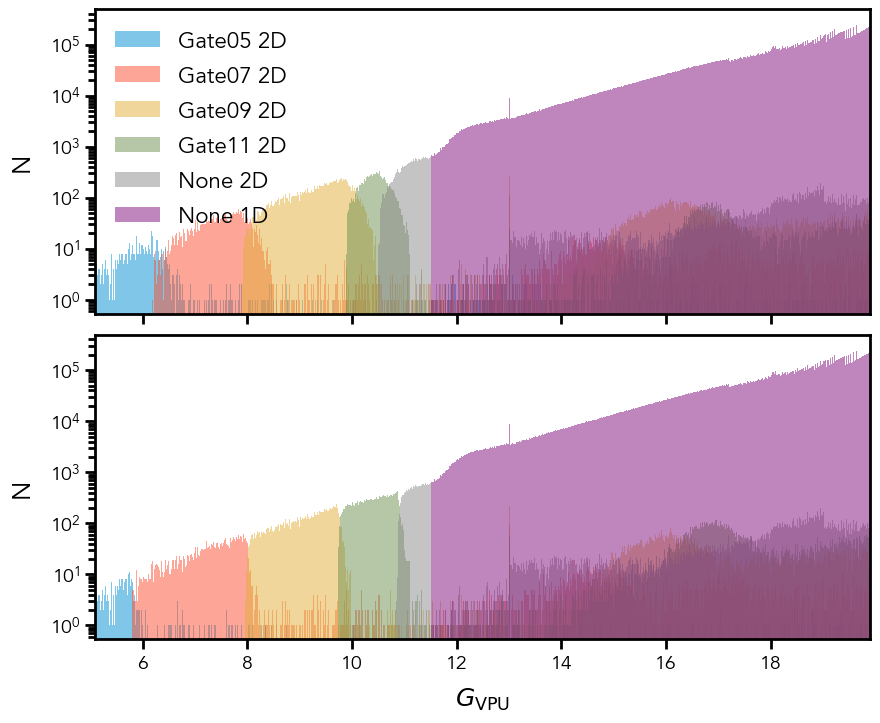}
    \caption{Distribution of the number of \xp observations acquired in BP (top panel) and RP (bottom panel) with a given gate and WC configuration vs on-board magnitude labelled as $G_{\rm VPU}$. The gated observations for sources fainter than $\approx 11.5$~mag in the \gband are due to occasional alignment of these sources with brighter objects triggering the activation of a gate.}
    \label{fig:gates}
\end{figure}

An \textit{ad hoc} strategy is also available to prevent saturation when observing bright sources. Different gates can be activated at different locations in the CCD to limit the section of the CCD where the charges are accumulated and therefore effectively reduce the exposure time. 
The exposure time of an \textit{ungated} observation is approximately 4.4 seconds, the shortest gate active in \xp (\texttt{Gate05}) reduces this to 0.06 seconds. Each gate is activated on-board as required based on a configured set of magnitude ranges and the on-board magnitude estimated for each transit. The configuration changes for different instruments (\xp) and across the focal plane (even within a CCD). See Fig. \ref{fig:gates} to see the distribution of different gate and WC configuration vs on-board magnitude for BP and RP. As already mentioned, the selection of the appropriate gate configuration is based on the on-board magnitude estimate which can show up to 0.5 magnitude uncertainties at the bright end. This implies that a given source may be observed in different gate configurations in different transits, some of these gate configurations will be sub-optimal and therefore some saturation cannot be excluded. Moreover the activation of a gate will affect all observations taken at the same time (within $60$ pixels or $0.06$ s AL) in the same CCD, thus generating gated observations for faint sources that would normally be observed without any gate. This can also cause what are called complex gate cases, where different gates are active in different sections of a window. Complex gate cases are also not included in the processing leading to \gdr3.

\begin{figure}[!ht]
    \centering
    \includegraphics[width=0.48\textwidth]{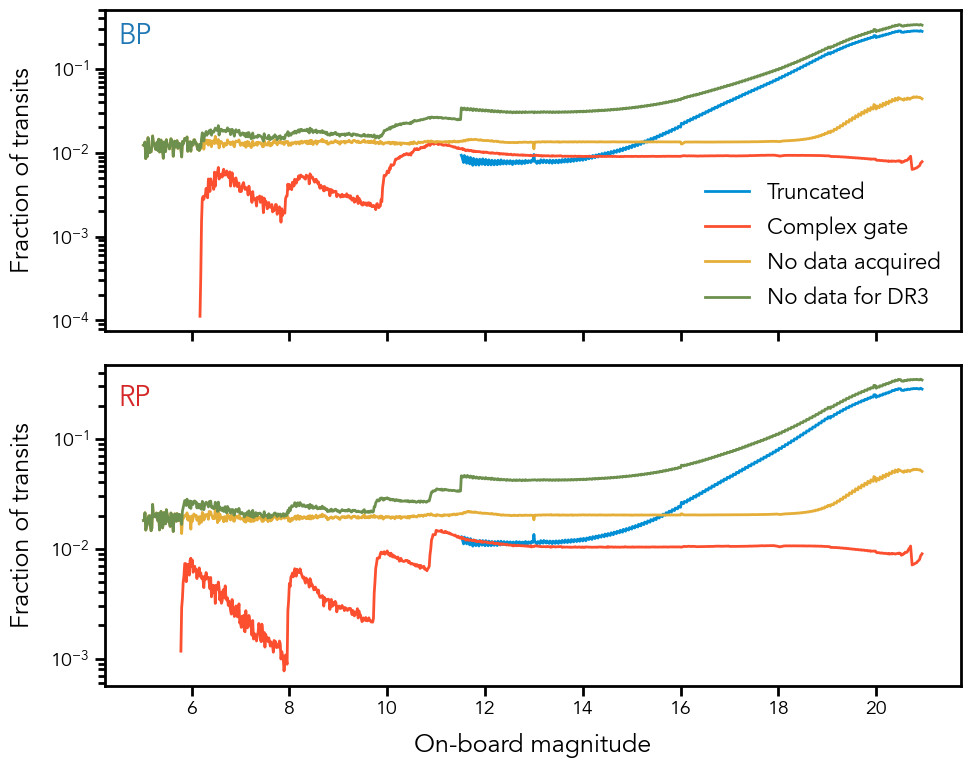}
    \caption{Fraction of transits that will not contribute a BP observation to the processing leading to \gdr{3} due to either the BP window not having been acquired (orange line), or to the BP window being truncated (blue line), or to the BP window having been observed with multiple gates active within the window (red line). The green line shows the total effect. This is shown as a function of the on-board magnitude estimate as this is the parameter that defines the observation strategy applied to each observation. Truncation for instance is only applied to 1D windows and therefore the corresponding fraction is 0 for on-board magnitude brighter than 11.5~mag.}
    \label{fig:xpavailability}
\end{figure}

Figure \ref{fig:xpavailability} shows the implications for the fraction of BP (top) and RP (bottom) transits available for processing of some of the mission aspects mentioned in this section (size of \xp windows, gates and truncation). The different curves show the fraction of transits that will not contribute a BP/RP observation to the processing leading to the \gdr{3} catalogue for various reasons: the blue curve shows the fraction of BP transits affected by truncation, the red line those acquired with a complex gate, the orange line shows the fraction of transits that do not have a BP or RP window acquired, and the green line simply shows the sum of the three previous quantities and therefore the fraction of transits that will not have an observation that can be processed at this stage. Both fractions of truncated and not-acquired windows increase significantly at the faint end as expected. 

The total number of transits acquired in the period covered by \gdr{3} was almost 78 billion. The processing of the \xp spectral data produced calibrated \xp epoch spectra (i.e. spectra generated from one single observation) for about 65 billion transits, and mean \xp spectra (i.e. spectra averaged over the many observations for a given source) for 
$2,094,515,608$ 
more than 2 billion sources. Not all transits nor sources had a complete set of BP and RP spectra. Please refer to Sec. \ref{sec:outputs} for more information on the selection criteria that lead to the composition of the \gdr{3} catalogue.

\section{Processing}\label{sec:processing}

When calibrating the \xp data, the characteristics of the various CCDs, the effects introduced by the different optical paths for the two FoVs and by the configuration activated for each observation and the variation in time of all these elements will need to be taken into account. We refer to a set of validity time range (i.e. interval in time where a given calibration is applicable), CCD, FoV, WC and gate as a configuration or calibration unit. A set of calibrations per calibration unit (for a total of several tens of thousands configurations) will be produced as part of the instrument calibration process to describe each effect that needs calibrating. Due to the complexity of the system (effectively equivalent to many instruments), the calibration of the data cannot rely on any existing catalogue of standards (all too limited in number and quality), but needs to be solved for internally in the first instance, using a large subset of the \xp data itself. This subset is selected to contain data for a sufficiently large catalogue of sources (referred to as calibrators, see \secref{processing:calibrators}) covering all calibration units as homogeneously as possible within the limits imposed by nature (e.g.\ in terms of magnitude and colour distribution). The goal of the \textit{internal calibration} is to define a reference instrument which is homogeneous across all configurations and time. It is then the responsibility of the \textit{external or absolute calibration} to define the link between the internal system and the absolute system using a carefully assembled catalogue of spectro-photometric calibrators \citep{Pancino2021,Marinoni2016,Altavilla2015,Altavilla2021} and other objects that present features in their spectra that are useful to calibrate specific aspects of the instrument and for which suitable absolute spectra are already available.
The internal reference system is in essence defined by the  calibrations, i.e. the actual calibration coefficients. Once the reference system is established, all the data can be brought to the same system by applying the calibrations.
The same approach has been followed for the processing of the \gaia photometric data \citep{Carrasco2016}.
In this paper we focus on the internal calibration of the \xp spectral data, while the external calibration is the subject of \cite{Montegriffo2022}.

\begin{figure*}[!ht]
    \centering
    \includegraphics[width=\textwidth]{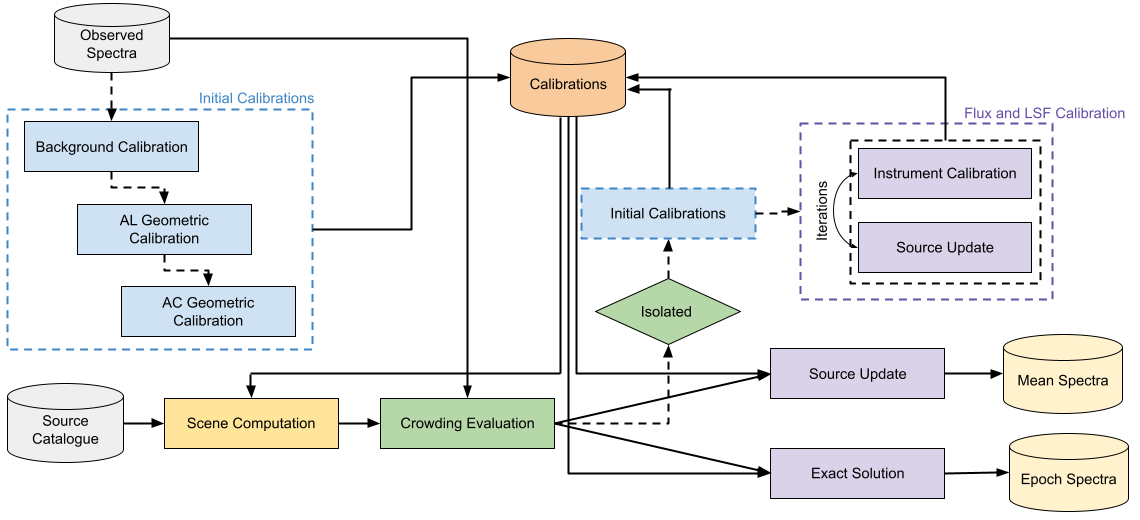}
    \caption{Schematic view of the processing leading to the generation of the \xp mean spectra in \gdr3.}
    \label{fig:flow}
\end{figure*}

The internal calibration includes many different individual calibration steps that are solved for in separate stages of the data processing, often relying on different subsets of calibrators and requiring different strategies for accessing the data in an optimal way.
Figure \ref{fig:flow} shows a schematic overview of the major steps and dependencies of the process starting from the input raw observed spectra until the output mean spectra.
The two main inputs to the process are the \xp observed spectra and the source catalogue containing astrometry and photometry information for all sources observed so far. In the flow diagram, dashed lines are used to represent data flow for calibrators only, while solid lines are used to indicate that the entire set of the observed spectra is used as input into a given stage. The process flows from left to right and top to bottom: so the first calibrations are the ones grouped in the \textit{Initial Calibrations} block (see \secref{processing:initial}), which are repeated after the crowding assessment
to ensure only the best suited data is used. The output of these calibrations are part of a database of calibrations that are needed in various stages of the process. The other block of calibrations is the one labelled \textit{Flux and LSF Calibration} (see \secref{processing:intcal}), which can only start after the initial calibrations are finalised. This is an iterative process that calibrates effect of different response, varying LSF across the focal plane and small deviations from the nominal differential dispersion functions. When all calibrations are defined, the final steps in the process produce the output catalogues of internally calibrated spectra. In this paper we focus on the mean source spectra (see \secref{processing:mean:spec}), produced using all the observations for a given source, while the process producing the epoch spectra (one calibrated spectrum per observed spectrum) is only briefly described in \secref{processing:exact:solution}. While the epoch spectra are not directly available in \gdr{3}, they contributed to the generation of mean reflectances for Solar System objects. 

\subsection{Initial calibrations}\label{sec:processing:initial}

Starting at the top left corner from the raw \xp data we find a first block of calibrations labelled \textit{Initial Calibrations}. Some of these have been described in previous papers \citep{Riello2018, Riello2021} because they are required also for the photometric processing: the computation of integrated \xp fluxes and Spectrum Shape Coefficients (SSCs) which are the input to the photometric processing together with the corresponding \gband fluxes require the application of the background and AL geometric calibrations. 

The background calibration for \gdr3 is a two-stage process: high resolution straylight maps are first generated to remove the effects due to diffraction from loose fibres in the sunshield \citep{Fabricius2016}; a k-nearest neighbour approach is then applied to the map residuals to describe the local astrophysical background (e.g. non-resolved sources, diffuse light from nearby objects, zodiacal light) at a resolution of about 25 arcsec.
More details about this calibration and a validation of the results are provided in \citealt{Riello2021} (see their Sect. 3.2).

Due to small inaccuracies in on-board detection and window assignment, sources are usually not perfectly centred within the acquired windows.
In order to be able to align spectra taken at different times and in different configurations for a given source we need to rely on a detailed geometric calibration, an accurate attitude reconstruction and high accuracy astrometry for all observed sources. Attitude and astrometry are inputs to the \xp processing, while the geometric calibration is a product of one of the calibration steps (\textit{AL} and \textit{AC Geometric Calibration} in Fig. \ref{fig:flow}).
The AL geometric calibration provides a correction in the AL direction to the location of a reference wavelength within the observed window as computed using our pre-launch knowledge of the CCD geometry. Once the reference wavelength is located within the window, this can also be used as reference position for the application of nominal differential dispersion functions that mitigates the difference in dispersion across the focal plane. More details about the AL geometric calibration can be found in \cite{Riello2018} and \cite{Carrasco2016}.
The AC geometric calibration is similarly defined as a correction to the predicted location on the source centroid in the AC direction as obtained from the pre-launch knowledge of the CCD geometry, the satellite attitude and the source astrometry. 

The two geometric calibrations (AL and AC) are required for the generation of accurate \xp transit time and AC coordinate predictions for all sources in the catalogue, the \textit{Scene Computation} in Fig. \ref{fig:flow}. An assessment of the crowding status of a given transit (the assessment needs to be done per transit rather than per source due to the overlapping of the two FoVs on the focal plane and the varying scan direction) cannot be purely based on the acquired surrounding windows. As we have already mentioned, crowding and priorities imply that a given source may not be assigned a window in the \xp CCDs, therefore such an assessment would be incomplete. This is why the scene is generated starting from the source catalogue containing objects that have been observed at all times during the mission operations so far. The astrometric information from the source catalogue is combined with the satellite attitude and with the geometric calibrations of the CCD of interest to generate the predictions. 
A detailed description of the scene computation and crowding evaluation has been included in \citep{Riello2021} due to its relevance in the generation of crowding information included in \gedr.

As shown in the schematic view, the Initial Calibrations are repeated after the Crowding Evaluation to include only data that has been assessed as not significantly affected by crowding thus minimising the disturbing effects of crowding on the calibrations. 
After this second run of the Initial Calibrations, the spectra are used to generate integrated \xp fluxes and Spectrum Shape Coefficients \citep[a set of \textit{ad hoc} filters designed for the photometric calibration, see][]{Riello2021}. At this point 2D spectra are marginalised in the AC direction to form 1D spectra and all subsequent processing only deals with 1D spectra.

\subsection{Internal calibrators}\label{sec:processing:calibrators}

Each calibration step normally relies on a specifically designed set of calibration data. For the background calibration, for instance, only Virtual Objects (empty windows acquired on a predefined pattern for calibration purposes) and observations of objects fainter than $G=18.95$ mag were used to avoid systematic effects due to the target source flux biasing the background measurement obtained from the first and last few samples in the window. For the AL geometric calibration, the need to find the best alignment of the spectra implies a requirement on their shape being approximately similar and therefore on the colour range of calibrators being quite narrow. For the AC geometric calibration finally, 2D spectra are essential to resolve the location of the peak in the flux distribution in the AC direction.

In the case of the Flux and Line Spread Function (LSF) calibration, the most important requirement is that all configurations are well covered by the set of calibrators. Calibrators covering more than one configuration are particularly valuable. This is naturally the case for time, FoV and CCD (sources are observed an average of about 40 times in the time range covered by \gdr{3}, in different FoVs and CCDs), while in the case of gates and WC only a limited subset of the calibrators will have observations in two or more observation configurations. These will be sources that have a magnitude close to the boundary of the magnitude range where that strategy is active and that, due to inaccuracies of the on-board magnitude estimate, may therefore be observed in different configurations in subsequent transits. 
The following criteria were tailored to ensure a clean but well populated set of calibrators. Only sources in the colour range $-2.0<(G_{\rm BP}-G_{\rm RP})<5.0$~mag and magnitude range $5.0<G<17.0$~mag based on the \gdr2 photometry were considered. Sources with \gband magnitude brighter than $11.5$~mag were selected as long as they had more than 10 transits in \xp, this is to ensure that the magnitude range where gates are activated is well covered. Sources fainter than $11.5$~mag with at least one 2D or gated observation were selected as long as their number of usable transits was larger than the median of the distribution of the number of transits in the same HEALPix pixel of level 6 minus the uncertainty estimated as the median absolute deviation of the distribution. This particular criterion was designed to avoid cases of faint sources that happened to be observed in a gated configuration because of their proximity to a bright object: in these cases, a large fraction of the transits of the faint source would be acquired with multiple gates (a case that is not currently processed) and would therefore not be usable. Only the few transits acquired when the two sources were observed at the same time would be usable. These would be likely to be significantly disturbed by the nearby bright source and therefore hardly suitable for calibration purposes. 
Finally, to enhance the fraction of sources with extreme colours (within the allowed range) with respect to sources of intermediate colours, the distribution of sources fainter than 11.5~mag that are only observed in ungated configuration and in 1D window strategy is flattened in colour as much as possible. Blue sources in particular are essential to constrain the internal calibration at short wavelengths and a poor calibration for blue sources may affect the absolute calibration given that the catalogue of external calibrators contains a large fraction of white dwarfs.
The colour flattening is achieved in ranges of magnitude and HEALPix pixels by considering the distribution in colour of the possible calibrators and selecting calibrators from the least populated colour ranges first: each time a number of calibrators are added to the list of selected calibrators from the least populated colour bin an equal number of calibrators are selected from each of the other colour ranges, giving priority to the sources with the largest number of transits. The process is repeated until the number of selected calibrators has reached the desired number of calibrators per HEALPix.
These criteria generated a list of internal calibrators including about 7.6M objects.

As very blue sources are naturally rare, during the calibration process measurements coming from sources from less populated areas of the colour magnitude diagram were given larger weight in the least squares solution of the calibration. These additional source-based weights were computed from the density of calibrators in the colour-magnitude diagram and were only applied for the calibration of the BP instrument.

\subsection{Flux and LSF calibration}\label{sec:processing:intcal}

The flux and LSF calibration model has been described in detail in \cite{Carrasco2021}. This calibration has been defined to take into account sensitivity differences, LSF variations, deviations from the nominal dispersion function and AC flux loss. However, flux loss terms were not activated for the processing that lead to \gdr3. The calibration model describes the over all effect of these different aspects on the \xp spectra.

It is useful to recall here Eq. 9 from \cite{Carrasco2021} as the basic formulation of the Flux and LSF calibration: 
\begin{equation}
    h_{s,k}(u_i) = \sum_{n=0}^{N-1} b_{s,n} \sum_{j=-J}^J A_{k}(u_i, u_{i+j}) \ \varphi_n(u_{i+j})\label{eq:intcal}
\end{equation}
which describes the observed spectrum of source $s$ in calibration unit $k$, $h_{s,k}$, as a discrete convolution via the instrument model $A_{k}$ of the mean spectrum. The mean spectrum is in turn defined as a linear combination of some basis functions $\sum_{n=0}^N b_{s,n}\varphi_n$. In the following, basis functions and bases will be used interchangeably. Here $u$ refers to a pseudo-wavelength system, close to the AL coordinate of the samples within a window but adjusted for AL geometry and differential nominal dispersion function. We use $u_i$ to indicate the coordinate of sample $i$ in the pseudo-wavelength system and consequently $h_{s,k}(u_i)$ is the flux measured in the sample $i$, corrected for effects calibrated in the initial calibration stage (see Sect. \ref{sec:processing:initial}). In this formulation, all the information about the individual source \xp spectra is encoded in the $\vec{b_{s}}$ coefficients, while the $A_{k}$ describes the instrument properties. The spectra available in \gdr{3} are in this format (see \secref{outputs} for more details on the archive content).

The discrete convolution kernel $A_k$, the actual calibration, describes the transformation to be applied to the mean spectrum to predict an observation in calibration unit $k$. Only differential effects between the reference system and the calibration unit it refers to are calibrated in this process. These include contributions from LSF, response and dispersion. The calibration $A_k$ depends on both the pseudo-wavelength of the sample $i$ that the model is trying to predict and the pseudo-wavelength of the sample $i+j$ that is contributing to the discrete convolution. As explained in \cite{Carrasco2021}, given the expected smooth behaviour of $A_k$ across the pseudo-wavelength range, the discrete kernel is replaced by a linear combination of polynomial bases. A smooth variation of the calibration with AC coordinate (within a CCD) is ensured by defining the coefficients of the polynomial in pseudo-wavelength to be a polynomial in AC coordinate \citep[see Eq. 13 in][]{Carrasco2021}. A quadratic dependency with the pseudo-wavelength and a cubic dependency in AC coordinate were used for \gdr3, where the AC coordinate refers to the centre of the window for both 1D and 2D spectra.
Given the size of the LSF \citep[see Fig. 5 in][]{Carrasco2021} and of the expected deviations from the nominal dispersion function, only contributions from neighbouring samples are expected to be significant. Two adjacent neighbours on each side (i.e. $J=2$ in \equref{intcal}) were considered in the processing leading to \gdr3. 
The number of neighbours and the possibility of introducing a step between neighbours have been adjusted during trial runs to offer the best balance between residuals and number of calibration parameters.

At the start of the calibration process, both the mean spectra for the internal calibrators (the $\vec{b_{s}}$ coefficients) and the instrument calibrations ($A_k$) are unknown. An identity calibration is therefore assumed to compute a first set of reference mean spectra for the internal calibrators, effectively solving for the $b_{s,n}$ parameters the simplified equation 
\begin{equation}
    h_{s,k}(u_i) = \sum_{n=0}^N b_{s,n} \ \varphi_n(u_{i})\label{eq:meanspec}
\end{equation}
The resulting mean spectra are then used to solve for a first set of calibrations $A_k$, using \equref{intcal}. With these in hand we can update the reference mean spectra, by solving again the same \equref{intcal} for the $\vec{b}_s$ coefficients. The process then proceeds via iterations. The step where the mean spectra are solved for is called \textit{Source Update}, while the one where the calibrations are computed is the \textit{Instrument Calibration}. When solving for the BP or RP mean spectrum for a given calibrator, all its observed spectra in that calibration unit need to be collected and used to set up the least squares problem. When solving for the instrument calibration of a specific calibration unit instead, all observed spectra for the calibrators that happened to be observed in that calibration unit and their corresponding mean spectra need to be combined to form the least squares problem. This iterative algorithm had been developed using the Map/Reduce paradigm \citep{MapReduce} which provides a simple parallelisation model; the Hadoop implementation provided a very efficient horizontally scalable I/O and processing capacity \citep[see e.g.][]{Riello2018}. Since the algorithm described above requires grouping the data in two different ways (by source, when producing the mean spectrum, and by calibration unit when solving the instrument model), the implementation required two
Map/Reduce jobs to perform a single iteration. Although the execution time of individual iterations was quite reasonable, the cost of running a large number of iterations and testing different configuration parameters for the instrument model proved to be the main limitation of this approach. For iterative algorithms, such as the one required for the instrument model computation, a better alternative to Map/Reduce has proven to be Apache Spark\footnote{\url{https://spark.apache.org}} which was used for the \gedr photometric processing. For \gdr{4} the iterative instrument model solution will be ported to Spark, allowing for in-memory iterations between source update and instrument model which will dramatically reduce the cost of running large number of iterations.

\begin{figure}[!ht]
    \centering
    \includegraphics[width=0.48\textwidth]{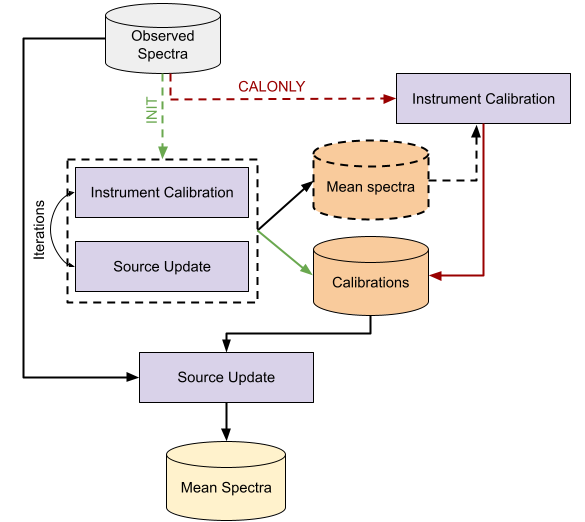}
    \caption{Flow diagram of the flux and LSF calibration process. Dashed arrows show the flow of calibrators data (also the corresponding mean spectra dataset is shown with dashed borders). When applicable the labels INIT or CALONLY have been added to indicate that only data from the corresponding time ranges is being used by a given process.}
    \label{fig:flow:intcal}
\end{figure}

Given the large systematic effects present in the data due to water-based contamination in the payload \citep{Prusti2016}, particularly at the start of the mission, and the discontinuities caused by the various decontamination campaigns aimed at reducing those effects, also for the \xp processing and in analogy to what was done for the \gedr photometry \citep{Riello2021}, the iterations aimed at initialising the reference system were restricted to use only data collected during a specific time range, chosen to have the lowest and most stable contamination level. We refer to this time period as INIT. The ranges adopted are approximately $[2574.7, 2811.7]$ and $[4121.4, 5230.1]$ in \obmtrev \citep[these are the same used for the photometric processing, see][]{Riello2021}. This effectively implies that the set of calibrators is defined not as a list of sources but as a list of observations, restricted in a specific time range and to a specific subset of sources. 
A consequence of this is that at the end of the iterative process described above, only instrument calibrations covering the INIT period will be available. Calibrations for all the other periods (collectively called CALONLY) can be computed with a final Instrument Calibration step using all the observed spectra from the CALONLY time ranges for the sources used as calibrators combined with their reference mean spectra. This is shown in the flow diagram in Fig. \ref{fig:flow:intcal} where dashed lines are used for calibrators' data and the labels INIT and CALONLY indicate the time ranges covered by each calibration step.
When calibrations are available to cover the entire time range, a final Source Update using all observed spectra for all sources, not only calibrators, produces the catalogue of mean spectra.

It should be mentioned that in all steps of this process, weighted least squares solutions are obtained via QR-decomposition using Householder reflection to ensure numerical robustness \citep{fvl2007}. Each solution is computed iteratively: at a given iteration, we use the solution computed at the previous iteration to reject observations that have residuals larger than $5\sigma$. Sample flux measurements are weighted by the inverse variance computed from the flux error for each sample. In the last run of the source update, the one that applies the instrument calibration to all observations to generate the catalogue of mean spectra, sample flux errors are re-scaled taking into account the scatter in the normalised residuals to mitigate the effects of error underestimation in the wings of the spectra.

\subsubsection{Exact solution}\label{sec:processing:exact:solution}

Calibrations can also be \textit{applied} to a single observed spectrum to obtain an internally calibrated epoch spectrum. This process appears as \textit{Exact Solution} in the schematic overview in \figref{flow}. In this case the system of equations to be solved is 
\begin{equation}
    h_{s,k}(u_i) = \sum_{j=-J}^J A_{k}(u_i, u_{i+j}) \ g_s(u_{i+j})\label{eq:exactsol}
\end{equation}
where $\vec{g_s}$ is the output internally calibrated epoch spectrum and $A_k$ is the instrument calibration for the calibration unit $k$ of the observed spectrum $\vec{h_{s,k}}$ being calibrated. In this case, the solution is simply obtained by inverting the matrix representing the instrument calibration and the resulting spectrum has the same sampling (in terms of number of samples and their location in pseudo-wavelength space) as the observed spectrum, as opposed to the mean spectrum that being defined as a linear combination of some analytic bases is effectively a continuous function in pseudo-wavelength.
The instrument calibration matrix $A_k$ was generally non-singular and the inversion could be done successfully. Only very few epoch spectra could not be calibrated using this procedure.

Epoch spectra are particularly valuable for objects that vary in time (either due to intrinsic variability or due to different distance or orientation such as is the case for Solar System objects). For these type of objects the mean spectrum will be ill-defined.
Although epoch spectra are not included in \gdr3, they are relevant here because they have been the input to the generation of the reflectances for Solar System objects.

\subsubsection{Calibrations}

Calibrations are obtained in time intervals or scopes of about $20$ \obmtrev (corresponding to about 5 days) for most calibration units. Only for the shortest-exposure configurations, with Gate 05 or Gate 07 active, due to the much smaller number of calibrators in these magnitude ranges, it has been necessary to extend the length of the time intervals to about $100$ \obmtrev ($\sim25$ days). The length of the time intervals will vary slightly between calibrations due to the few events that cause discontinuities in the calibrations \citep[such as decontamination campaigns and refocus events, see also][]{Riello2021}. As within a time scope the calibration is assumed to be constant in time, time scopes need to be defined so that such events happen at the boundary between two subsequent intervals.

A set of calibration parameters was solved for each of the $31860$ calibration units. For Gate 05 and Gate 07 the number of nominal calibration units was 1064 per gate configuration, while for other gate configurations or in the ungated case the number of nominal calibration units was 5708 (the ungated case having twice as many as the others because of the two possible window strategies active for objects with magnitude fainter than 11.5~mag). This implies a total of $24960$ nominal calibration units, however there are often cases of non-nominal configurations that get a sufficient number of observations to allow a robust calibration. These are cases of faint sources being observed with a gate triggered by a nearby bright source being observed at the same time (see also Fig. \ref{fig:gates}). 

\begin{figure}[!ht]
    \centering
    \includegraphics[width=0.4\textwidth]{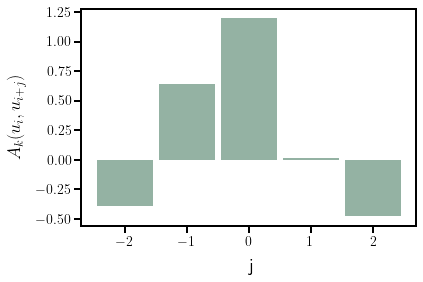}
    \caption{The $A_{k}(u_i, u_{i+j})$ values defining the instrument calibration for one specific configuration (RP, CCD row 1, preceding FoV, ungated, 1D) in the time range including \obmtrev $5000$ evaluated at $u_i=30.0$ and AC coordinate 1000.}
    \label{fig:aij:example}
\end{figure}

Displaying detailed information for such a large number of calibrations is challenging. To facilitate this we have defined two parameters describing each calibration: 
\begin{itemize}
    \item One is defined as the sum over $j$ of the $A_{k}(u_i, u_{i+j})$ values weighted by the distance between $u_i$ and $u_{i+j}$. In the case of a perfectly symmetric calibration (seen here as a convolution kernel) this would be 0. In general it indicates the location of the \textit{peak} of the kernel. A skewed kernel might be caused by small deviations from the nominal dispersion.
    \item The second parameter is given by the sum over $j$ of all $A_{k}(u_i, u_{i+j})$ values, i.e. the \textit{integral} of the kernel. Variations in this parameter show differences in the response across the focal plane and between different calibration units. 
\end{itemize}
Figure \ref{fig:aij:example} shows an example of the calibration for a given calibration unit, evaluated in the central part of the spectrum and of the CCD. This particular case has the peak parameter equal to $-0.80$ and the integral parameter equal to $0.98$.

\begin{figure*}[!h]
    \centering
    \includegraphics[width=0.235\textwidth]{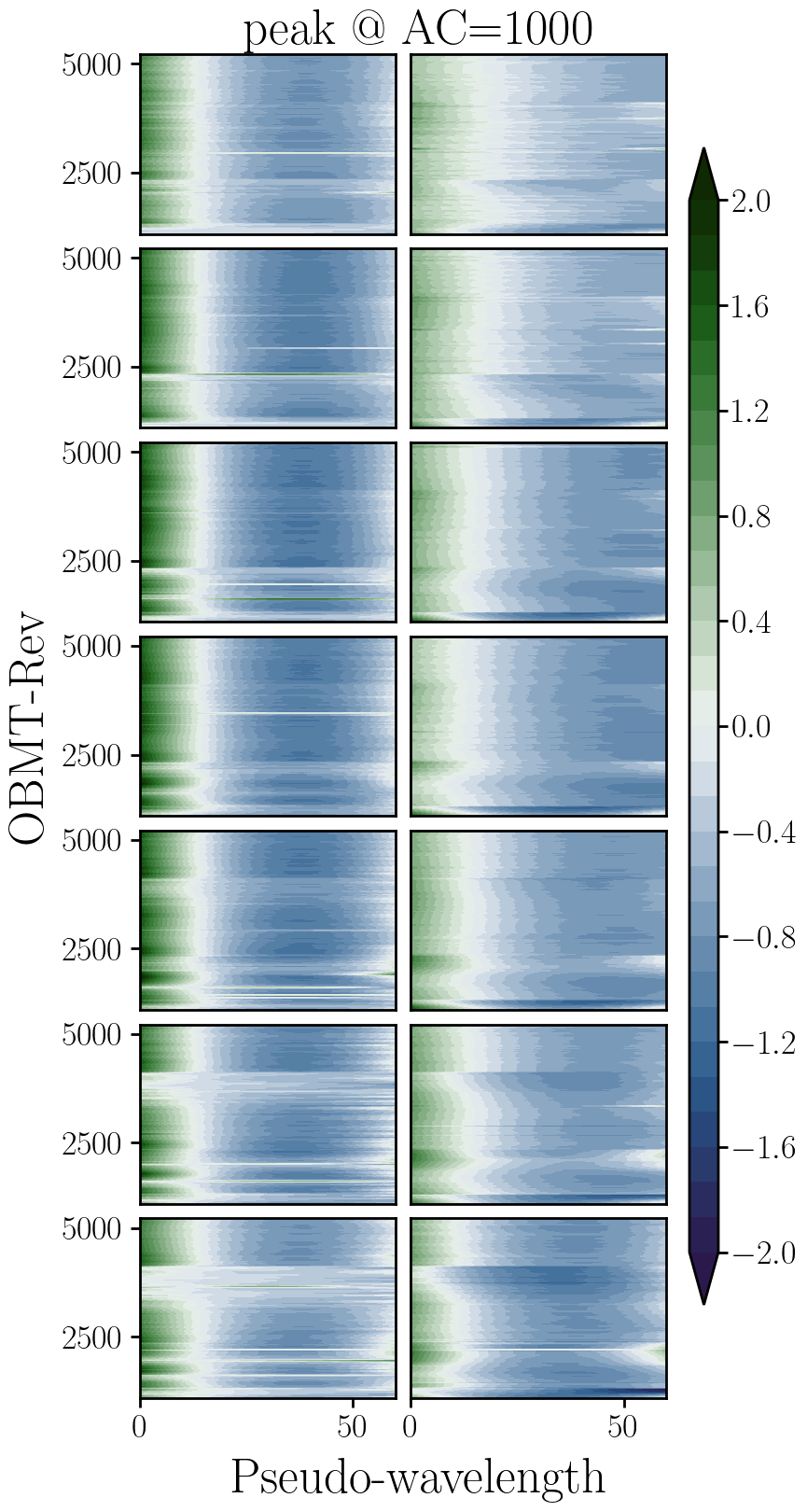}
    \includegraphics[width=0.235\textwidth]{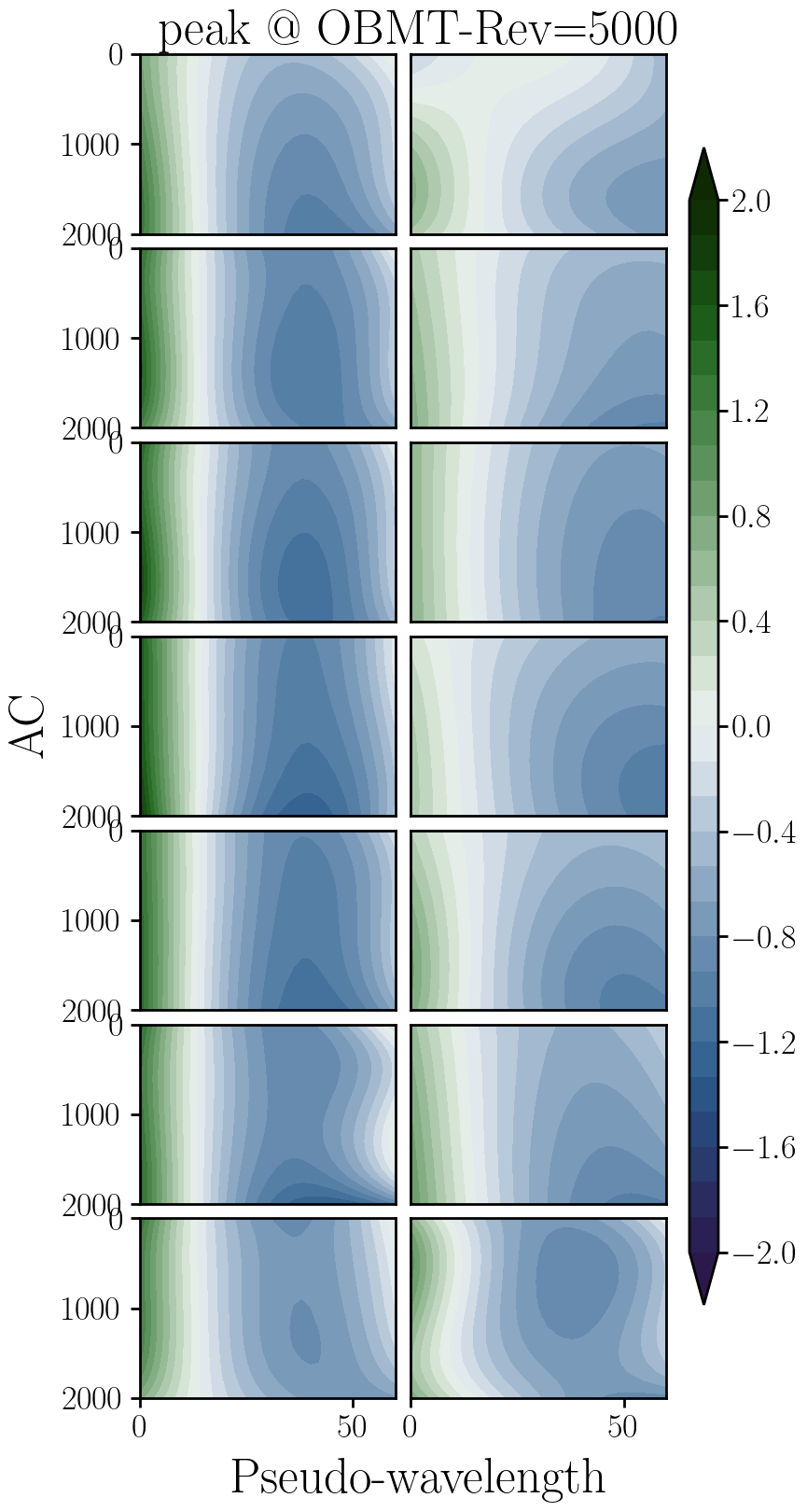}
    \includegraphics[width=0.23\textwidth]{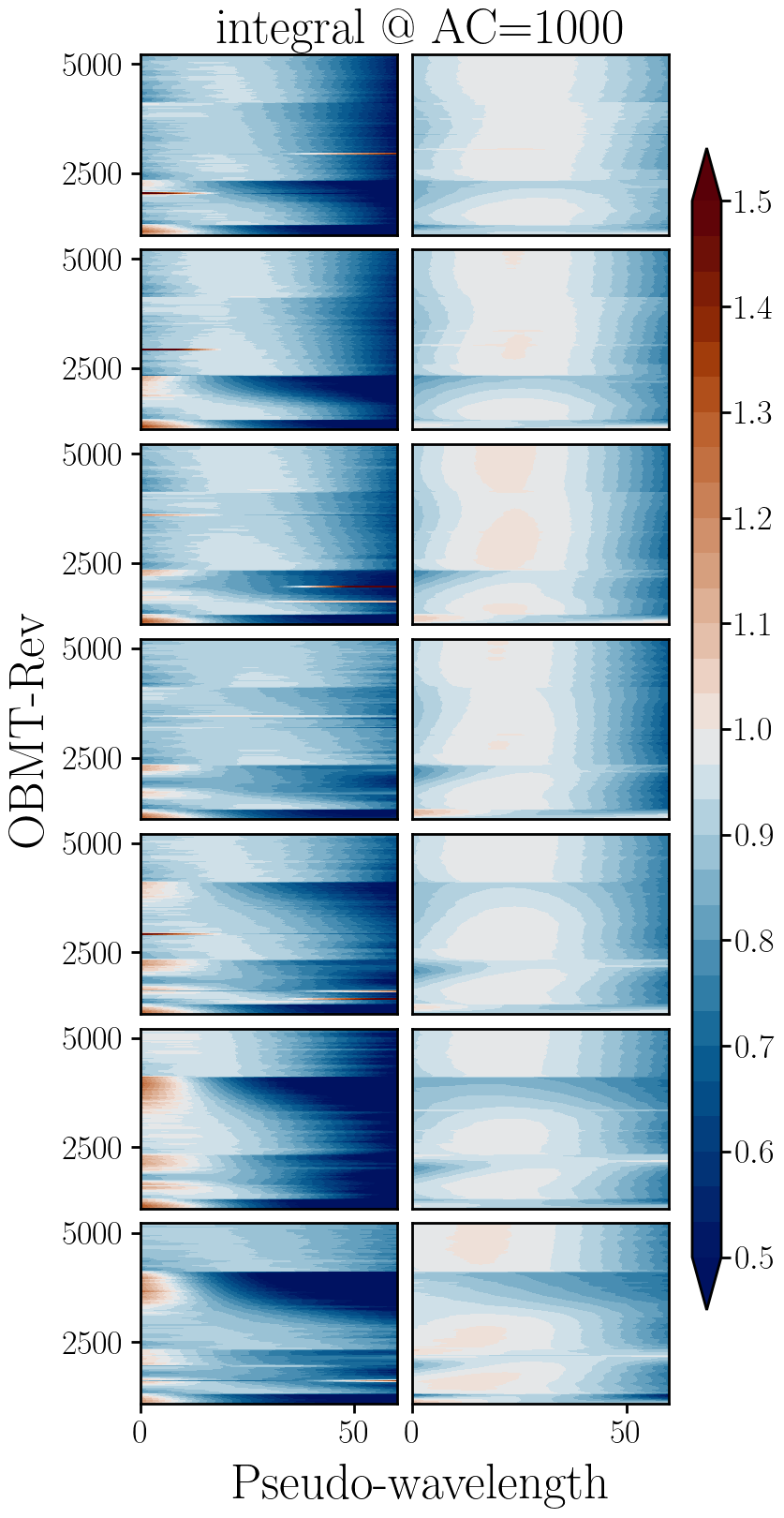}
    \includegraphics[width=0.23\textwidth]{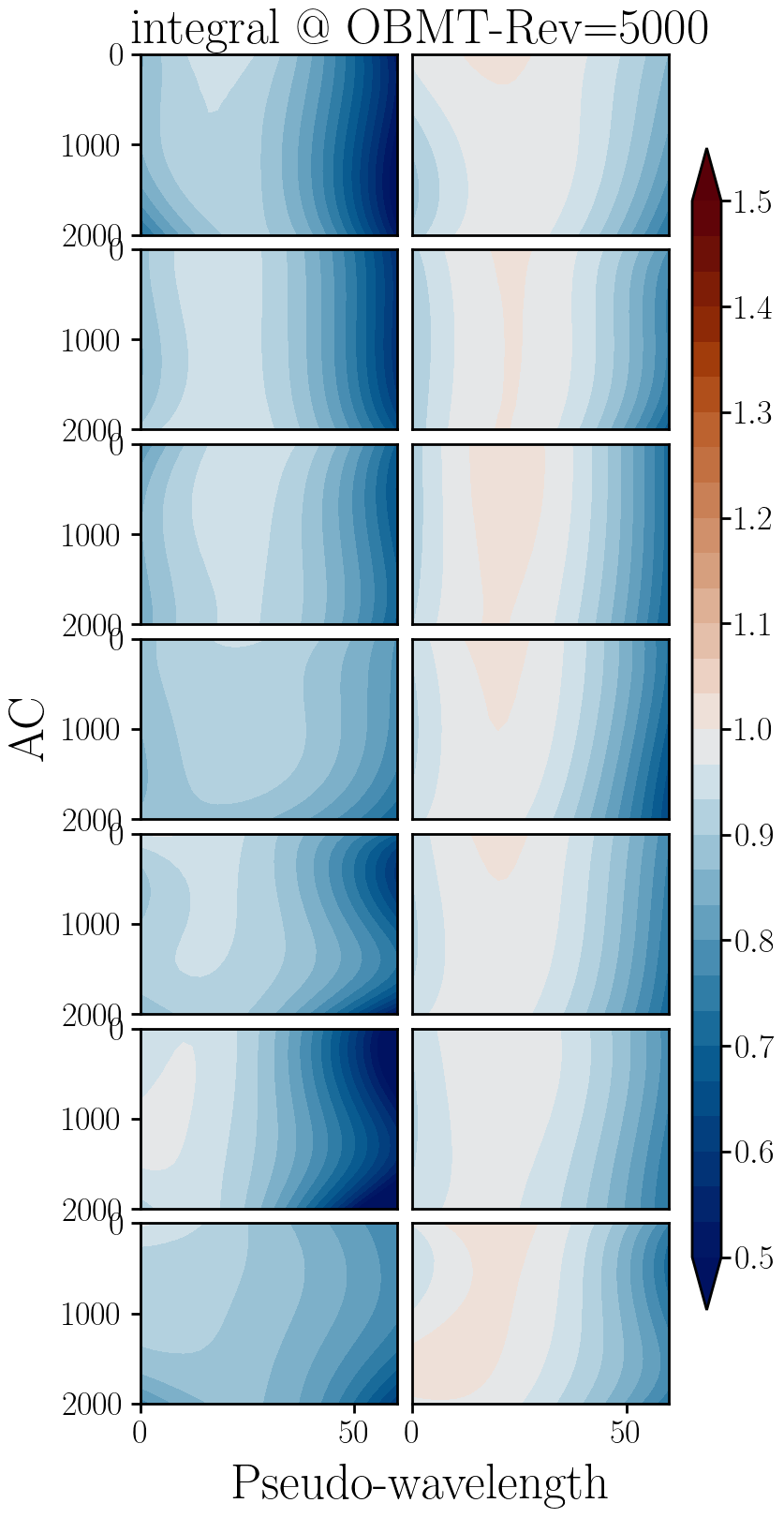}
    \includegraphics[width=0.235\textwidth]{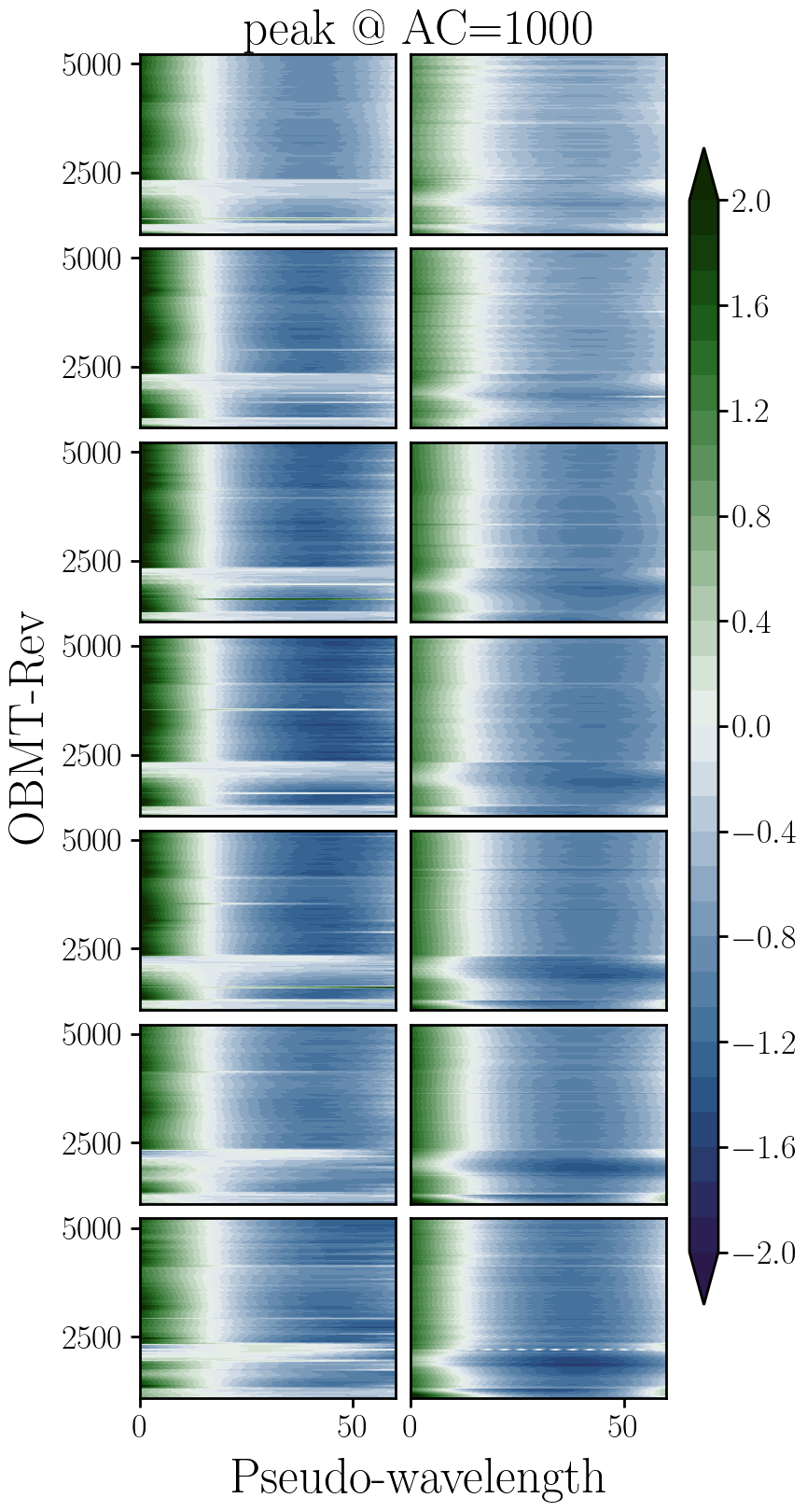}
    \includegraphics[width=0.235\textwidth]{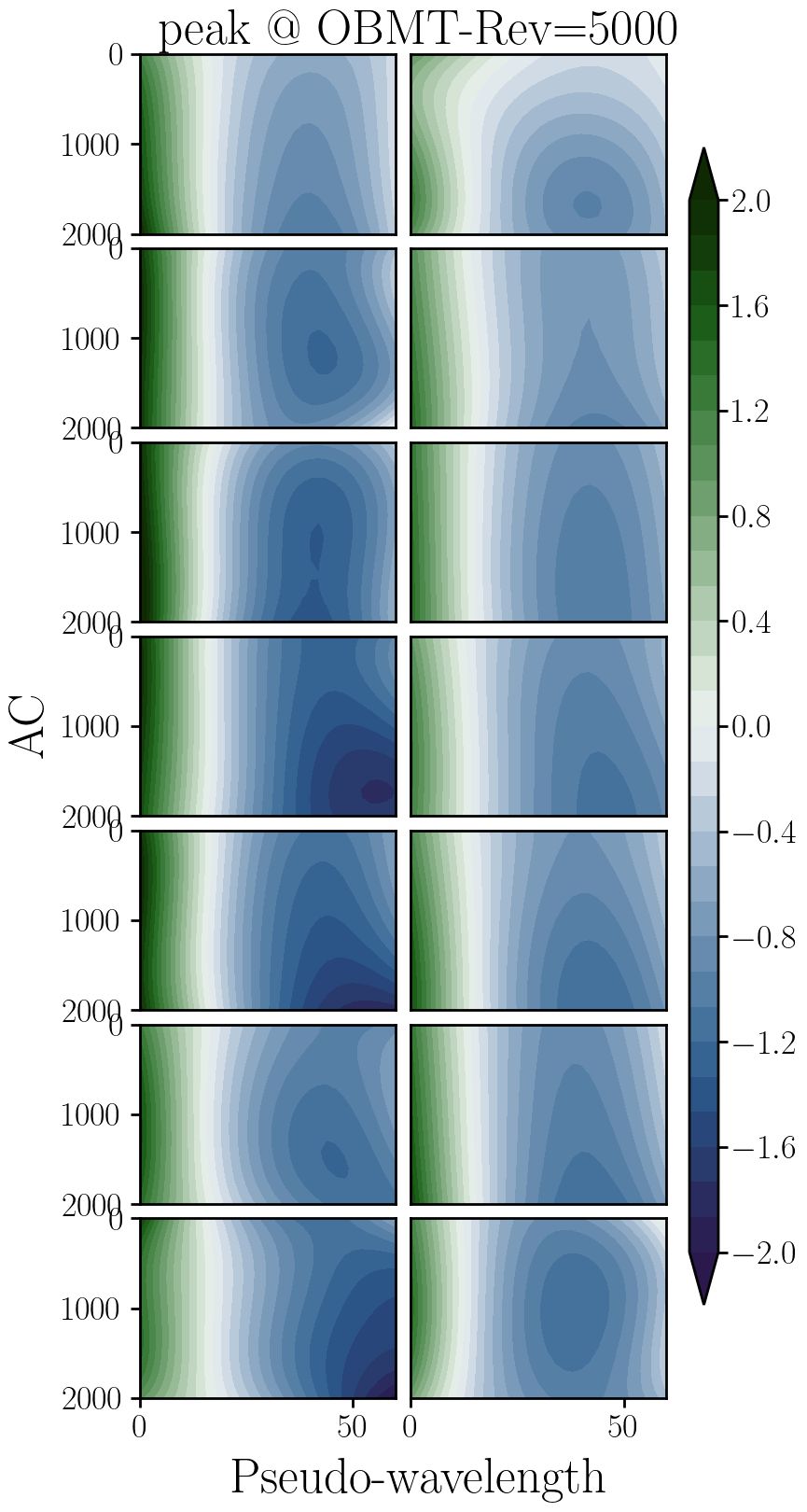}
    \includegraphics[width=0.23\textwidth]{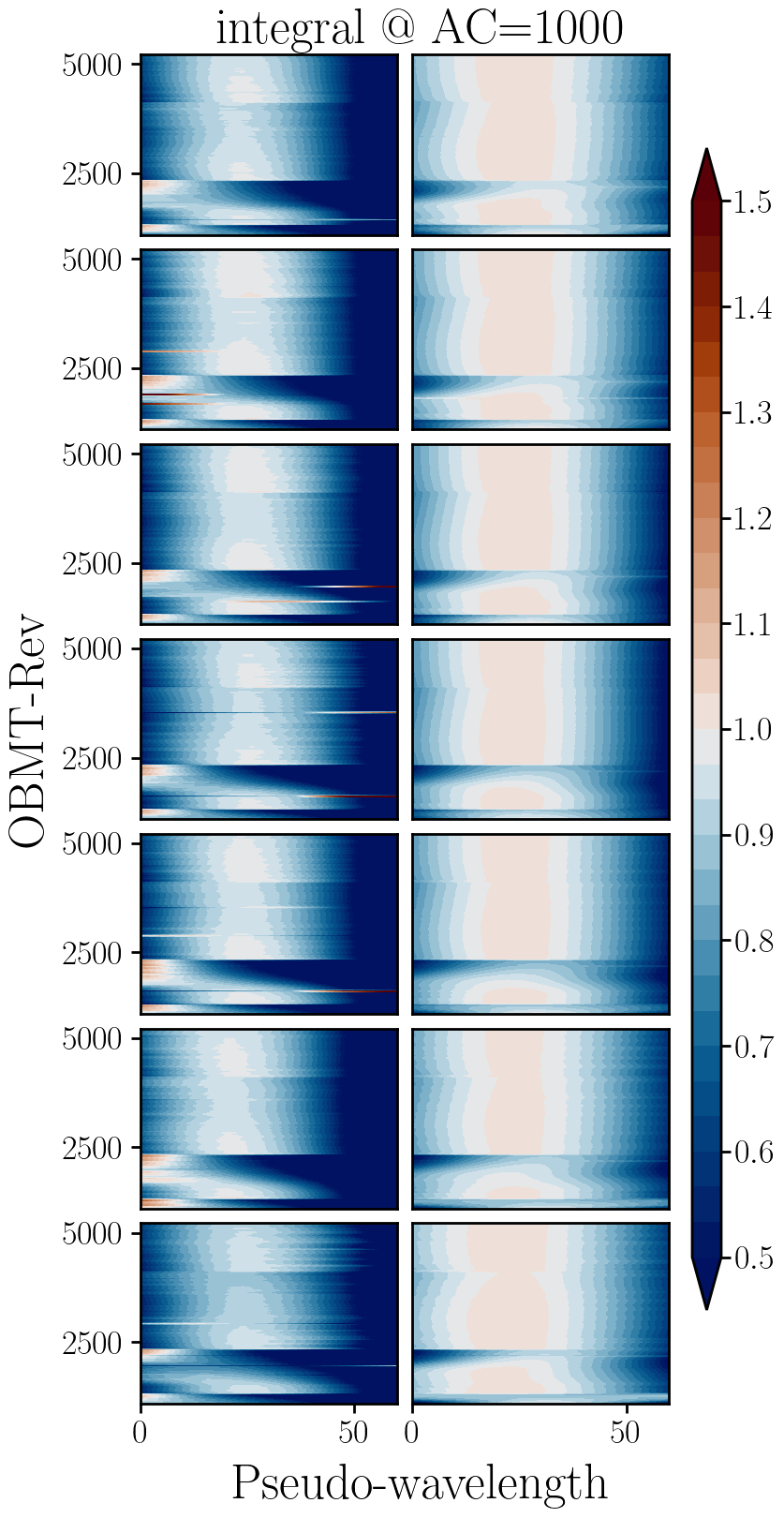}
    \includegraphics[width=0.23\textwidth]{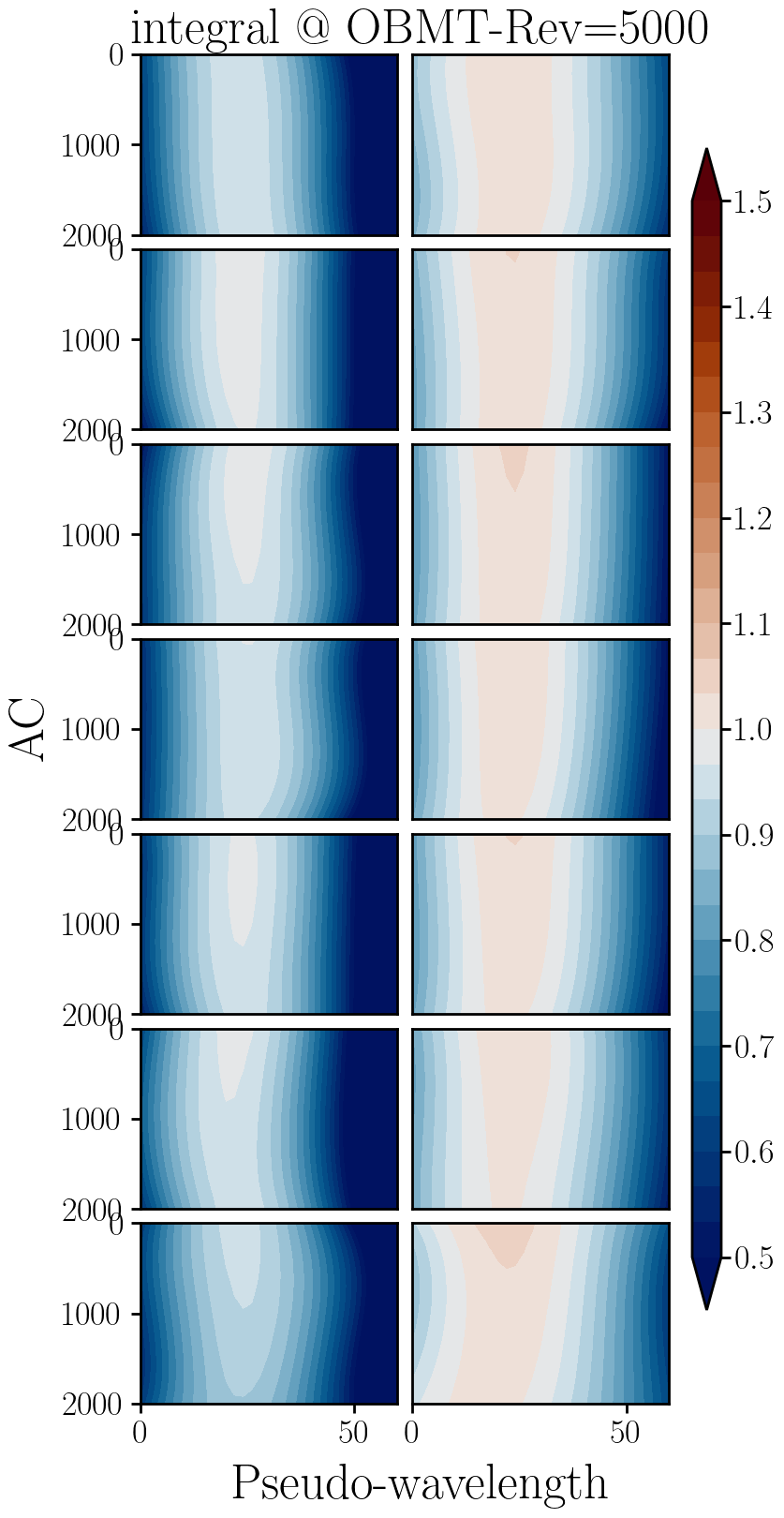}
    \caption{Overview of the BP and RP calibrations for the preceding (first row of plots) and following (second row) FoVs, ungated 1D configuration: peak and integral parameter variations vs wavelength, time and AC coordinate are shown for each CCD. Each set of 14 panels show the peak (first two sets) and integral (second two sets) variations (see the top title label and colour bar next to each set) as a function of different parameters: the first set shows the variation of the peak parameter in time (expressed in OBMT-Rev) and pseudo-wavelength, while the second set shows the variation of the same parameter in AC coordinate and pseudo-wavelength, the third and fourth sets show the same dependencies for the integral parameter. When showing the dependency in time and pseudo-wavelength the parameters have been evaluated at the centre of each CCD in the AC direction (i.e. AC=1000), while when showing variations with AC coordinate and pseudo-wavelength the reference time \obmtrev=5000 was used. Within each set, the 14 panels show the BP case in the left column of 7 panels (one per CCD) and the RP case in the right column.}
    \label{fig:aij:ungated1D}
\end{figure*}

The plots shown in Fig. \ref{fig:aij:ungated1D} offer a quick view of the calibrations computed for the preceding and following FoVs, ungated and 1D configuration in terms of the two parameters just defined.
The first row of plots refer to the preceding FoV while the second shows the following FoV calibrations. Starting from left, the first two sets of 14 panels show the variation of the peak parameter with the AL position $u_i$ (and therefore wavelength) and time in OBMT-Rev or AC coordinate in all the BP (first 7 panels, one panel per CCD) and RP (second column of 7 panels) CCDs; the following two sets of 14 panels show the variation of the integral parameter with respect to the same dependencies.
Several discontinuities can be observed in the time variation of these parameters. Most of these can be traced back to particular events during the mission, such as decontamination campaigns and refocus activities.
The strong variations in the BP calibrations and in particular in the integral parameter vs AL position and time are mostly linked to the varying level of contamination due to water-based contaminant present in the payload \citep{Prusti2016}, that affects BP more strongly than other instruments \citep[see Fig. 8 in][where the effect of contamination on the photometry in \gband, \gbp and \grp is compared]{Riello2018}.

\begin{figure*}[!h]
    \centering
    \includegraphics[width=0.25\textwidth]{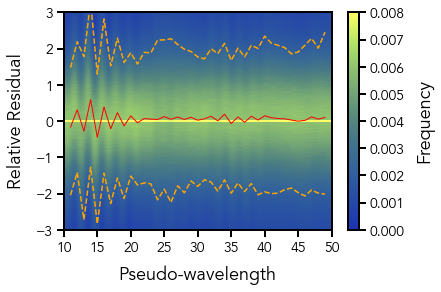}
    \includegraphics[width=0.25\textwidth]{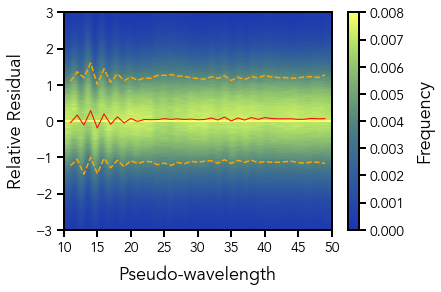}
    \includegraphics[width=0.24\textwidth]{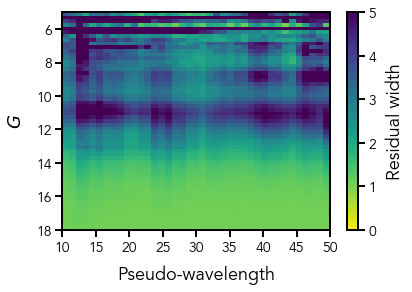}
    \includegraphics[width=0.24\textwidth]{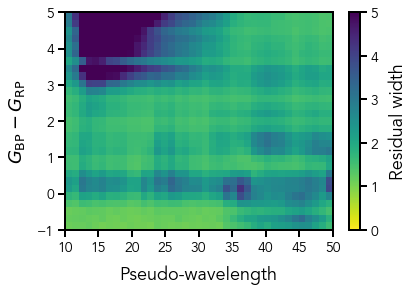}
    \includegraphics[width=0.25\textwidth]{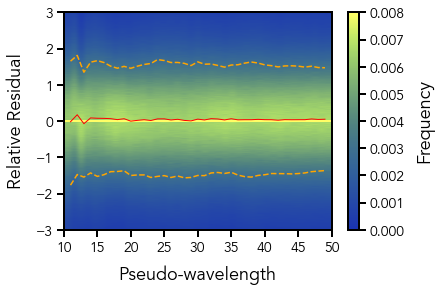}
    \includegraphics[width=0.25\textwidth]{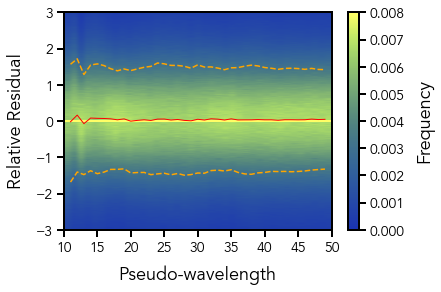}
    \includegraphics[width=0.24\textwidth]{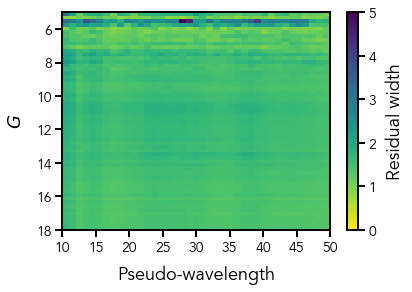}
    \includegraphics[width=0.24\textwidth]{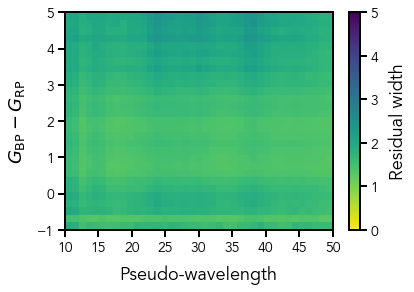}
    \caption{Relative residual distribution for a subset of the calibrators covering the \gband magnitude range $[5, 18]$. The first row of plots shows the BP results, while the bottom row shows RP. In each row the first plot shows the distribution of relative residuals vs AL coordinate in the range $[10, 50]$ where most of the flux is observed. In the second plot the same distribution is shown including only data from sources in the magnitude range $[13, 17]$. In these first two plots the 2D histogram is normalised to the number of measurements in each column and the relative number of sources is shown by the colour bar. The red line shows the median value, while the orange dashed lines show the 15.865 and 84.134 percentiles. The following two plots show the robust width of the distribution of relative residuals defined as the difference between the 84.134 and 15.865 percentiles divided by 2 vs \gband magnitude and \bprp colour and AL coordinate for the entire magnitude range covered by this subset.}
    \label{fig:aij:res}
\end{figure*}

Relative residuals computed for a random subset of the calibrators (about $50$ thousand sources) are shown in Fig. \ref{fig:aij:res} for BP and RP. For each observed spectrum, relative residuals are computed as the difference between the observed flux value and the predicted value (computed applying the calibration to the source mean spectrum) divided by the observation flux error. Residuals from all observations and all sources in this dataset are accumulated in a grid in $u_i$, magnitude and colour to analyse residual dependencies.
From these plots it is evident that the performance of the internal calibration for the BP data varies significantly over the wavelength range covered and with magnitude and colour. Sources brighter than $G=12.5-13$ and in particular red bright sources show a much larger spread in relative residuals. Performances in RP have a much smoother behaviour across all parameters. The additional weights based on the relative frequency of sources in the colour magnitude diagram (see Sect. \ref{sec:processing:calibrators}) is likely to be the cause of this. We remind readers that source-based weights were only adopted for the BP calibration to boost the leverage of rare blue sources and help calibrating the bluest wavelength range where only very blue sources have significant flux. 
This may have affected the calibration process, particularly in magnitude ranges where the number of blue sources is very small, due to the natural magnitude and colour distribution of sources in the sky: in these cases a few blue outliers might adversely affect the solution.

\begin{figure}[!h]
    \centering
    \includegraphics[width=0.45\textwidth]{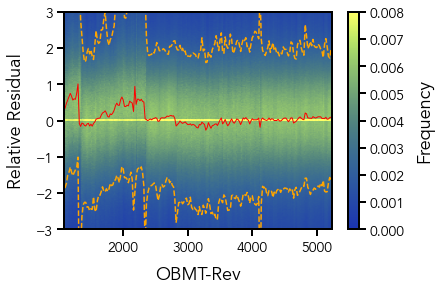}
    \includegraphics[width=0.45\textwidth]{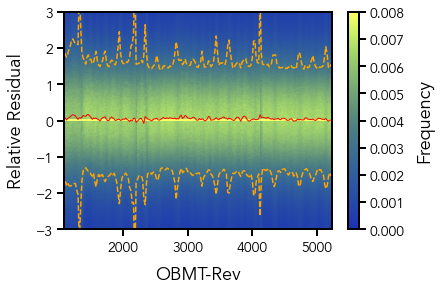}
    \caption{Relative residual distribution for a subset of the calibrators covering the magnitude range $[5, 18]$. The top panel shows the BP residuals, while the bottom one shows the RP residuals. Only samples with AL coordinate in the range $[10, 50]$ are included in this plot. The 2D histogram is normalised to the number of measurements in each column.}
    \label{fig:aij:res:time}
\end{figure}

The plots in Fig. \ref{fig:aij:res} include only data and calibrations for the INIT period. As explained before, once a stable set of calibrations for the INIT period have been obtained and a reference set of mean spectra for the calibrators is established, this is used to generate consistent calibrations covering all the rest of the mission data collected so far. The distribution in time of the relative residuals covering the whole period included in \gdr{3} is shown in Fig. \ref{fig:aij:res:time} for BP and RP in the top and bottom panels respectively.
The top panel shows that the calibration algorithm was not able to fully remove the large systematics affecting the BP data due to the contamination in the early phases of the mission. Considering the long period of time with minimal contamination available, it was decided to ignore all BP data collected before the decontamination event that took place shortly before \obmtrev 2340 when generating the final catalogue of mean spectra.

\subsubsection{Convergence}

Convergence of the iterative process was monitored looking at different parameters: the median standard deviation of the solutions, the overall absolute change in parameters and the average $\chi^2$ of the residuals for a subset of the calibrators were all considered.

\begin{figure*}[!h]
    \centering
    \includegraphics[width=0.95\textwidth]{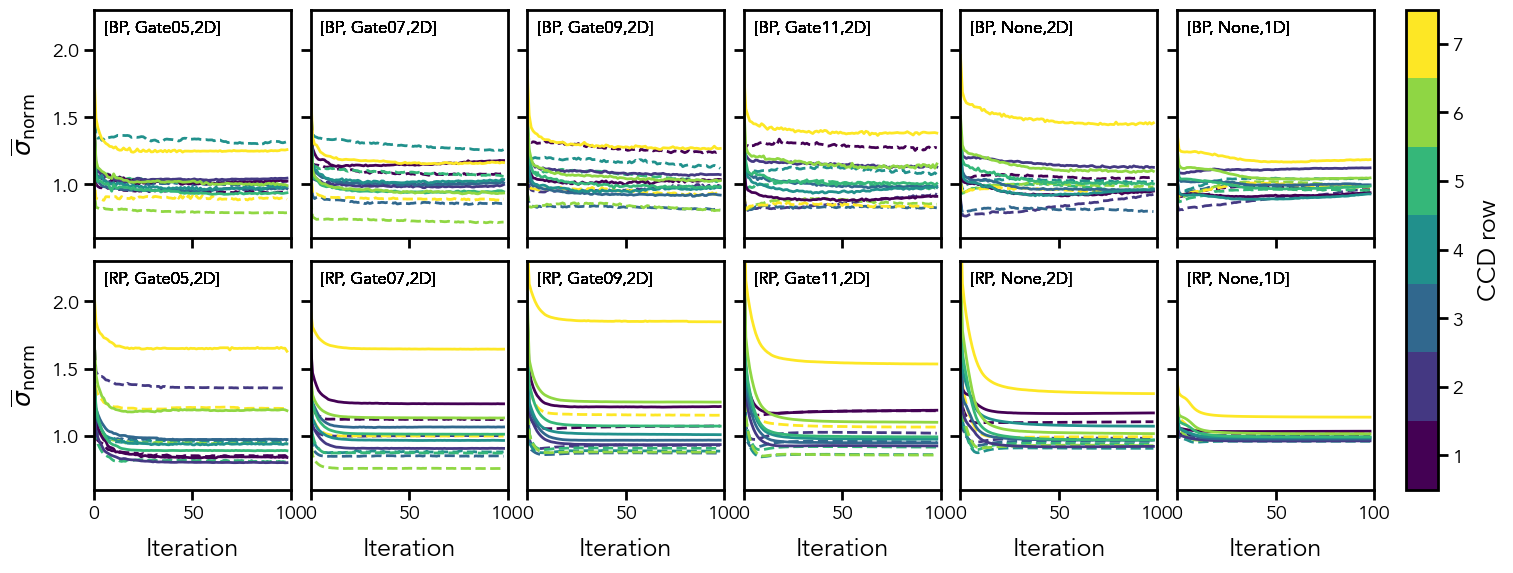}
    \caption{Median standard deviation for all solutions covering the OBMT-Rev range [3000, 4000], normalised to the median standard deviation of all calibrations obtained for the same photometer (BP/RP), gate and window class at iteration 50 (by that iteration the system seems to have become quite stable). The top panels show the BP solutions, one panel per nominal combination of gate and window class. The bottom panels show the RP solutions. Different colours indicate different CCD rows and solid and dashed lines are used for the preceding and following FoV respectively.}
    \label{fig:aij:conv:std}
\end{figure*}

Each least squares solution for a calibration unit is assigned a standard deviation. 
The normalised median standard deviation of all least squares solutions over the OBMT-Rev range [3000, 4000] grouped by gate and window class combination versus iteration number is shown in Fig. \ref{fig:aij:conv:std}. Each panel shows a combination of photometer (BP/RP), gate and window class as indicated in the label. 
There are some configurations where the evolution of the median standard deviation is not monotonically decreasing, particularly in the first few iterations. If the calibration of each configuration was solved independently, one would expect the corresponding standard deviation to decrease in subsequent iterations.
However, in the iterative process described in Sect. \ref{sec:processing:intcal}, all calibration units are linked together by the common catalogue of reference spectra that is updated at each source update. For this reason, the fact that the standard deviation does not decrease for all configurations is not showing lack of convergence over all. 

\begin{figure*}[!h]
    \centering
    \includegraphics[width=0.95\textwidth]{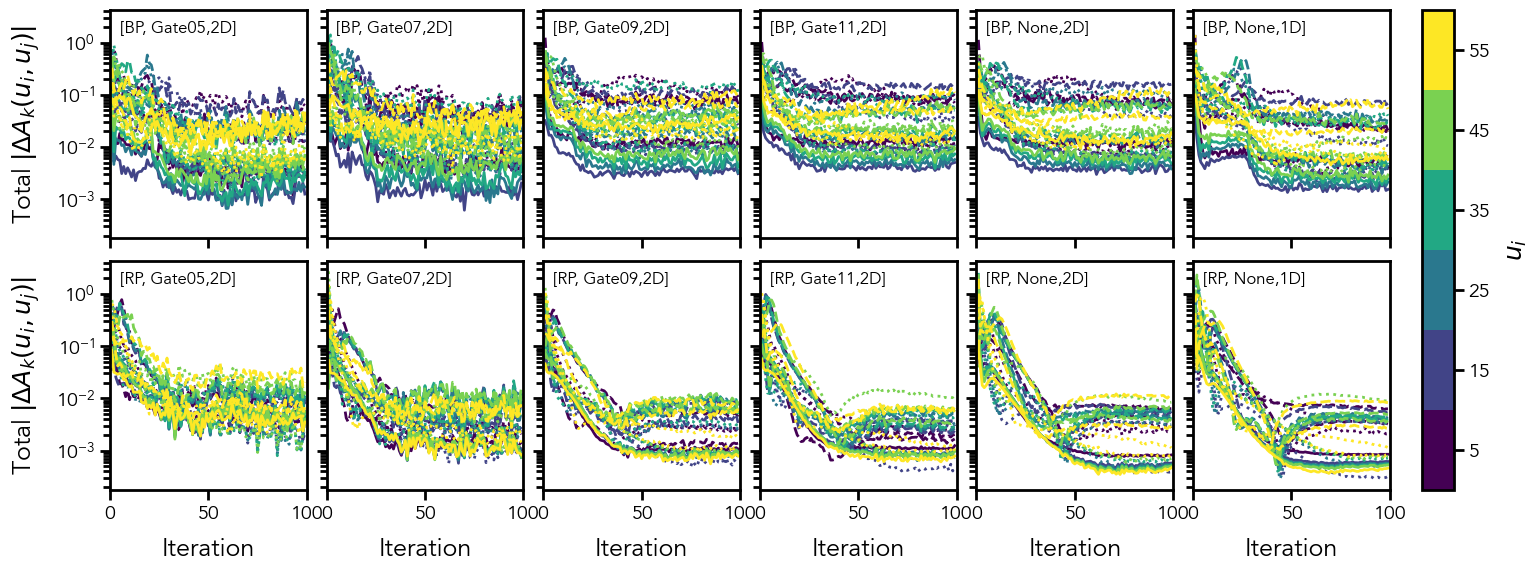}
    \caption{Absolute relative change in the values of model parameters between two subsequent iterations for all solutions covering the OBMT-Rev range [3000, 4000] in a logarithmic scale. The relative change for each parameter is computed as the absolute difference between the values at two subsequent iterations, normalised by the value of the same parameter at the preceding iteration.
    The top panels show the BP solutions, one panel per nominal combination of gate and window class. The bottom panels show the RP solutions. Different colours indicate different ranges of $u_i$ and solid, dashed and dotted lines are used for the central sample ($j=0$) and neighbouring samples ($j=\pm1$ and $j=\pm2$) respectively.}
    \label{fig:aij:conv:delta}
\end{figure*}

Overall convergence is assessed looking at the absolute relative change in the values of model parameters $A_k$ between two subsequent iterations. Figure \ref{fig:aij:conv:delta} shows how these evolve during the iterations for different nominal combinations of gate and window class in BP (top panels) and RP (bottom panels). Given the large number of parameters, only results for ROW4 are shown here, other rows showing similar trends. Different colours are used for different pseudo-wavelength and different line styles for different neighbours (the solid line being used for the central sample).
The absolute relative change in calibration parameters are over all at or below $1\%$ well before iteration 50, particularly for the central part of the spectra and for $j=0$. For BP there seem to be larger relative changes (at about the $10\%$ level) in the wings of the spectra and for $j\ne0$. This is not completely unexpected and is probably due to correlations between the parameters. 

\begin{figure}[!h]
    \centering
    \includegraphics[width=0.45\textwidth]{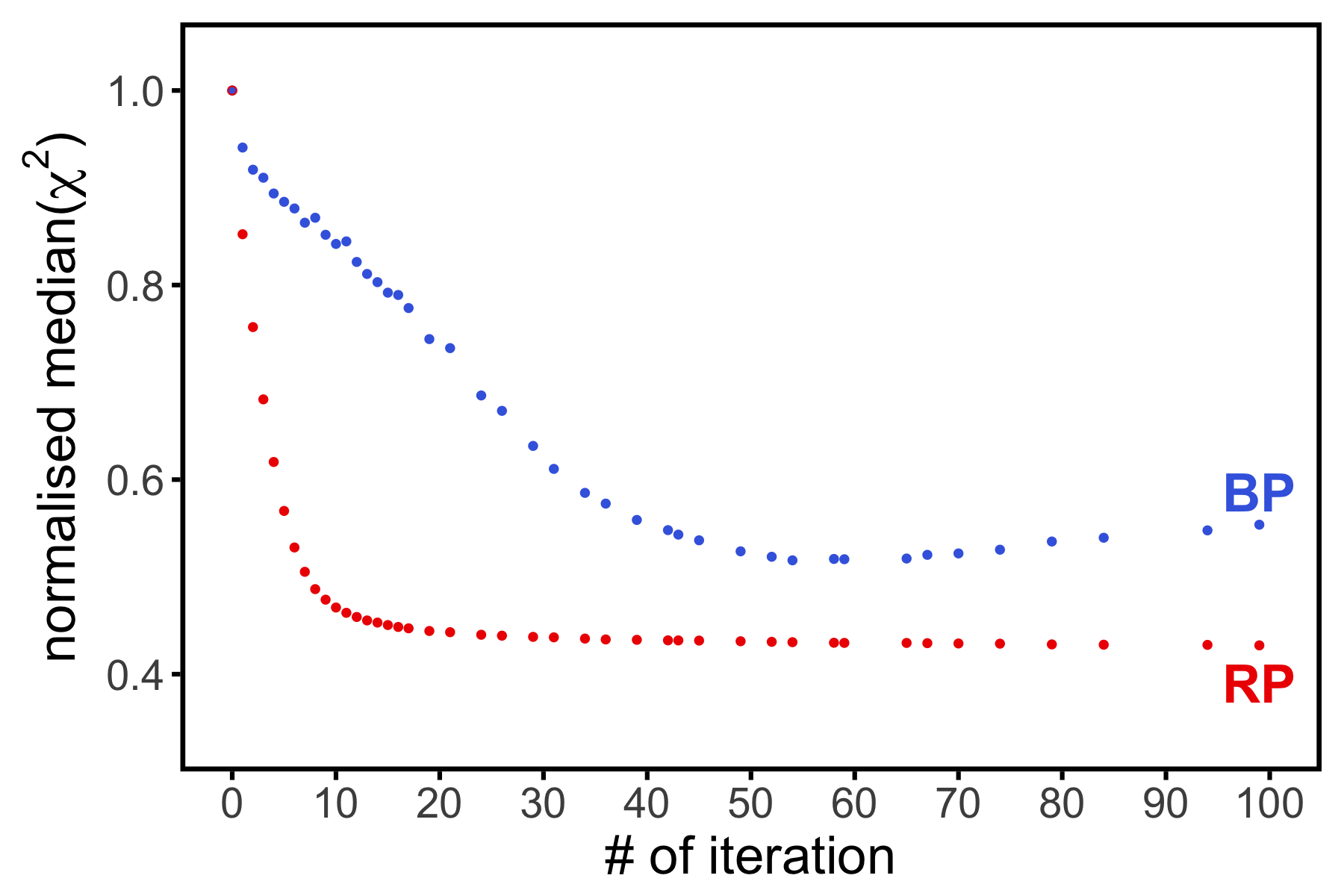}
    \caption{Normalised median $\chi^2$ for a subset of about 50K calibrator sources with respect to the iteration number. Blue and red symbols show the BP and RP residuals respectively.}
    \label{fig:aij:conv:chisq}
\end{figure}

Finally, Fig. \ref{fig:aij:conv:chisq} shows the evolution through the iterations of the normalised median $\chi^2$ for the same random subset of the calibrators used for which residuals where shown in Sect. \ref{sec:processing:intcal}, in blue and red symbols for BP and RP respectively. In this plot the normalised median $\chi^2$ value
at each iteration is obtained by dividing the corresponding median $\chi^2$ to the value at the first iteration. The $\chi^2$ value for each epoch spectrum is given as the sum of squared residuals between the observed spectrum and the predicted spectrum divided by the observed flux error.
It is important to point out that the normalised median $\chi^2$ shown here is not the quantity that is being minimised within the iterative process, which will be the sum of squared residuals for all observations of all calibrators within each calibration unit when solving the instrument model and the sum of squared residuals for all observations of each calibrator when solving the source update step. The increase in late iterations for BP shown in Fig. \ref{fig:aij:conv:chisq} could be due to changes in the distribution of $\chi^2$ due to the iterations trying to catch a few extreme outliers at the expense of slightly degrading the residuals for other sources.

There are indications from both the standard deviation and $\chi^2$ analyses that in late iterations the solutions start diverging. We have mentioned a possible cause but this is not fully understood. The additional weighting introduced to give more leverage to blue sources seems to have an effect in this respect. Alternative strategies are being considered in future data releases. 
From the analysis of all criteria, iterations 55 and 40 were finally adopted for BP and RP respectively to proceed with the generation of a reference catalogue of mean spectra to be used for the calibration of the CALONLY data. 

\subsection{Mean spectra representation}\label{sec:processing:mean:spec}

Once the internal reference system has been established by the flux and LSF calibration and calibration solutions are available covering all calibration units, a final source update is run including all observed spectra to generate the catalogue of mean spectra that are released as part of \gdr{3}.
The algorithms described in this section have been applied only to this last run of the source update.

\subsubsection{Internal reference system}

The flux and LSF calibration procedure described in Sect. \ref{sec:processing:intcal} leads to the definition of an internal reference system. This can be seen as an average instrument. The monitoring of intermediate results during the iterative process showed that in late iterations some of the spectral features in mean spectra assumed a smoother, shallower shape with respect to what is observed in the predicted and observed spectra. In order to maintain the reference system and the corresponding mean spectra as close as possible to the actual instrument and to the actual data, we have decided to use instead a specific \textit{epoch instrument} and to represent the final mean spectra as observed in this system. The epoch instrument was chosen somewhat arbitrarily to be the one corresponding to CCD row 7 for BP and row 5 for RP at a time equal to $4500$ in OBMT-Rev.

To avoid having to invert the instrument model to derive mean spectra directly in this new system, we have computed a transformation matrix $T$ where each row $k$ contains the coefficients that need to be applied to the canonical Hermite function bases to reproduce the prediction of the $k$-th basis in the chosen epoch instrument. 
These are the result of a fit of each predicted basis function, obtained by applying Eq. \ref{eq:intcal} to a mean spectrum where only one coefficient is equal to 1 while all others are 0, with the same set of 55 Hermite function bases.
In the new system, the mean spectra are defined by the array of coefficients $\vec{b'}$ computed by multiplying the transformation matrix by the array of coefficients in the starting reference system $\vec{b}$, i.e. $\vec{b'}=T\,\vec{b}$. The covariance matrix of the source update least squares solution needs also to be converted by computing $C'=T\,C\,T^T$ where $C$ is the covariance matrix in the starting reference system and $C'$ is the covariance matrix in the new system.

\subsubsection{Bases function optimisation}\label{sec:processing:optimisedbases}

As described in \cite[see Sect. 5]{Carrasco2021} and introduced in Sect. \ref{sec:processing:intcal}, the source mean \xp spectrum is described as a combination of basis functions. At the start of the calibration process little is known about the instrument and therefore a generic set of basis functions is used throughout the initialization phase. Hermite functions, i.e. Hermite polynomials multiplied by a Gaussian, were used in this stage: they provide an orthonormal set of basis functions, are centred around 0 and allow to increase details and range by adding higher order bases. They also tend to 0 for sufficiently high absolute values of the independent variable. This resembles the behaviour of \xp spectra where the combination of CCD efficiency and response ensures that the measured flux tends to 0 for increasing distance from the source location. 

We denote the $n-$th Hermite function $\varphi_n(x)$. In order to make the Hermite functions efficient in representing the \xp spectra, a linear transformation between the pseudo-wavelength and the argument of the Hermite functions is required. This transformation includes a shift $\Delta \theta$ such that the Hermite functions are centered approximately on the centre of the spectra, and a scaling factor $\Theta$ that adjusts the width of the Hermite functions to the width of the spectra to be represented. Furthermore, a suitable number of Hermite functions needs to be chosen. The \xp spectrum of a source $s$, $f_s(u)$, is then represented by the linear combination
\begin{equation}
    f_s(u) = \sum\limits_{n=0}^{N-1} b_{s,n} \, 
    \varphi_n \left(\frac{u - \Delta \theta}{\Theta}\right)
    \label{eq:mean:spec:bases}
\end{equation}
In Eq. \ref{eq:intcal} the mean spectrum $f_s(u)$ appeared as $\sum_{n=0}^{N-1} b_{s,n}\,\varphi_n$. Here we have made explicit the transformation of the pseudo-wavelength $u$ into the argument of the Hermite functions $\varphi_n$.
The values of $\Theta$, $\Delta \theta$, and $N$ cannot be chosen independently from each other. Since the pseudo-wavelength range covered by most \xp spectra is $[0,60]$, a value of $\Delta \theta$ around 30 is required to center the Hermite functions on the spectra. Furthermore, the linear combination of Hermite functions need to cover the range from $-30$ to $30$. Increasing the number of Hermite functions used in the representation results in the coverage of a wider range of arguments, while increasing the scaling factor results in a reduction of the range of arguments \citep{Carrasco2021}. To find a suitable combination, we first determined the values of $N$ for $\Delta \theta = 30$, for values of $\Theta$ from 2 to 3.5, such that the local minimum or maximum at the largest value of $u$ of the $N-1$-th basis function is closed to 30. For all resulting combinations of $\Theta$ and $N$, a fixed number of five iterations of the instrument calibration was performed. A random subset of approximately 50 thousand internal calibrators was used for this purpose. The total residuals in the epoch spectra were then computed and compared for different combinations of parameters. We selected the combination of parameters that resulted in the smallest value for the summed squared residuals. In both, BP and RP, $N = 55$ is used, implying that $55$ coefficients will be available for each \xp spectrum in \gdr{3}. The values for $\Theta$ and $\Delta \theta$ are slightly different for BP and RP, with $\Theta = 3.062231$ for BP and $3.020529$ for RP, and $\Delta \theta = 30.00986$ for BP and $30.00292$ for RP. The slight deviations from round numbers result from adjusting the parameters to the smallest and largest values in pseudo-wavelength in the set of internal calibrators used. 

Once the catalogue of mean spectra for the calibrators is established based on the set of standard Hermite functions, the set of bases can be optimised to improve the efficiency of the representation. This is achieved when most of the information is contained in the coefficients for the lowest-index bases and allows reducing the number of coefficients required to describe each spectrum by dropping coefficients that are within the noise. 

The optimisation algorithm used normalised mean spectra for the subset of calibrators already used to define the best configuration for the standard Hermite functions. 
L2 normalisation was used to ensure equal weights for sources of different magnitude in the decomposition.
The $N$ coefficients representing each of these sources in the canonical set of bases are normalised with respect to their $l_2$-norm and are used to populate a matrix $M \times N$ where $M$ is the number of sources. Singular Value Decomposition of this matrix gives the orthogonal matrix $V$ that represents a rotation of the canonical Hermite bases into a new set of optimised bases.

\begin{figure}[!h]
    \centering
    \includegraphics[width=0.48\textwidth]{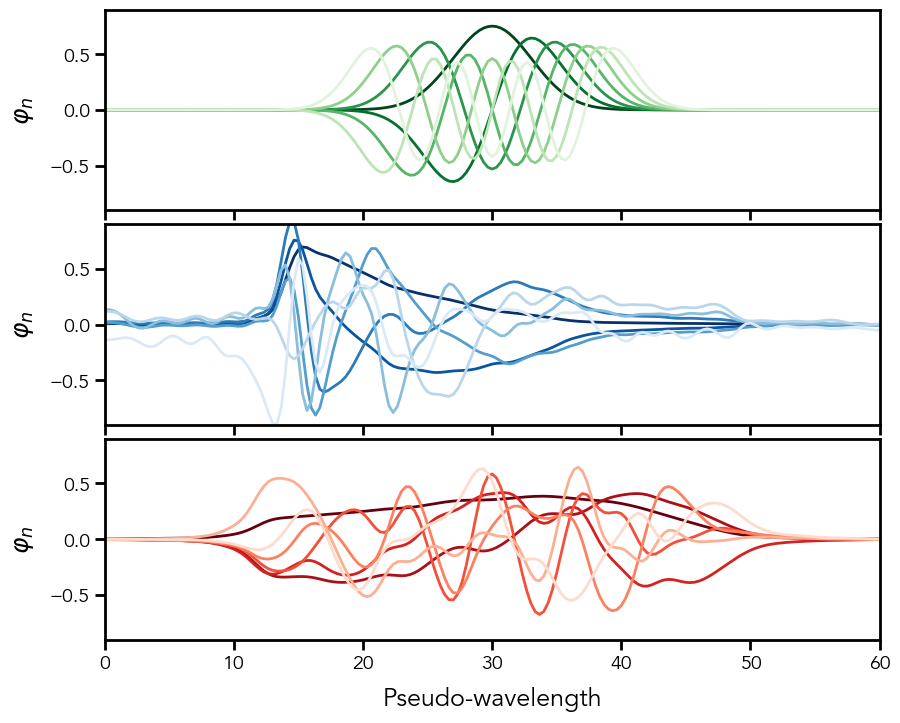}
    \caption{Comparison between the first few canonical Hermite function (top panel), BP (middle panel) and RP (bottom panel) optimised bases.}
    \label{fig:ms:bases}
\end{figure}

Figure \ref{fig:ms:bases} shows the first few bases in the canonical Hermite function set (in the top panel) and in the optimised BP and RP sets of bases (in the following two panels).
Darker shades are used for lower-index bases. The first optimised bases, being tailored to the actual spectra, reproduce the \textit{average} spectrum and exhibit the imprint of the transmission curve. Higher order bases become increasingly complex with narrower wavy structures required to fit the sharpest features in the spectra.  

\subsubsection{Truncation}\label{sec:truncation}

As explained in Sect. \ref{sec:processing:optimisedbases}, by expressing the mean spectra in terms of an optimised set of basis functions, a particular spectrum is essentially described by a small number of basis functions with low indices. The coefficients corresponding to higher order basis functions have small absolute values, and, taking their errors into account, are close to zero. Their effect in representing an \xp spectrum is therefore essentially adding noise, which manifests itself in wavy structures in the sampled spectrum. It is therefore of interest to suppress the insignificant high order coefficients and with it, reduce the noise on the spectra.\par
A simple criterion to decide whether a number of high-order coefficients is insignificant or not has been suggested by \cite{Carrasco2021}. The criterion is based on the standard deviation of the $M$ coefficients with the highest indices, i.e. the coefficients with indices ranging from $N-M$ to $N-1$. If the standard deviation of the $M$ coefficients with highest indices, normalised to their errors, is above a specified multiple of the standard error of the standard deviation above the expected mean standard deviation, it is assumed that the coefficients are not only random values consistent with a mean of zero, but are actually contributing significantly to the spectrum. Otherwise, they are considered insignificant and can be set to zero. For the standard deviation of a set of $M$ samples from a normal distribution with zero mean we assume the simplified expression of $1/\sqrt{2(M-1)}$, and a mean of one. Thus, if the standard deviation of the $M$ coefficients with highest indices, divided by their error, is smaller than $1+x/\sqrt{2(M-1)}$, with $x$ being an adjustable threshold, the coefficients are assumed to be consistent with being zero, and can be truncated. We used a value of $x=2$, and for each \xp spectrum, progressively increasing values of $M >2$ were tested for truncation until the truncation threshold is exceeded for the first time for some $M$. If the truncation threshold is never reached, i.e. all coefficients are considered being consistent with being zero, the full number of $N=55$ is kept. This happened for a small number of sources, in particular for BP spectra of faint and very red sources, where the flux in the BP spectrum is so low that it is indeed essentially consistent with being only noise.\par
This criterion makes two simplifications. First, the assumed mean and standard deviation is inaccurate for very small numbers of $M$. However, the resulting overestimation of the truncation threshold is on the level of a few per cent in the worst case, and has no significant impact on the truncation levels. Second, the truncation ignores correlations between the errors on the coefficients. For sources for which the optimised basis was constructed, the correlations are indeed very low, and the negligence is justified. This is by number the vast majority of sources. For sources for which the optimised basis is less efficient, correlations might however be larger, and the truncation unreliable. This is in particular the case for extremely red sources, or sources with spectral energy distributions that are very different from typical stellar spectral energy distributions, such as QSOs or sources with strong emission lines. In the latter case the truncation is to be used with caution, as it might affect the representation of narrow spectral features.\par
\begin{figure}[!h]
    \centering
    \includegraphics[width=0.48\textwidth]{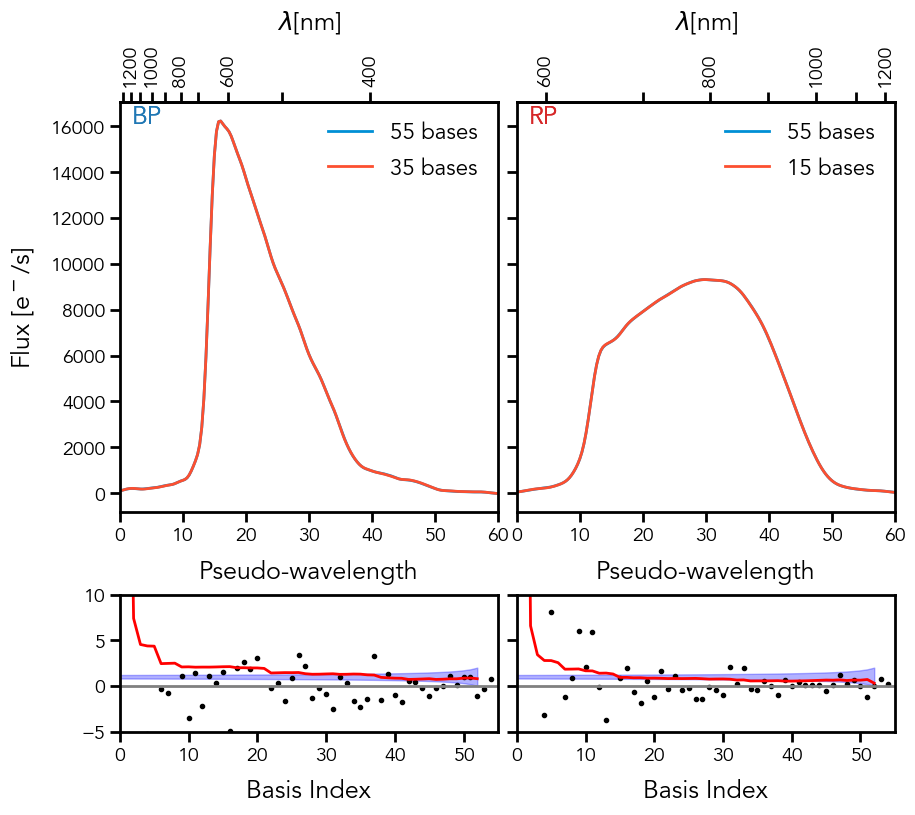}
    \caption{ 
    Sampled BP (left) and RP (right) spectra are shown in the top panels for source  \texttt{Gaia\,DR3\,6210089815971933056} ($G \approx 11.5$~mag and $\bprp\approx 1.0$~mag). Each panel contains two curves: a blue curve showing the non-truncated spectrum using all 55 coefficients, a red curve showing the truncated spectrum. The number of coefficients used for each spectrum is given in the label within the plot. The bottom panels show the truncation assessment. This is run independently for BP and RP. The black circles indicate the coefficients normalised by their formal errors, the red line shows the standard deviation of the $M$ normalised coefficients, starting from $M=3$ on the right hand side. The blue shaded region is the cone given by $1\pm 2/\sqrt{2(M-1)}$. }
    \label{fig:ms:truncation:bright}
\end{figure}

In the following, we illustrate the effect of truncation for four example cases. First, we consider the case of a typical, bright star ($G \approx 11.5$~mag and $\bprp\approx 1.0$~mag) in Fig.~\ref{fig:ms:truncation:bright}. The top panels compare the sampled BP and RP spectra, represented by all 55 coefficients, and by the number of coefficients considered significant according to the procedure described above. These numbers of coefficients are 35 and 15 for BP and RP, respectively, for this example source. No difference in the sampled spectra is visible to the eye, although the number of basis functions used in the representation of the sampled spectrum is significantly smaller. The bottom panels of the figure illustrate the truncation process. The black symbols show the values of the coefficients, normalised to their errors. The red curve shows the standard deviation of the $M$ normalised coefficients, starting from $M=3$ on the right hand side. The blue shaded region is the cone given by $1\pm 2/\sqrt{2(M-1)}$. When the red curve exceeds the blue cone, the corresponding number of coefficients is considered significant.\par
\begin{figure}[!h]
    \centering
    \includegraphics[width=0.48\textwidth]{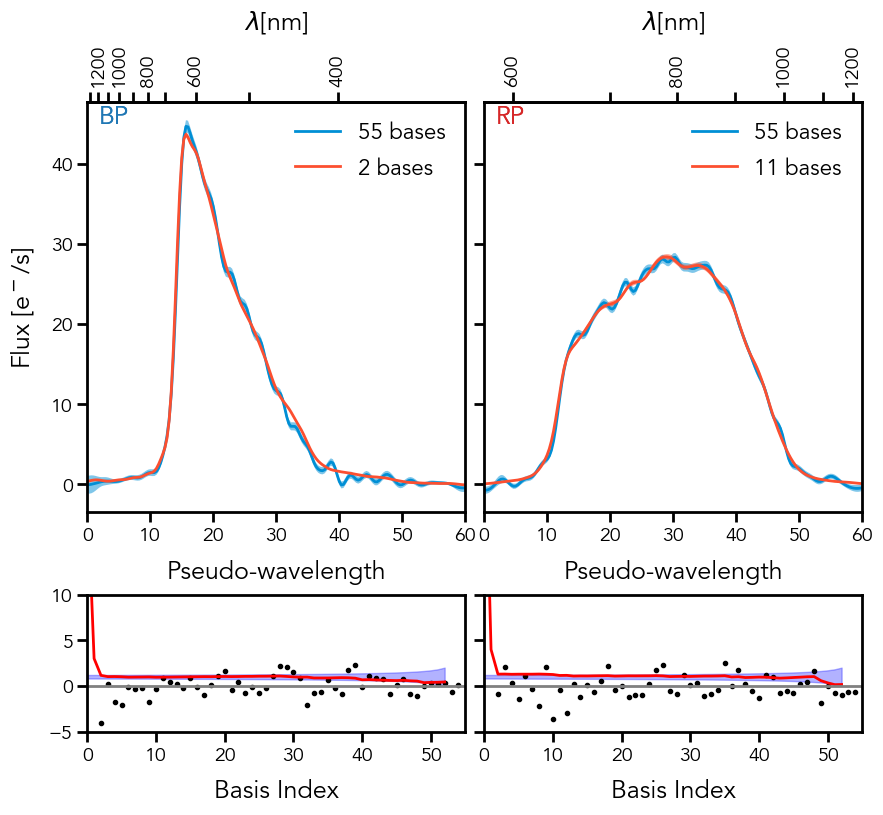}
    \caption{
    Illustration of the effects of truncation on the mean spectra of source \texttt{Gaia\,DR3\,6776463197626299392} ($G \approx 18.1$~mag and $\bprp\approx 1.0$~mag). Please see the caption of Fig. \ref{fig:ms:truncation:bright} and the text for details.}
    \label{fig:ms:truncation:faint}
\end{figure}

The truncation becomes more significant for noisier spectra. As a second example, we therefore consider a source with a similar colour as the first example, but fainter magnitude ($G \approx 18.1$~mag and $\bprp\approx 1.0$~mag). This case is shown in Fig.~\ref{fig:ms:truncation:faint}, which is analogous to Fig.~\ref{fig:ms:truncation:bright}. In this case, more coefficients are in agreement with being zero, and the number of significant coefficients is only 2 and 11 for BP and RP, respectively. Truncating the representation of the spectra at these numbers of basis functions maintains the general shape of the spectra, but suppresses the wavy patterns introduced by the noisy higher indices coefficients.\par
\begin{figure}[!h]
    \centering
    \includegraphics[width=0.48\textwidth]{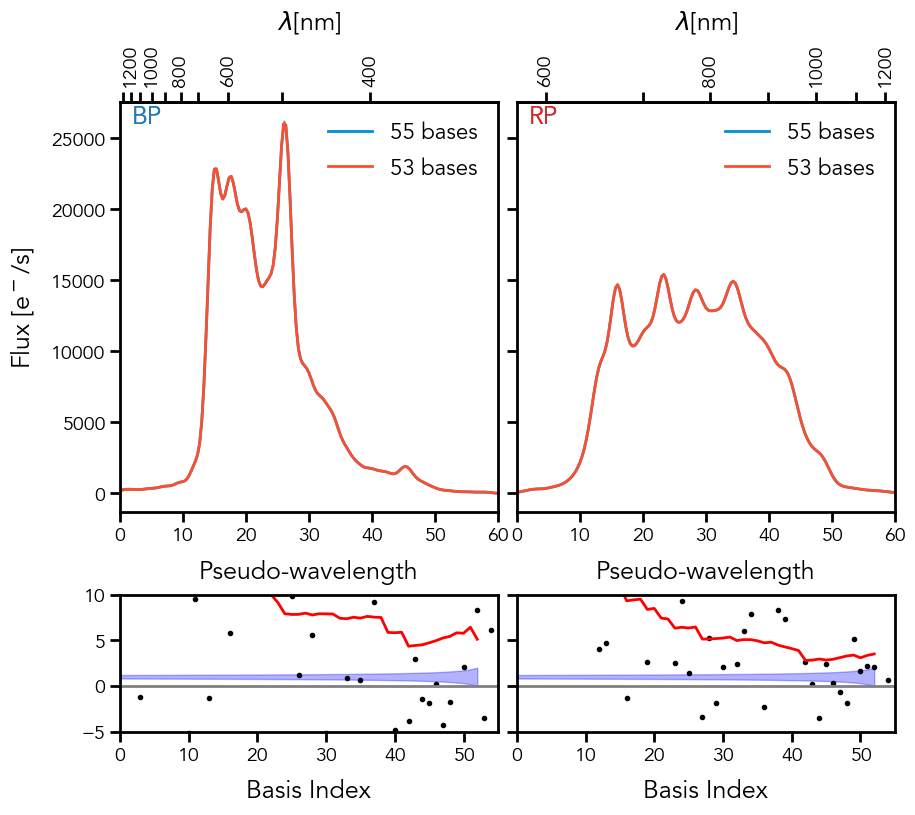}
    \caption{
    Illustration of the effects of truncation on the mean spectra of source  \texttt{Gaia\,DR3\,3032940844556081408} ($G \approx 11.5$~mag). Please see the caption of Fig. \ref{fig:ms:truncation:bright} and the text for details.}
    \label{fig:ms:truncation:emissionline}
\end{figure}

We furthermore show examples for sources with emission lines. The first case is a bright source ($G \approx 11.5$~mag) with multiple emission lines in BP and RP, shown in Fig.~\ref{fig:ms:truncation:emissionline}. Here, the truncation criterion is exceeded already for $M=3$, as all coefficients are required to represent the complex spectra for this source. The specified number of significant coefficients however is the number of coefficients when the truncation criterion is exceeded, which is 53 in this case. The use of all 55 coefficients is recommended in case of the number of significant coefficients is 53.\par
\begin{figure}[!h]
    \centering
    \includegraphics[width=0.48\textwidth]{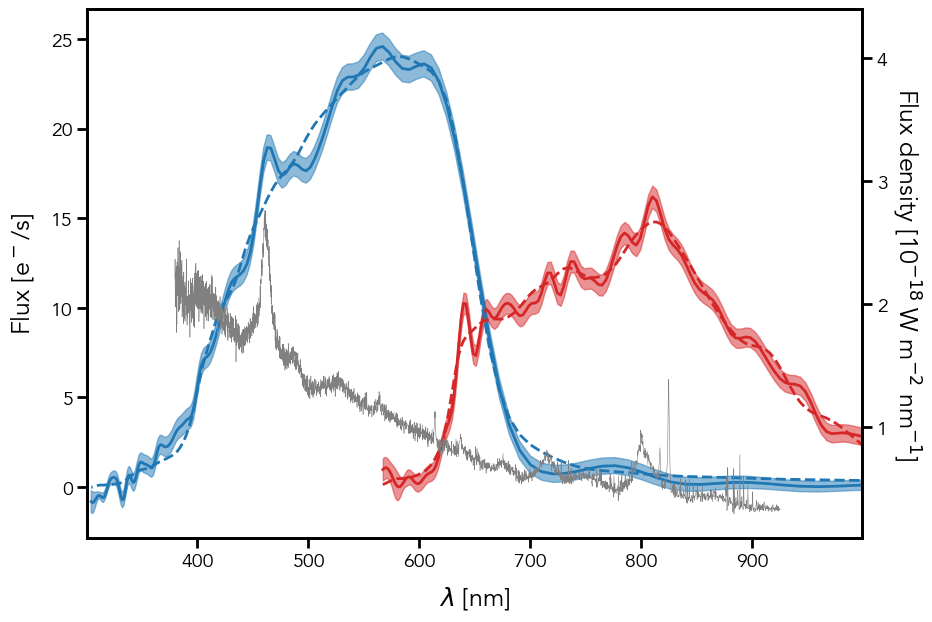}
    \caption{
    Comparison between the internally calibrated BP (in blue) and RP (in red) spectra vs. the SDSS (in grey) spectrum for QSO \texttt{Gaia\,DR3\,578415237301611520} (SDSS \texttt{thing\_id = 144680521}). Dashed lines are used for the truncated spectra (using only 3 bases for BP and 11 for RP), while continuous lines show the spectra obtained using the full set of $55$ coefficients.}
    \label{fig:ms:truncation:qso}
\end{figure}

Finally, we consider a faint QSO with emission lines as an example. Figure~\ref{fig:ms:truncation:qso} shows the BP and RP spectra of a QSO ($G=18.7$~mag and $\gbp-\grp=0.5$~mag), with all 55 coefficients, and with the truncated representation, using 3 and 11 coefficients in BP and RP, respectively. The spectral energy distribution from SDSS is shown for comparison. In particular the strong emission line visible in the SDSS spectrum coincides with a line in the BP spectrum. This line is removed by the truncation process. The truncation in the case of complex spectral shapes might therefore be too strong.

The truncation procedure was also tested by the sub-system dedicated to the estimation of astrophysical parameters within the DPAC analysis pipeline, referred to as Apsis \cite[see][]{DR3-DPACP-157}. 
Most Apsis modules found that the truncation would have negative impact on the quality of the scientific results, as far as emission lines in quasars or certain types of stars are concerned. These tests were conducted at a very early stage when no external calibration was available yet, such that the conclusions were uncertain and it was considered by most Apsis modules to be the safer option to not truncate the coefficients. In the extreme case of ultra-cool dwarfs, which are very red and very faint stars, the truncation was found to have a positive impact and has been employed specifically for the Apsis module ESP-UCD focusing on this type of stars. For these faint stars, the suppression of noise might aid the data analysis.\par
The result of the truncation assessment is provided as part of the \gdr{3} in the parameters \texttt{bp\_n\_relevant\_bases} and \texttt{rp\_n\_relevant\_bases} available in the \texttt{xp\_summary} table and in the mean continuous spectra available via Datalink (see also \secref{outputs}). In case of very faint and typical stars, the use of the truncated representation of BP and RP spectra might be useful. In particular for sources with unusual spectral energy distributions, such as sources with emission lines, the use of all 55 coefficients for BP and RP, respectively, is advised. The full array of $55$ coefficients is available via the archive. Users will need to decide if the suggested truncation is appropriate for their use case.

\section{Output data}\label{sec:outputs}

This Section describes the \xp data available via the \gaia archive\footnote{\href{https://gea.esac.esa.int/archive/}{https://gea.esac.esa.int/archive/}}.

The exact number of sources with \xp mean spectra in the \gdr{3} release
is 219,197,643.
This list is the result of several selection criteria. Sources with \gband magnitude brighter than 17.65~mag and more than 15 CCD transits contributing to the generation of the mean spectra for both \xp were automatically selected. The criterion based on the number of transits leads to a (slightly) non-uniform completeness across the sky (see the density sky distribution in Fig. \ref{fig:sky}). From this initial list, sources that had shown poor estimates of SSC values \citep[see Sect. 8.2 for more details]{Riello2021} were excluded unless they were part of one of the lists of specific objects (see below). 
An additional 35K sources were excluded to allow further processing and validation within DPAC which is likely to be finalised only after \gdr{3}.
A few lists of specific objects for which other criteria would not apply were defined: these included about 500 sources used for the calibration of the \xp data, a catalogue of about 100K WD candidates, 17K galaxies, about 100K quasars, about 19K ultra-cool dwarfs, 900 objects that were considered to be the most representative sources (or centroid) for each of the 900 neurons of the Self Organising Map used by the Outlier Analysis module
\citep{DR3-DPACP-157} and finally 19 solar analogues.
All these selections are specific to \gdr{3} and will not affect the content of future releases.

In \gdr{3}, there is one source (\texttt{Gaia\,DR3\,5405570973190252288}) that has only an RP spectrum.

The \texttt{gaia\_source} table in the archive contains a boolean column \texttt{has\_xp\_continuous} that is \texttt{true} if the corresponding source has \xp mean spectra available\footnote{When querying the \texttt{gaia\_source} table for sources fulfilling some criteria and having \xp spectra available, the user needs to add \,\texttt{WHERE has\_xp\_continuous='true'} to the ADQL query.}. 
After retrieving a list of \texttt{gaia\_source} entries, \xp spectra can be downloaded from the archive via Datalink\footnote{See \href{https://www.cosmos.esa.int/web/gaia-users/archive/ancillary-data}{https://www.cosmos.esa.int/web/gaia-users/archive/ancillary-data}} in various file formats. This can be done either from the archive web interface or programmatically. In appendix \ref{sec:archive} we provide instructions for downloading the data from Python. 

The spectra are provided in the continuous representation (see also App. \ref{sec:data_format} for more details): for each BP and RP the spectrum is defined as a set of coefficients (\texttt{bp}/\texttt{rp\_coefficients}), the corresponding array with the coefficient formal errors defined as the standard uncertainties from the least square solution multiplied by the standard deviation of the solution (\texttt{bp}/\texttt{rp\_coefficient\_errors}), the correlation matrix
\footnote{Given the symmetry of the correlation matrix, only the upper triangular elements (above and not including the diagonal elements which are 1 by definition) of the matrix are provided. The matrix elements are stored as a 1D array of size $n\,(n-1)/2$ where $n$ is the number of coefficients. The full correlation matrix would therefore be of size $n\times n$. The ordering of the elements in the array follows a column-major scheme.}
(\texttt{bp}/\texttt{rp\_coefficient\_correlations}) and various parameters from the source update process, such as number of measurements, number of degrees of freedom, $\chi^2$ and standard deviation of the solution.

In addition to the data available via Datalink, the \texttt{xp\_summary} table provides access to some of the parameters listed in the previous paragraph via queries (to enable for instance selecting sources based on the standard deviation of their mean spectrum solution) and to other relevant information. Users interested in retrieving the number of CCD transit spectra (and individual measurements) that contributed to the generation of the mean spectrum or that want to know how many of these were assessed as contaminated or blended should interrogate this table, not the main \texttt{gaia\_source} table which instead provides similar counters for the photometric data.
While \xp spectra and \gband and \xp photometry share part of the processing and filtering criteria, there are also some important differences that can lead to apparent inconsistencies in these counters.

The Python package \texttt{GaiaXPy}\footnote{\href{https://gaia-dpci.github.io/GaiaXPy-website/}{https://gaia-dpci.github.io/GaiaXPy-website/}} has been developed to help the users of \xp spectra. It offers the following functionalities: generation of a sampled version of the original continuous representation in both internal and absolute flux and wavelength systems, computation of synthetic photometry in various photometric systems and simulation of \gaia-like mean spectra from an input absolute spectral energy distribution (SED). For more information on these tools please refer to the package on-line documentation. 

\section{Validation}
\label{sec:validation}

\subsection{Errors}
\label{sec:validation:errors}

In order to test the performance of the calibration, a special validation dataset was generated where for each source the available transits were randomly divided into two groups and processed separately to generate two mean spectra for BP and two for RP. This allows us to compare the calibration results from two sets of transits for the same sources. We refer to this dataset as the BP/RP split-epoch validation dataset. Further details (including how to access the dataset) are available in App. \ref{sec:split:epoch}.

For this comparison we computed the Mahalanobis distance, $D_M$ between the two solutions for each source, given by
\begin{equation}
    D_M = \sqrt{ \left( c_1 - c_2 \right)^{\mathsf T} \, \left( \Sigma_1 + \Sigma_2 \right)^{-1} \, \left( c_1 - c_2 \right) } \; .
\end{equation}
Here, $c_1$ and $c_2$ denote the coefficient vectors for the two solutions, and $\Sigma_1$ and $\Sigma_2$ the corresponding covariance matrices. Under the idealized circumstances of normally distributed noise, correct covariance matrices, and absence of intrinsic photometric variability of the sources used in the test, $D_M$ follows a chi distribution with the degree of freedom corresponding to the length of $c_1$ and $c_2$. Deviations from a chi distribution therefore indicate unreliable covariance matrices $\Sigma_1$ or $\Sigma_2$.

\begin{figure*}[!htbp]
\center{
\includegraphics[width=0.45\textwidth]{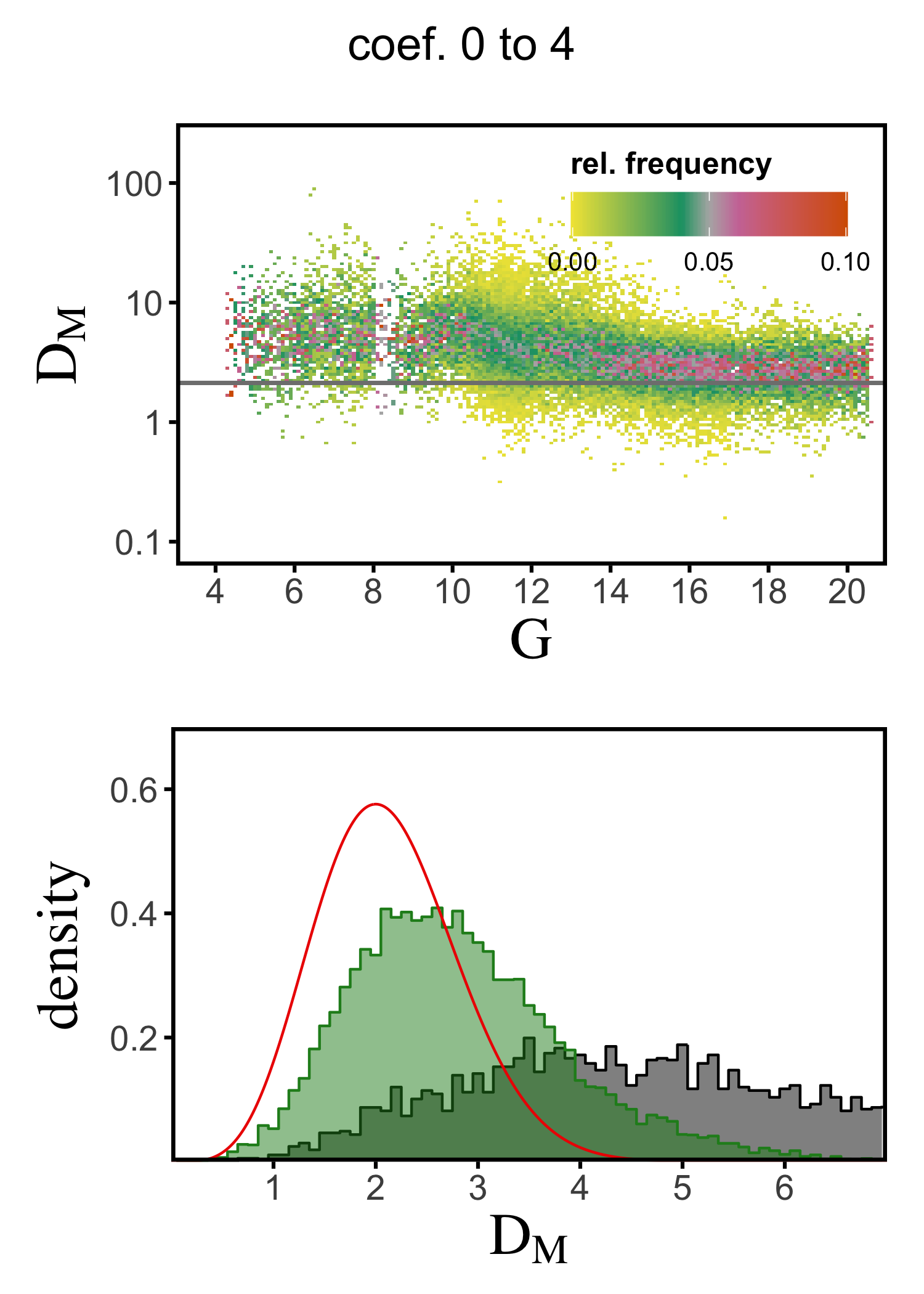}
\includegraphics[width=0.45\textwidth]{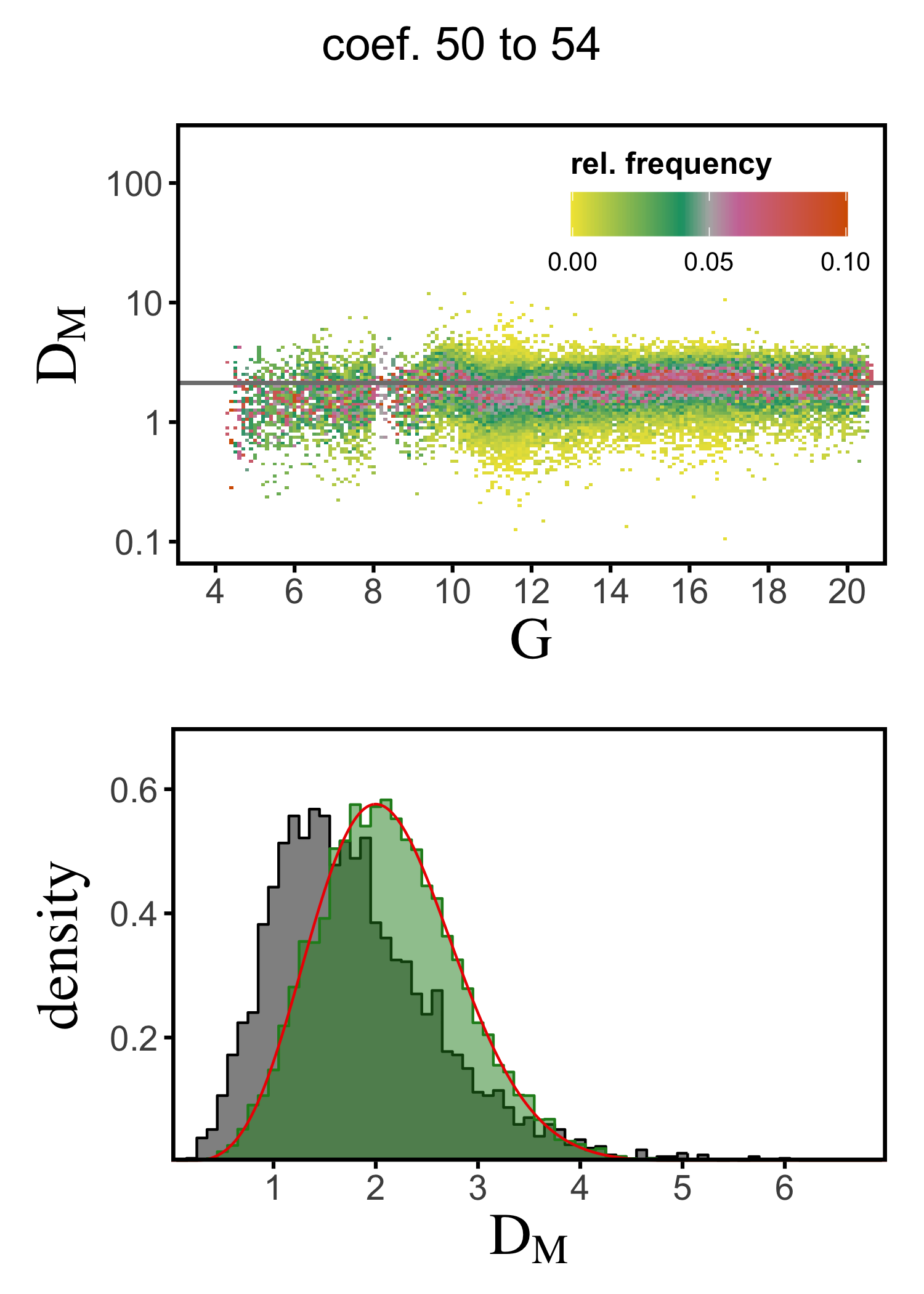}
}
\caption{Top panels: Distribution of the Mahalanobis distances of all test sources as a function of \gband magnitude. The grey horizontal line indicates the mean of the chi distribution.  Bottom panel: Histograms of the Mahalanobis distances for sources with $G<10$~mag (grey) and $G>16$~mag (green). The red line is the corresponding chi distribution. The left hand side plots are for the first five coefficients, with indices 0 to 4, the right hand side plots for the five coefficients of highest order, with indices 50 to 54.
\label{fig:splitEpoch1}
} 
\end{figure*}

We analysed the distribution of the $D_M$ in comparison to the chi distribution as a function of colour, magnitude, and indices of coefficients. The dependency on colour is only weak, with slightly larger values of $D_M$ for very red sources, with $\bprp\gtrsim3.0$~mag. The magnitude dependency is more pronounced, and depends on the indices of the coefficients. This is illustrated in Fig.~\ref{fig:splitEpoch1}. The top panels of this figure show the distribution of the $D_M$, normalised to the total number of sources in each magnitude bin, for all 40K test sources, for the first five and the last five coefficients in BP, respectively. For the first five coefficients, the values of $D_M$ are in general too large compared to what is expected from a chi distribution, an effect that is more pronounced for bright sources. For the five coefficients corresponding to the highest order basis functions, the magnitude dependency is weaker, with values being slightly smaller than expected from a chi distribution for the brighter sources.

The bottom panels of Fig.~\ref{fig:splitEpoch1} show the density histograms for bright sources, with $G<10$~mag in grey, and faint sources, with $G>16$~mag in green, respectively, in comparison with the chi distribution for five degrees of freedom. For the first five coefficients, the distribution is much wider than the chi distribution, in particular for the bright sources, and shifted to larger values. For the last five coefficients, the faint sources are in good agreement with a chi distribution, while the distribution for the bright sources is shifted towards smaller values of $D_M$.

An underestimation of the error results in larger $D_M$ than expected from a chi distribution, while an overestimation of the error results in smaller values. The differences in $D_M$ with respect to the chi distribution can therefore be interpreted as an underestimation of the errors for the coefficients with low indices, and an overestimation of the errors for coefficients of high indices for bright sources. For high indices and faint sources, the errors are however reliable. While the results shown here are from BP spectra, the situation for RP is similar.

\subsection{Specific cases}

Although most of the spectra have a good behaviour, there are few cases where we have some peculiar shapes due to several factors. We analyse in the following a few of the most common situations.

\begin{figure}[!htbp]
\center{
\includegraphics[width=0.45\textwidth]{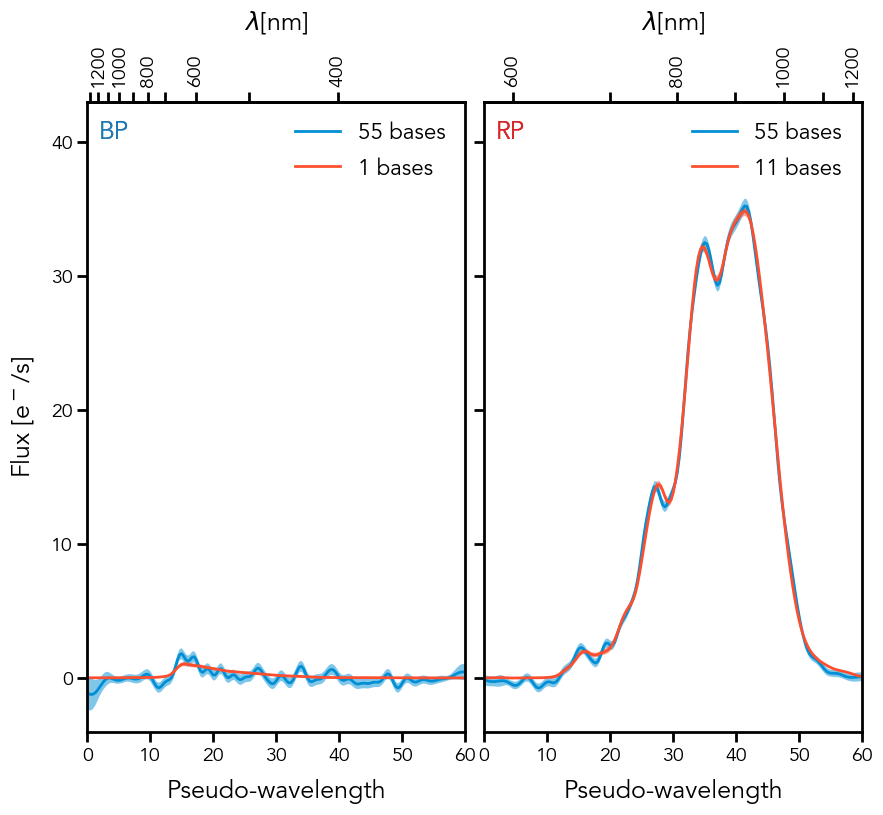}
}
\caption{BP (left) and RP (right) spectra for the faint red source \texttt{Gaia\,DR3\,1252666141462905344} ($\gbp=21.6$~mag and $\grp=17.8$~mag). The blue curves show the spectra defined by the $55$ coefficients (errors are shown as a shaded area). The red curves show the truncated spectra where only the first \texttt{bp}/\texttt{rp\_n\_relevant\_bases} have been used.
\label{fig:faintBP}
} 
\end{figure}

In the case of very faint sources, the fitting procedure generating the mean spectrum will be poorly constrained and may produce unrealistic features. For example, \figref{faintBP} shows the spectra of a faint red source (with $\gbp=21.6$~mag and $\grp=17.8$~mag). For this type of spectra the parameters  \verb|bp_n_relevant_bases| and \verb|rp_n_relevant_bases| in the \verb|xp_summary| table in the \gdr{3} archive are particularly relevant, as they indicate the number of coefficients that are significant considering the noise level \citep[see][and Sect. \ref{sec:truncation} in this paper for more details]{Carrasco2021}. In this case, only $1$ of the $55$ coefficients defining the BP spectrum is considered significant. Our adopted truncation procedure suggests that for BP all coefficients beyond the first one are only fitting the noise fluctuations rather than real spectral features and can be ignored when using the mean spectra for further investigations. For RP the number increases to $11$ thanks to the larger signal to noise ratio. 

\begin{figure}[!htbp]
\center{
\includegraphics[width=0.45\textwidth]{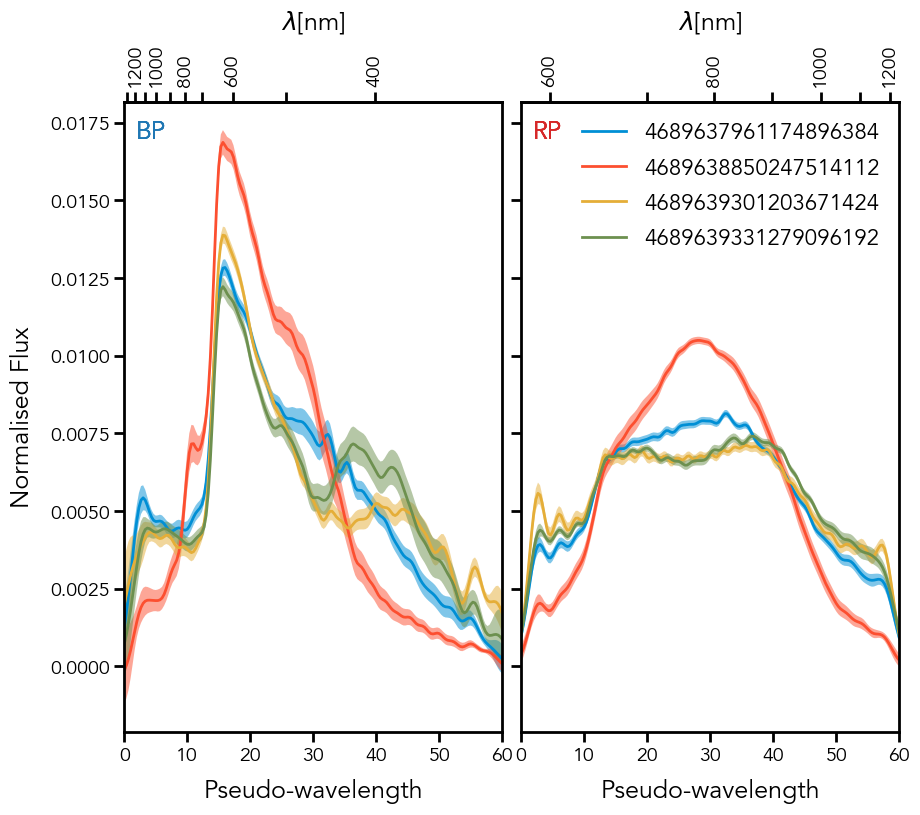}
}
\caption{BP (left) and RP (right) normalised internal spectra of some sources with all transits blended by other nearby sources in the 47\,Tuc cluster.
\label{fig:blending47Tuc}
} 
\end{figure}

In crowded areas, it is possible that two or more sources are so close in the sky to cause their observations to be always or often contaminated or blended. We refer to blended spectra when two or more sources fall within the observed window, while contamination refers to flux belonging to a source that is located outside the window. If this happens in a large fraction of the observations of a given source, then the mean spectra for that source will be affected. To enable users assess the reliability of \xp mean spectra the \verb|xp_summary| table in the archive includes several parameters (\verb|bp|/\verb|rp_n_blended_transits| and \verb|bp|/\verb|rp_n_contaminated_transits|) indicating the number of transits affected by blending or contamination for all sources for which \xp spectra are published. 
Figure \ref{fig:blending47Tuc} shows the case of four sources in the globular cluster 47\,Tuc that have all their observations flagged as blended. Users are strongly encouraged to make use of the available crowding flags to detect problematic cases.

\begin{figure}[!htbp]
\center{
\includegraphics[width=0.45\textwidth]{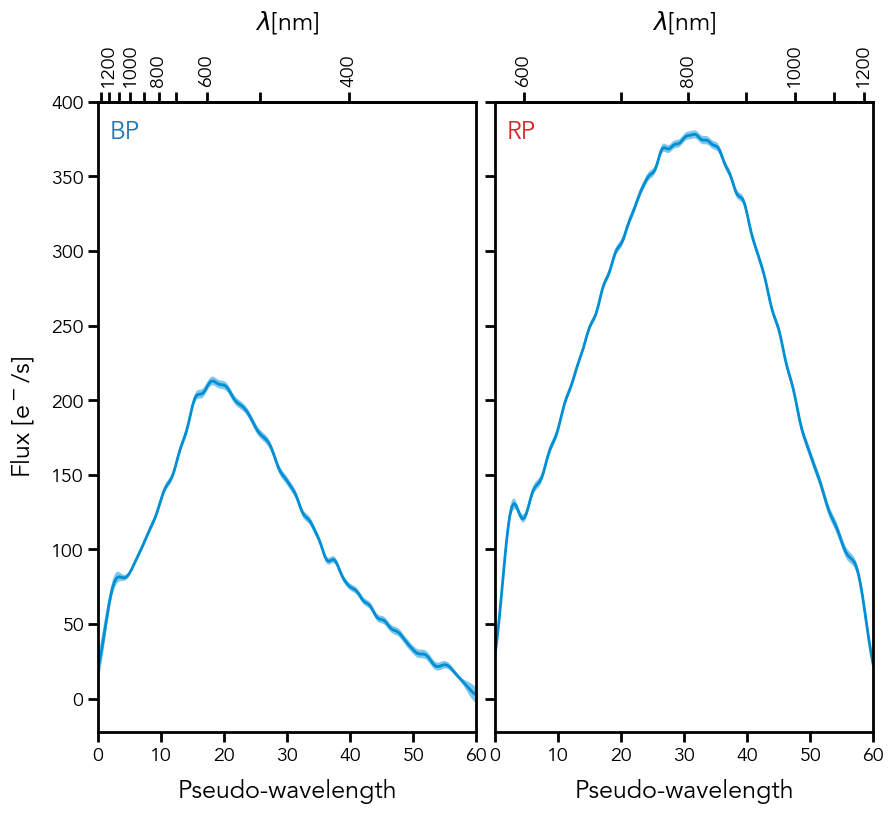}
}
\caption{BP (left) and RP (right) internally calibrated spectra of a source (\texttt{Gaia\,DR3\,1252344813484742272}) flagged as \texttt{galaxy} in the \texttt{gaia\_source} table. The spectra are broader than expected and the corresponding integrated magnitudes are much brighter compared with the \gband photometry.
\label{fig:galaxy}
} 
\end{figure}

The wings of the spectra should normally have low flux level, due to the combined action of LSF, dispersion and response. If this is not the case it could be due to the presence of residual background flux not fully removed in the background calibration stage or diffused flux due to the source being extended. For example, \figref{galaxy} shows the BP and RP internal spectra for a source with the \verb|in_galaxy_candidates| flag in the \verb|gaia_source| table set to \texttt{true}. Both spectra present a larger than normal flux in the wings. This source shows also a significant mismatch between the photometry in the different bands ($G=18.7$~mag, $\gbp=15.7$~mag and $\grp=14.3$~mag), the two \xp integrated flux values being much brighter than the value in the \gband, due to the much larger size of the \xp windows with respect to the AF ones. 

\begin{figure}[!htbp]
\center{
\includegraphics[width=0.45\textwidth]{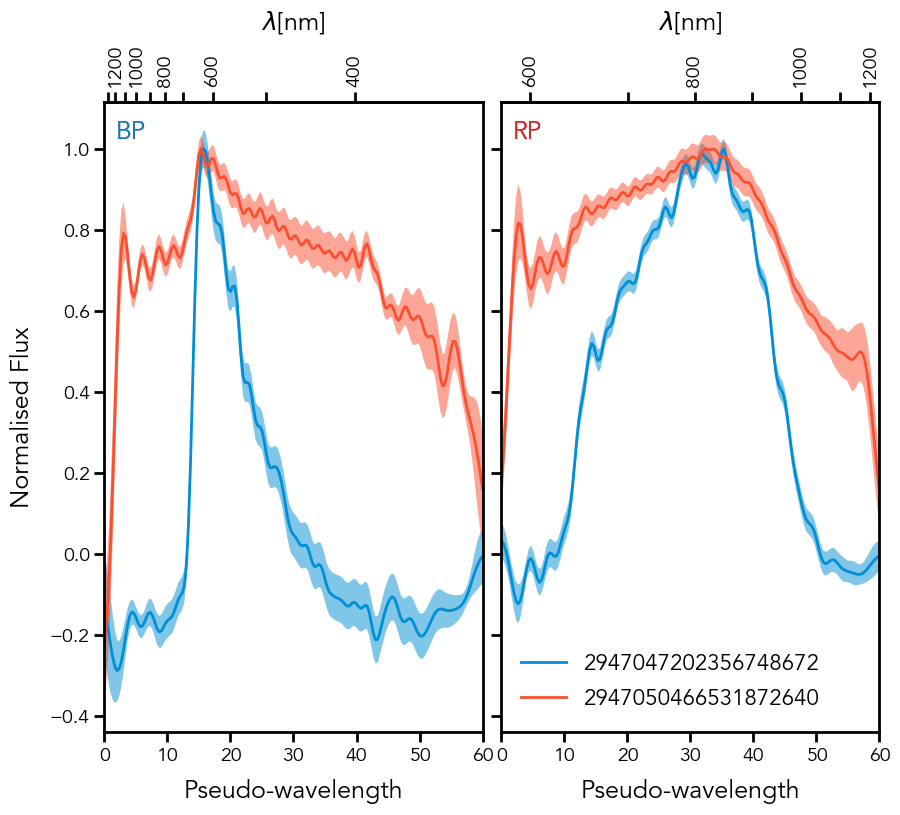}
}
\caption{BP (left) and RP (right) internally calibrated spectra of two sources near Sirius: one located at $30$ arcsec (in red) and the other at $3$ arcmin (in blue). The source closest to Sirius shows clear signs of contamination from the nearby object.
\label{fig:sirius}
} 
\end{figure}

A similar effect is seen when considering objects that are close to a very bright source. Their spectra will appear to be contaminated by flux coming from the nearby bright object. The resolution of the background calibration is not sufficient to remove completely this effect and may actually over/under-estimate the background in the regions surrounding very bright sources. 
Figure \ref{fig:sirius} shows the {\xp} spectra for two sources near Sirius. Source \texttt{Gaia\,DR3\,2947050466531872640}, at $30$ arcsec from Sirius, is clearly contaminated by diffuse flux coming from the nearby bright source. Also in this case, the photometry indicates a much brighter source in the \xp integrated bands than in the \gband: $G=15.7$~mag, $\gbp=13.2$~mag and $\grp=13.2$~mag. The second source  (\texttt{Gaia\,DR3\,2947047202356748672}) is located further away at about $3$ arcmin. In this case the background seems to have been overestimated causing negative flux values in the wing of the spectra in both BP and RP.

\subsection{Signal to noise ratio}

\begin{figure}[!htbp]
\center{
\includegraphics[width=1.\columnwidth]{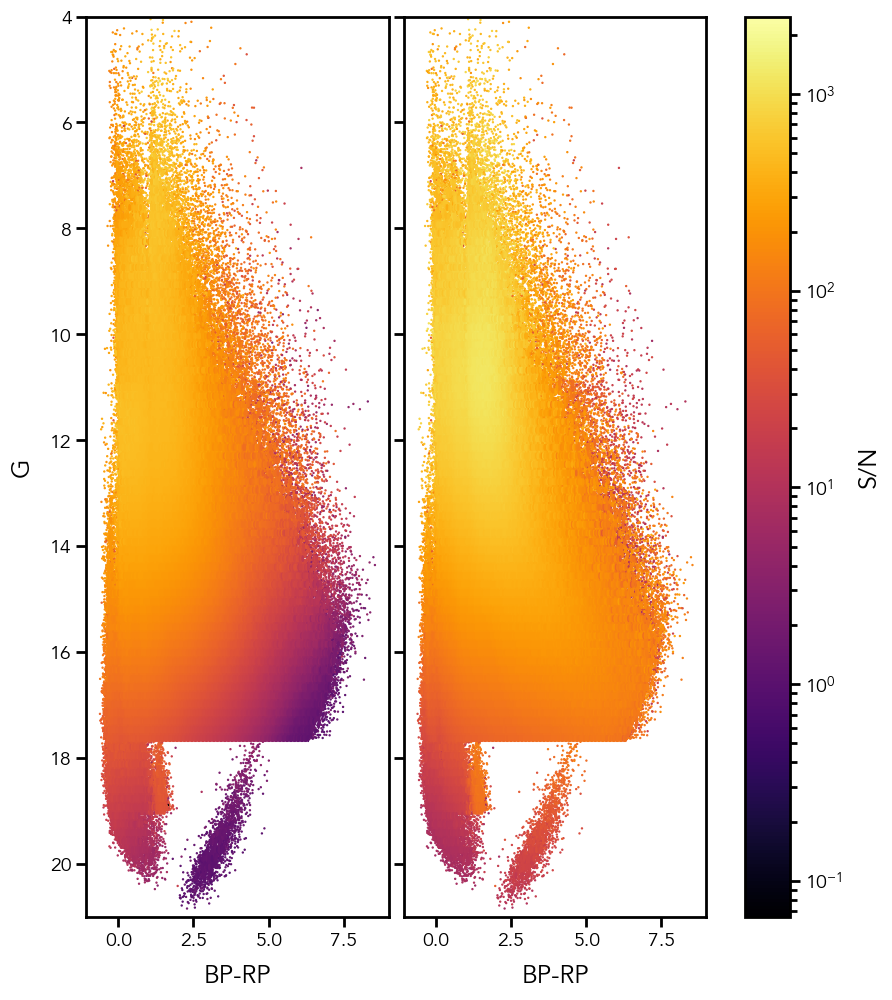}
}
\caption{Colour magnitude diagram of a random $10\%$ of the sources for which \xp spectra are available in \gdr{3}, colour coded in a logarithmic scale by the global S/N as computed directly from the continuous representation coefficients and their errors. BP S/N is shown in the left panel while RP S/N is used in the right one.
\label{fig:snr:cmd}
} 
\end{figure}

An overall indication of the signal to noise ratio (S/N) for a given source and photometer can be obtained directly from the coefficients by dividing the 
L2-norm of the vector of coefficients by the L2-norm of the vector of errors on the coefficients. Figure \ref{fig:snr:cmd} shows a colour magnitude diagram of the sources with \xp spectra in \gdr{3} colour coded by this global S/N in the BP and RP photometers in the left and right panel.

\begin{figure*}[!htbp]
\center{
\includegraphics[width=0.9\textwidth]{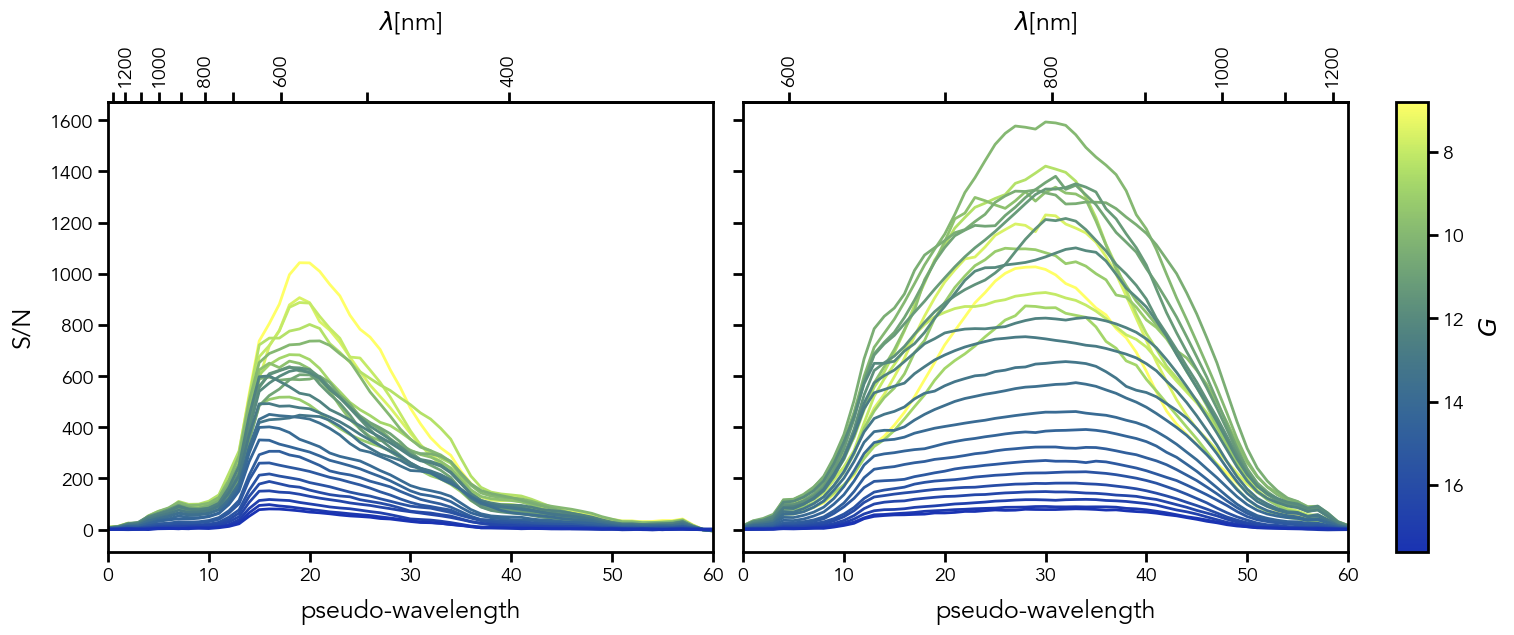}
\includegraphics[width=0.9\textwidth]{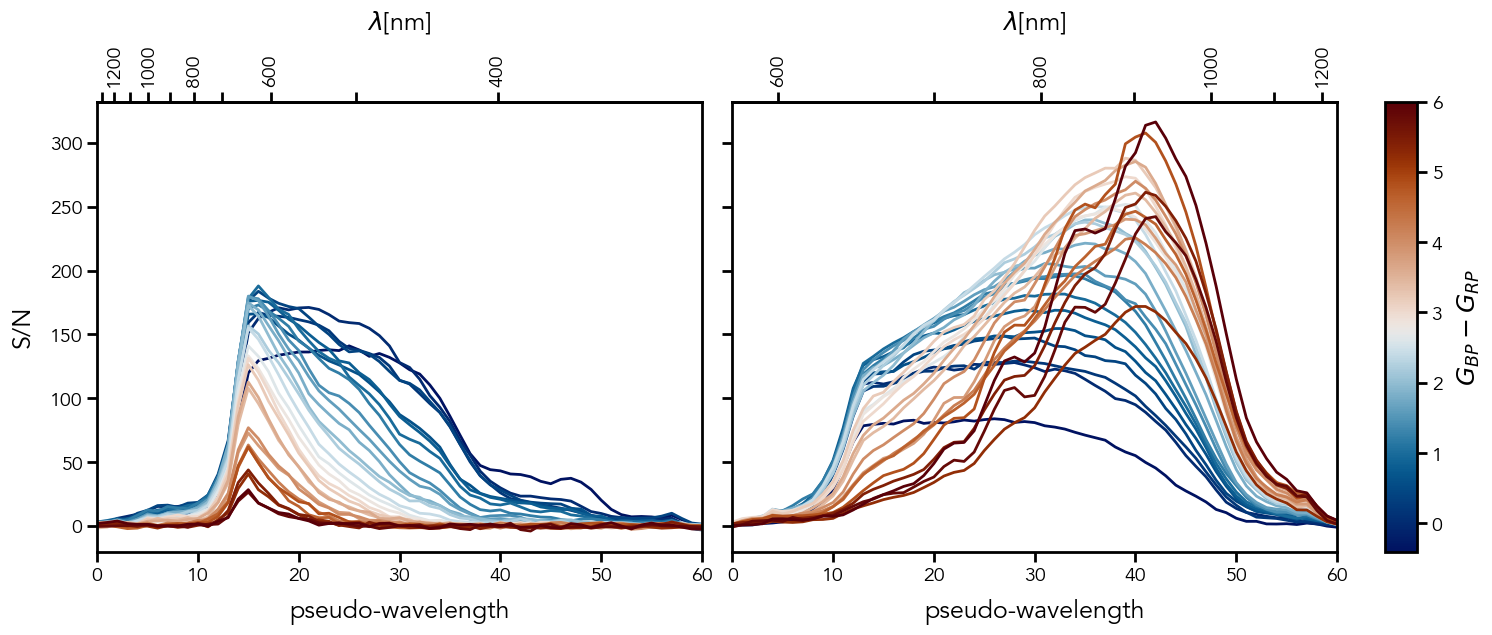}
}
\caption{S/N vs pseudo-wavelength (and approximate absolute wavelength) for internally calibrated spectra. The top panels show the S/N for sources of different magnitude and similar colour (close to 1.0), while the bottom panels focus on sources with similar \gband magnitude (close to 16.0) and a range of colours.
\label{fig:snr:sampled}
} 
\end{figure*}

A user that is interested in the S/N at different wavelengths will have to consider the representation of the spectrum by the linear combination of basis functions that do have an explicit wavelength dependency rather than rely on the mere coefficients. The panels in Fig. \ref{fig:snr:sampled} show typical S/N distribution of internally calibrated spectra over the BP (left panels) and RP (right panels) pseudo-wavelength ranges covered by the \xp spectra. In the top two panels each curve shows the S/N for sources of different magnitude, as reported in the colour bar, \bprp colour close to 1.0 and with typical global S/N (for sources of similar magnitude and colour). In the bottom two panels instead, each curve shows the S/N for sources of different colour, as reported in the colour bar, \gband magnitude close to 16.0 and with typical global S/N (for sources of similar magnitude and colour). Only sources with $|\cstar|< 0.02$ have been considered for these plots, \cstar being the corrected \xp flux excess factor as defined in \cite{Riello2021}. As in previous figures the top axes showing the correspondence with absolute wavelengths are only indicative.

Due to the fact that the mean \xp spectra are a combination of many single observations for each object, intrinsic variability will result in larger uncertainties in the mean spectra. This is confirmed by the fact that the S/N for a sample of known RR~Lyrae \citep[extracted from][]{DR3-DPACP-168} is significantly lower than the S/N for a sample of random (mostly non-variable) sources with similar apparent $G$.

The dependency of the S/N from pseudo-wavelength is linked to the spectrum itself. Looking at the right top panel of Fig. \ref{fig:snr:sampled}, the maximum S/N ratio in RP is achieved for sources with $G~9--10$. Saturation and occasional gate mis-configuration could be responsible for this: while the mean spectra of very bright sources do not show clear signatures of saturation, the presence of some saturated epoch spectra among those contributing to the mean spectrum, possibly due to gate mis-configuration caused by large on-board magnitude errors at the bright end, could lead to a larger scatter around the peak and therefore a larger error and a smaller-than-expected S/N ratio.

\section{Recommendations}\label{sec:recommendations}

The mean spectra are available in the archive in the form of a set of coefficients that define a continuous function over the pseudo-wavelength range. This is the fundamental product of the \xp spectral data processing. When sampling the spectra on a discrete grid in pseudo-wavelength (or wavelength if working in the absolute system), some information is unavoidably lost.
In particular, the continuous representation comes with full covariance information, whereas a spectrum sampled on a (pseudo-)wavelength grid with more points than the number of coefficients in the continuous representation cannot.
Users are therefore strongly encouraged to consider using the continuous representation to exploit at best the \xp spectra in \gdr{3} (e.g. to derive astrophysical parameters or analyse the presence of spectral features) and avoid sampling the spectra or deriving synthetic photometry from them, losing information in the process.

\begin{figure}[!htbp]
\center{
\includegraphics[width=0.45\textwidth]{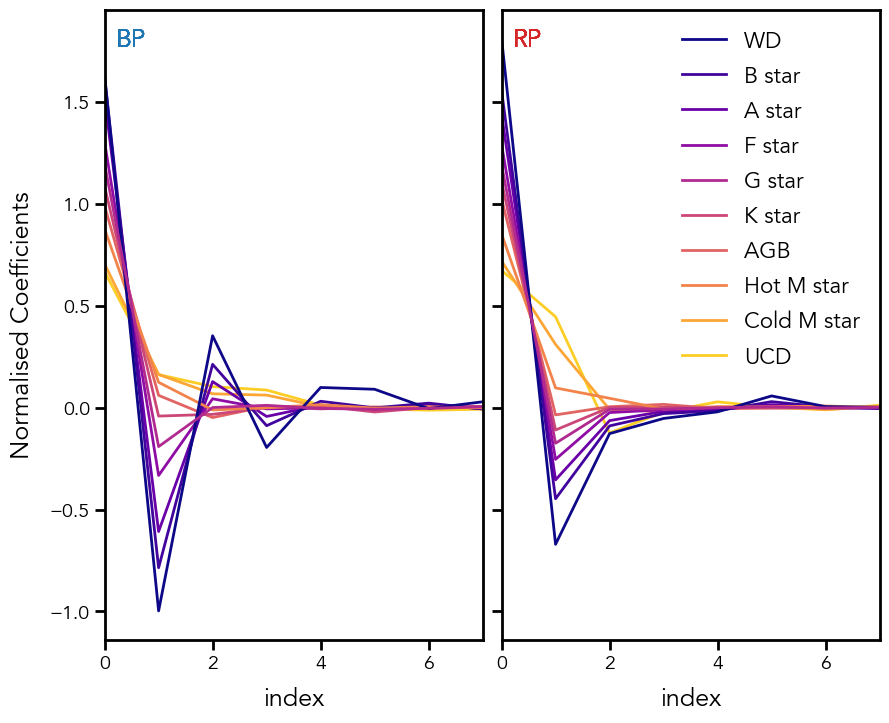}
}
\caption{First eight coefficients of the continuous representation in BP (left) and RP (right) for some sources with different astrophysical parameters.
\label{fig:coefs_SpT}
} 
\end{figure}

\begin{figure}[!htbp]
\center{
\includegraphics[width=0.45\textwidth]{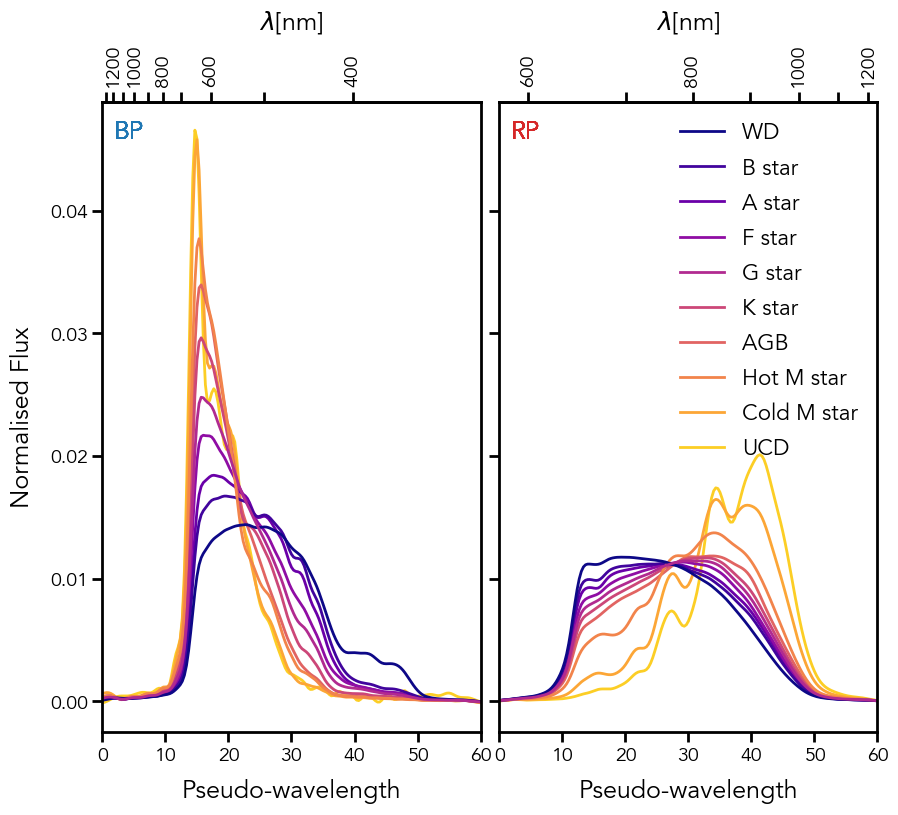}
}
\caption{Normalised internal mean spectra in BP (left) and RP (right) for the same sources shown in \figref{coefs_SpT}.
\label{fig:spectra_SpT}
} 
\end{figure}

Figure \ref{fig:coefs_SpT} shows that the coefficients can be used to successfully classify sources in different regions of the Hertzsprung-Russell diagram. At least for the few cases shown in the plot, most of the information required for classification is already available in the first few coefficients of the continuous representation. Figure \ref{fig:spectra_SpT} shows the corresponding plot with the more familiar sampled spectra.

\begin{figure}[!h]
\center{
\includegraphics[width=0.45\textwidth]{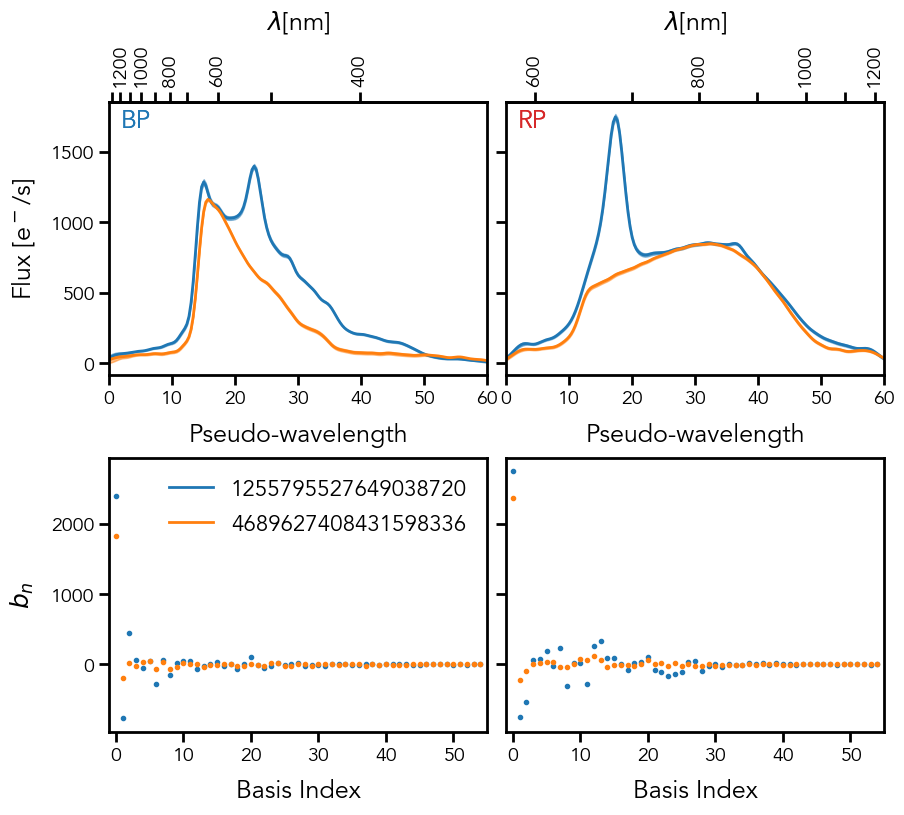}
}
\caption{Comparison of the mean spectra obtained for a QSO with a strong emission line (\texttt{Gaia\,DR3\,1255795527649038720} in blue), and another source with similar shape and flux level but without strong features (\texttt{Gaia\,DR3\,4689627408431598336} in orange). BP is shown in the left panels and RP in the right ones. Sampled spectra are shown in the top panels, while the bottom panels show the corresponding coefficients. 
\label{fig:pair:with:qso}
} 
\end{figure}

Figure \ref{fig:ms:bases} clearly shows that narrow spectral features in the spectra can only be reproduced with larger higher order coefficients. For example, \figref{pair:with:qso} shows an example of two sources with rather similar RP spectra except for
the presence of a strong emission line. One of the two sources is a QSO. As it can be seen in the bottom right panel, higher order coefficients for the QSO have larger values. 

\subsection{Effects of noise}\label{sec:recommendations:noise}

\begin{figure*}[!htbp]
\center{
\includegraphics[width=0.95\textwidth]{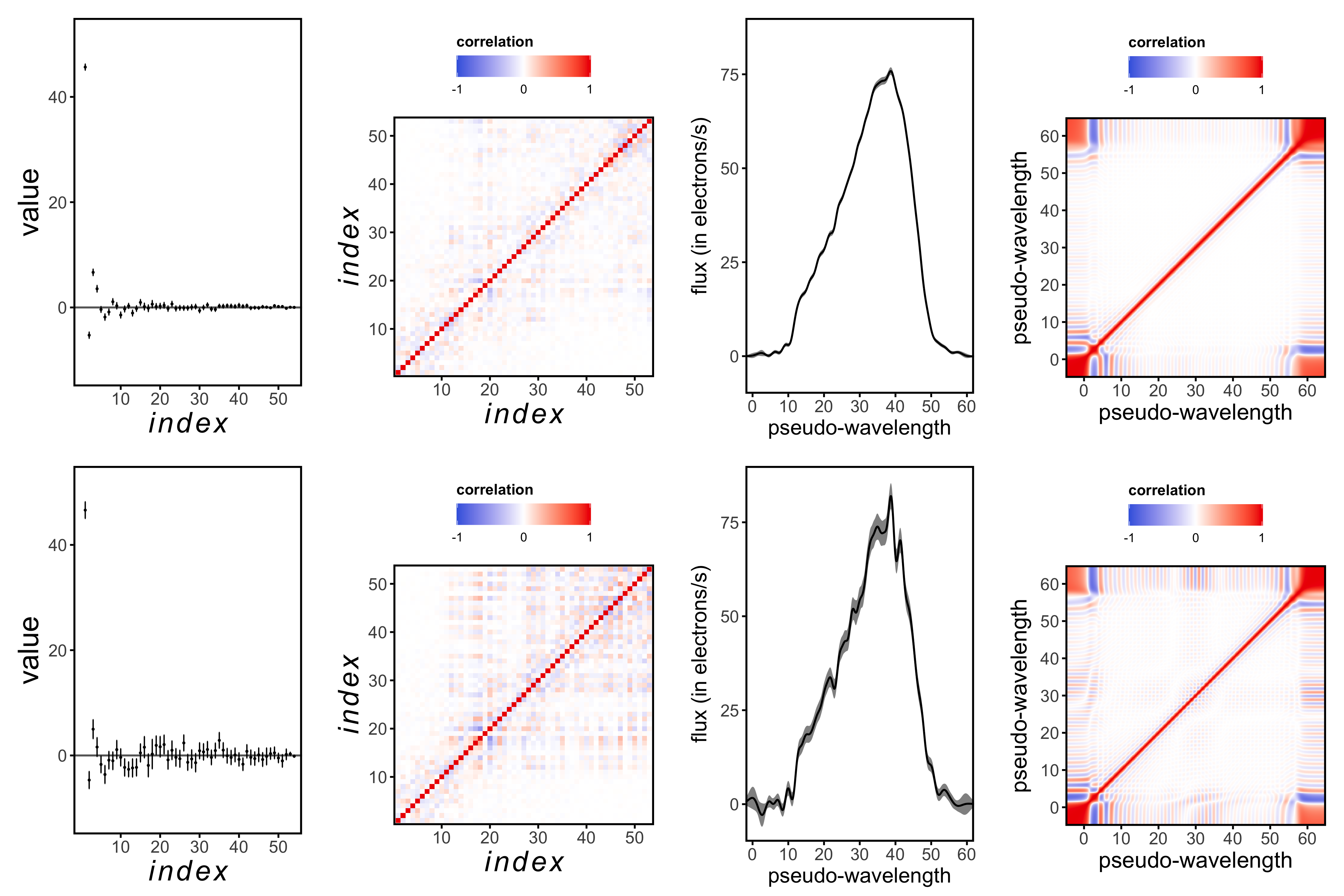}
}
\caption{Example of the effect of noise for an RP spectrum. First column: RP coefficients with errors. Second column: correlation matrices for the coefficients. Third column: sampled RP spectrum (black line) with 1-sigma uncertainty interval (grey shaded region). Forth column: correlation matrix for the sampled RP spectrum. The top row is for 18 transits, the bottom row for 3 transits, for the same source.
\label{fig:noiseEffect}
} 
\end{figure*}

The correlations between the coefficients of a source, both, for BP and RP, are in general rather low, with median correlation coefficients well below 0.1 in both, BP and RP. When constructing the sampled spectrum as a function of pseudo-wavelength (or wavelength), the correlations might become much more important. Since there are only 55 basis functions for BP and RP, respectively, any sampled spectrum with more than 55 sample points needs to have linear dependencies among the samples. Furthermore, even if the coefficients would be uncorrelated, the non-local character of the basis function representation introduces correlations between different pseudo-wavelengths. This effect is illustrated in Fig.~\ref{fig:noiseEffect} for the RP spectrum of one particular source, with $G = 17.89$, $\bprp = 2.74$. The \xp split epoch validation dataset (see App. \ref{sec:split:epoch}) has been used for this analysis. The two sets of transits for this source contain 18 transits and 3 transits, respectively. Consequently, the signal-to-noise ratio in the first set is higher than in the second one. This is seen in the first column of Fig.~\ref{fig:noiseEffect}, where the coefficients for the calibration using only three transits are noisier and have larger error bars than for the 18 transits set. The second column in this figure shows the correlation matrices for the two cases. In general the correlations are low, with little structure in the off-diagonal entries. For the noisier case, the correlations are however larger. The third column shows the sampled RP spectra for the 18 transits and the three transits cases. The larger noise in the latter case manifests itself in a wavy structure in the sampled spectrum. In the correlation matrix for the sampled spectrum, shown in the fourth column, this manifests itself in the form of alternating short-scale patterns of positive and negative correlations. These patterns are again more pronounced when the signal-to-noise ratio is lower.

Since random noise in the \xp spectra manifests itself in the sampled spectra as wavy structures, and correlations within the sampled spectra are not negligible, the interpretation of the coefficients, being much less affected by correlations, might be more convenient.

\section{Conclusions}\label{sec:conclusions}

In this paper we have focused on the processing that generated the internally calibrated \xp spectra contributing to \gdr{3} starting from the raw satellite data. The released data are time-averaged source spectra that result from the combination of all single observations of a given source. 
Only a selection of all generated spectra will be included in the release at this stage, but several other new products are based on the entire dataset. The main challenges faced by this step in the data processing are due to the vast amount of data (about 65 billion single BP/RP transits were processed), to the nature of the low-resolution aperture prism spectroscopy with the additional complications added by the TDI mode, to the large number of different observing configuration effectively corresponding to different instruments that need to be calibrated onto the same homogeneous system. We have explained how we have dealt with these challenges and have shown how we have been monitoring the intermediate performances of our calibration procedures. 
We also described the somewhat unfamiliar format of the BP/RP spectral data in the archive. Rather than providing spectra defined as a flux value corresponding to a sample covering a given wavelength range, the \xp spectra are represented by an array of coefficients, their errors and correlations, to be applied to a set of basis functions to obtain a continuous function. This approach allows combining multiple transit spectra, each having its own pixel/wavelength sampling, dispersion and LSF \citep{Carrasco2021}. The set of bases has been optimised to ensure maximum efficiency thus focusing most of the flux in the first few coefficients and leaving higher-order coefficients to be constrained by narrow spectral features.

\begin{figure*}[!h]
    \centering
    \includegraphics[width=0.48\textwidth]{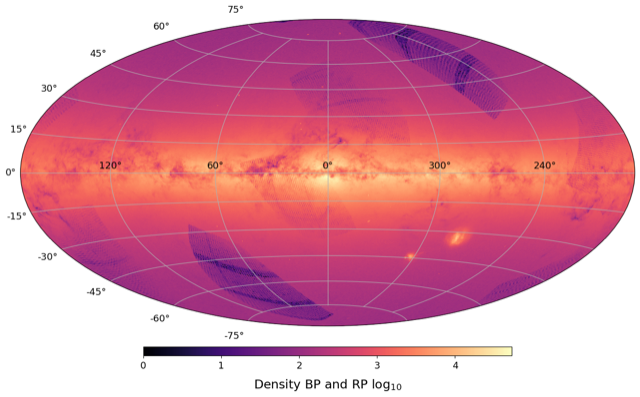}
    \includegraphics[width=0.48\textwidth]{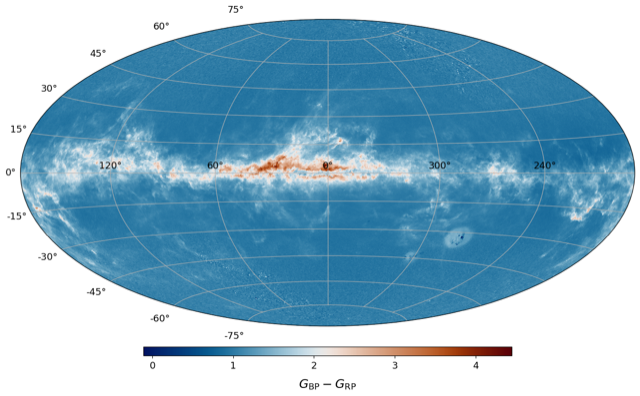}\\
    \includegraphics[width=0.48\textwidth]{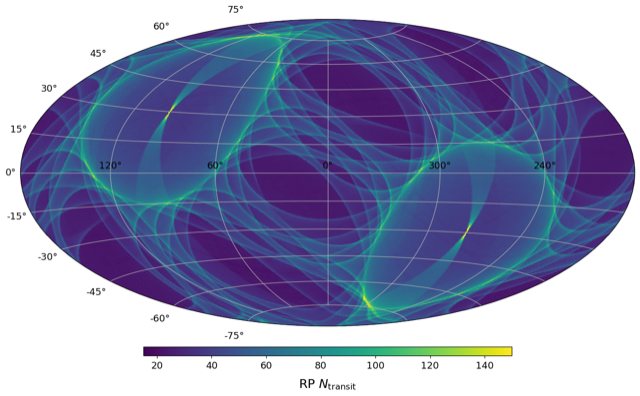}
    \includegraphics[width=0.48\textwidth]{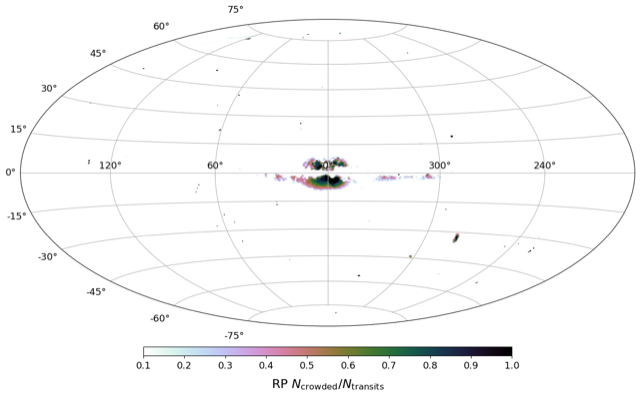}\\
    \includegraphics[width=0.48\textwidth]{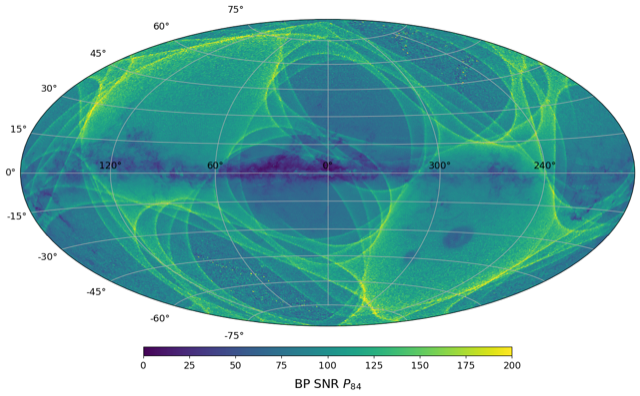}
    \includegraphics[width=0.48\textwidth]{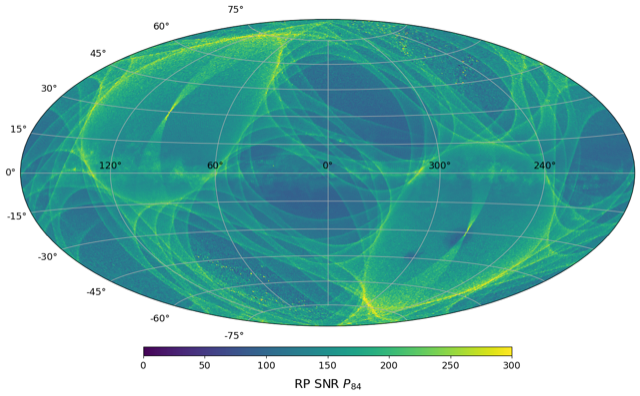}
    \caption{Sky distribution (in Galactic coordinates in Hammer-Aitoff projection, with resolution equivalent to HEALPiX level 7) of various parameters related to the \xp data: from the top left to the bottom right the maps show the sky density of objects with \xp spectral data, the median \bprp colour, the median number of transits in RP contributing to the mean spectra, the median crowding level, the median of the $84^{\rm th}$ percentile of the S/N over the BP and RP ranges. The colour scales do not cover the full range covered by the data.}
    \label{fig:sky}
\end{figure*}

We want to conclude this paper by showing some sky distributions related to the \xp data in Fig. \ref{fig:sky}.
All maps are in Galactic coordinates and show the entire catalogue of sources with \xp spectra in \gdr{3}.
The first map shows the density distribution in the sky. As expected most of the sources are concentrated along the Galactic plane. The two Magellanic Clouds also stand out as well as a few clusters. The darkest areas close to the Galactic plane in the map correspond to both regions obscured by dust and regions with extremely high density where the \xp data are particularly affected by strong crowding (both in the acquisition and in the processing). Some regions with lower density off the Galactic plane still show imprints of the scanning law (compare this with the map showing the median number of transits). These are expected to disappear with the addition of more observations in future releases.
The second map shows the distribution of \bprp colour.
The third map shows the median number of transits per source (in RP). This is clearly defined by the satellite scanning law. A similar map of BP would be very similar with the exception of the occurrences of larger number of transits near the Ecliptic poles. These are due to the first month of operations in Ecliptic scanning law. This period was not included in the generation of average source BP spectra as explained in Sect. \ref{sec:processing:intcal}.
The fourth map shows the median fraction of contaminated or blended transits with respect to the number of transits per source for RP. The equivalent maps for BP would look very similar. The areas showing higher density in the first map stand out also in this map as regions where the mean spectra are more affected by crowding. This is justified by the fact that the crowding evaluation is limited to the \gaia source catalogue itself.
Finally the last two maps show the distribution in the sky of the median of the $84^{\rm th}$ percentile of the S/N distribution over the BP and RP wavelength ranges. As expected the scanning law signature is very evident in these maps with errors being lower in the most observed regions. Areas at low Galactic latitude show lower S/N in the BP spectra due to the abundance of red-colour sources.
The S/N distribution of the internally calibrated spectra shows values larger than 1000 for bright sources in some wavelength ranges (see Fig. \ref{fig:snr:sampled}). \gdr{3} will contain about 700 thousand BP spectra and 4.3 million RP spectra with the $84^{\rm th}$ percentile of the S/N above 500.

\begin{figure}[h]
\centering
    \includegraphics[width=0.48\textwidth]{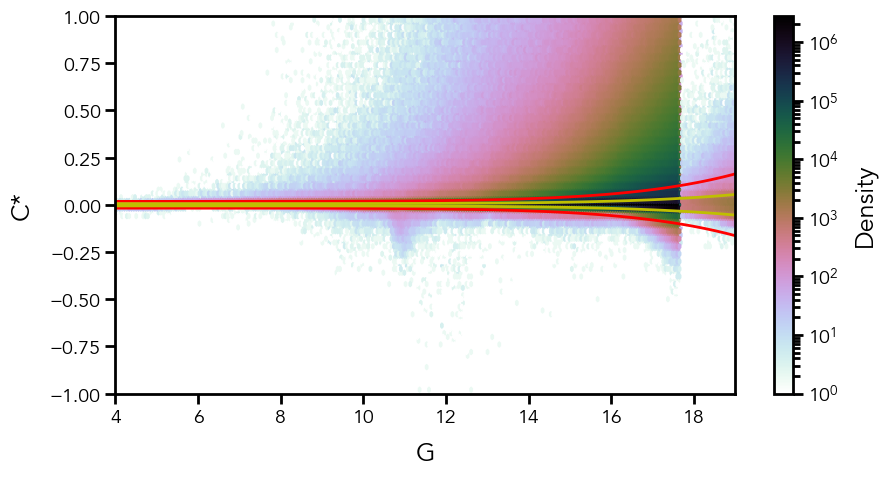}
    \caption{Distribution of $C^{\star}$ vs. magnitude for all sources with \xp spectra in \gdr{3}. Also shown are the 1- and 3-$\sigma$ curves  in yellow and red respectively as defined in \cite{Riello2021}.}
    \label{fig:cstar}
\end{figure}

Various parameters available from the archive can be useful to clean the catalogue from disturbed spectra. A very useful quantity already introduced for \gdr{2} is the  \texttt{phot\_bp\_rp\_excess\_factor}. This parameter is available from the \texttt{gaia\_source} table and is defined as the ratio between the sum of BP and RP integrated fluxes and the \gband flux for the same source. Due to the shape of the $G$, \gbp and \grp passbands some colour dependency of this ratio is expected and may bias selections based on \texttt{phot\_bp\_rp\_excess\_factor}. To correct for the expected colour trends, users should apply the equation recommended in \cite{Riello2021} to form what is known as $C^{\star}$\footnote{$C^{\star}$ is obtained from the \texttt{phot\_bp\_rp\_excess\_factor} $C$ as $C- f(\bprp)$ where $f(\bprp)$ is a polynomial in colour defined as
\[
f(x) = \begin{cases}
    1.154360 + 0.033772\,x + 0.032277\,x^2 & \text{for } x<0.5\\
    1.162004 + 0.011464\,x + 0.049255\,x^2 - & \\
    \phantom{xxxxxxxxxxxxx} 0.005879\,x^3 & \text{for } 0.5\leq x<4.0\\
    1.057572+0.140537\,x & \text{for } x\geq 4.0\\
\end{cases}
\]


where $x=\bprp$.

The corrected parameter (\texttt{c\_star}) will be available for all sources included in the Gaia Synthetic Photometric Catalogue from the archive, see \cite{DR3-DPACP-93}.
}.
The deviation of this parameter from 0.0 indicates the presence of inconsistencies between the flux measured in the \xp windows with respect to the flux in the \gband. These inconsistencies can be due to different source properties (e.g. in the case of extended sources) or systematic errors in the calibration procedures (e.g in the case of residual background due to nearby bright sources). Section 9.4 in \citet{Riello2021} provides also a function reproducing the $1\sigma$ scatter for a sample of well behaved isolated stellar sources with good quality photometry. Users wishing to use $C^{\star}$ and its $1\sigma$ scatter to select the most reliable spectra, would find that $90\%$ of the sources have $C^{\star}<3\sigma$ while $79\%$ fulfil the criterion $C^{\star}<1\sigma$. Figure \ref{fig:cstar} shows the distribution of $C^{\star}$ together with the $1$- and $3$-$\sigma$ limits.

In terms of \xp spectral data, future releases will see a vast increase in the number of average source spectra and the addition of calibrated epoch spectra, i.e. spectra derived from one single observation in \xp. From the processing and validation point of view this will focus the attention on calibrations that deviate from the average behaviour. While when generating mean spectra robust techniques help mitigate these problems, the application of noisy calibrations can generate unreliable data. This needs to be mitigated to ensure the quality of calibrated \xp epoch spectra, planned to be included in future releases. One other area where some improvement is being sought, is in the bluest wavelength range covered by BP ($350--400$ nm) where the small fraction of calibrators makes the flux and LSF calibration particularly challenging.
The effect of this can be seen in some systematics offsets in the bluest part of the wavelength range covered by \xp data. These can be quantified when comparing BP/RP spectra with external absolute spectra \cite{Montegriffo2022} and/or synthetic photometry generated from BP/RP spectra in various bands and photometric systems vs existing catalogues \cite{DR3-DPACP-93}. In particular, in the latter work, the comparison of synthetic photometry from externally calibrated BP/RP spectra with state-of-the-art ground based photometric standard stars suggests that in the wavelength range spanned by SDSS $u$ band (and/or Johnson-Kron-Cousins $U$) differences can be as large as $20\%$, for some spectral types  and in some color ranges. In the range covered by SDSS $g$ band (and/or Johnson-Kron-Cousins $B$ band) systematics reach the $5\%$ level at most, while for redder passbands are typically below the $2\%$ level.


\begin{acknowledgements}

We are very grateful to the anonymous Referee for a careful and constructive report, that improved the quality of the manuscript. 
We would also like to thank R.~Blomme for kindly reviewing an earlier version of this manuscript.

This publication made extensive use of the online authoring Overleaf platform (\url{https://www.overleaf.com/}).
The data processing and analysis made use of
matplotlib \citep{Hunter:2007},
NumPy \citep{harris2020array},
the IPython package \citep{PER-GRA:2007},
TOPCAT \citep{Taylor2005}.

This work presents results from the European Space Agency (ESA) space mission \gaia. \gaia\ data are being processed by the \gaia\ Data Processing and Analysis Consortium (DPAC). Funding for the DPAC is provided by national institutions, in particular the institutions participating in the \gaia\ MultiLateral Agreement (MLA). The \gaia\ mission website is \url{https://www.cosmos.esa.int/gaia}. The \gaia\ archive website is \url{https://archives.esac.esa.int/gaia}.
Acknowledgements are given in App.~\ref{app:acknowledgements}.
\end{acknowledgements}

\bibliographystyle{aa} 
\bibliography{refs} 

\begin{thebibliography}{29}
\expandafter\ifx\csname natexlab\endcsname\relax\def\natexlab#1{#1}\fi

\bibitem[{{Akeson} {et~al.}(2019){Akeson}, {Armus}, {Bachelet}, {Bailey},
  {Bartusek}, {Bellini}, {Benford}, {Bennett}, {Bhattacharya}, {Bohlin},
  {Boyer}, {Bozza}, {Bryden}, {Calchi Novati}, {Carpenter}, {Casertano},
  {Choi}, {Content}, {Dayal}, {Dressler}, {Dor{\'e}}, {Fall}, {Fan}, {Fang},
  {Filippenko}, {Finkelstein}, {Foley}, {Furlanetto}, {Kalirai}, {Gaudi},
  {Gilbert}, {Girard}, {Grady}, {Greene}, {Guhathakurta}, {Heinrich},
  {Hemmati}, {Hendel}, {Henderson}, {Henning}, {Hirata}, {Ho}, {Huff},
  {Hutter}, {Jansen}, {Jha}, {Johnson}, {Jones}, {Kasdin}, {Kelly}, {Kirshner},
  {Koekemoer}, {Kruk}, {Lewis}, {Macintosh}, {Madau}, {Malhotra}, {Mandel},
  {Massara}, {Masters}, {McEnery}, {McQuinn}, {Melchior}, {Melton},
  {Mennesson}, {Peeples}, {Penny}, {Perlmutter}, {Pisani}, {Plazas}, {Poleski},
  {Postman}, {Ranc}, {Rauscher}, {Rest}, {Roberge}, {Robertson}, {Rodney},
  {Rhoads}, {Rhodes}, {Ryan}, {Sahu}, {Sand}, {Scolnic}, {Seth}, {Shvartzvald},
  {Siellez}, {Smith}, {Spergel}, {Stassun}, {Street}, {Strolger}, {Szalay},
  {Trauger}, {Troxel}, {Turnbull}, {van der Marel}, {von der Linden}, {Wang},
  {Weinberg}, {Williams}, {Windhorst}, {Wollack}, {Wu}, {Yee}, \&
  {Zimmerman}}]{wfirst}
{Akeson}, R., {Armus}, L., {Bachelet}, E., {et~al.} 2019, arXiv e-prints,
  arXiv:1902.05569

\bibitem[{{Altavilla} {et~al.}(2021){Altavilla}, {Marinoni}, {Pancino},
  {Galleti}, {Bellazzini}, {Sanna}, {Rainer}, {Tessicini}, {Carrasco},
  {Bragaglia}, {Schuster}, {Cocozza}, {Gebran}, {Voss}, {Federici}, {Masana},
  {Jordi}, {Mongui{\'o}}, {Castro}, {Pe{\~n}a-Guerrero}, \&
  {P{\'e}rez-Villegas}}]{Altavilla2021}
{Altavilla}, G., {Marinoni}, S., {Pancino}, E., {et~al.} 2021, \mnras, 501,
  2848

\bibitem[{{Altavilla} {et~al.}(2015){Altavilla}, {Marinoni}, {Pancino},
  {Galleti}, {Ragaini}, {Bellazzini}, {Cocozza}, {Bragaglia}, {Carrasco},
  {Castro}, {Di Fabrizio}, {Federici}, {Figueras}, {Gebran}, {Jordi}, {Masana},
  {Schuster}, {Valentini}, \& {Voss}}]{Altavilla2015}
{Altavilla}, G., {Marinoni}, S., {Pancino}, E., {et~al.} 2015, Astronomische
  Nachrichten, 336, 515

\bibitem[{{Andrae, R. et al}(2022)}]{GSPPhot}
{Andrae, R. et al}. 2022, \aap\ in prep.

\bibitem[{{Babusiaux, C. et al}(2022)}]{DPACP-127}
{Babusiaux, C. et al}. 2022, \aap\ in prep.

\bibitem[{{Carrasco} {et~al.}(2016){Carrasco}, {Evans}, {Montegriffo}, {Jordi},
  {van Leeuwen}, {Riello}, {Voss}, {De Angeli}, {Busso}, {Fabricius},
  {Cacciari}, {Weiler}, {Pancino}, {Brown}, {Holland}, {Burgess}, {Osborne},
  {Altavilla}, {Gebran}, {Ragaini}, {Galleti}, {Cocozza}, {Marinoni},
  {Bellazzini}, {Bragaglia}, {Federici}, \&
  {Balaguer-N{\'u}{\~n}ez}}]{Carrasco2016}
{Carrasco}, J.~M., {Evans}, D.~W., {Montegriffo}, P., {et~al.} 2016, \aap, 595,
  A7

\bibitem[{{Carrasco} {et~al.}(2021){Carrasco}, {Weiler}, {Jordi}, {Fabricius},
  {De Angeli}, {Evans}, {van Leeuwen}, {Riello}, \&
  {Montegriffo}}]{Carrasco2021}
{Carrasco}, J.~M., {Weiler}, M., {Jordi}, C., {et~al.} 2021, \aap, 652, A86

\bibitem[{{Clementini, G. et al.}(2022)}]{DR3-DPACP-168}
{Clementini, G. et al.} 2022, \aap\ in prep.

\bibitem[{{Costille} {et~al.}(2016){Costille}, {Caillat}, {Rossin}, {Pascal},
  {Sanchez}, {Foulon}, \& {Vives}}]{euclid}
{Costille}, A., {Caillat}, A., {Rossin}, C., {et~al.} 2016, in Society of
  Photo-Optical Instrumentation Engineers (SPIE) Conference Series, Vol. 9912,
  Advances in Optical and Mechanical Technologies for Telescopes and
  Instrumentation II, ed. R.~{Navarro} \& J.~H. {Burge}, 99122C

\bibitem[{{Creevey, O. et al.}(2022)}]{DR3-DPACP-157}
{Creevey, O. et al.} 2022, \aap\ in prep.

\bibitem[{Dean \& Ghemawat(2008)}]{MapReduce}
Dean, J. \& Ghemawat, S. 2008, Commun. ACM, 51, 107

\bibitem[{{Evans, D.~W. et al.}(2022)}]{DR3-DPACP-142}
{Evans, D.~W. et al.} 2022, \aap\ in prep.

\bibitem[{{Fabricius} {et~al.}(2016){Fabricius}, {Bastian}, {Portell},
  {Casta{\~n}eda}, {Davidson}, {Hambly}, {Clotet}, {Biermann}, {Mora},
  {Busonero}, {Riva}, {Brown}, {Smart}, {Lammers}, {Torra}, {Drimmel},
  {Gracia}, {L{\"o}ffler}, {Spagna}, {Lindegren}, {Klioner}, {Andrei}, {Bach},
  {Bramante}, {Br{\"u}semeister}, {Busso}, {Carrasco}, {Gai}, {Garralda},
  {Gonz{\'a}lez-Vidal}, {Guerra}, {Hauser}, {Jordan}, {Jordi}, {Lenhardt},
  {Mignard}, {Messineo}, {Mulone}, {Serraller}, {Stampa}, {Tanga}, {van
  Elteren}, {van Reeven}, {Voss}, {Abbas}, {Allasia}, {Altmann}, {Anton},
  {Barache}, {Becciani}, {Berthier}, {Bianchi}, {Bombrun}, {Bouquillon},
  {Bourda}, {Bucciarelli}, {Butkevich}, {Buzzi}, {Cancelliere}, {Carlucci},
  {Charlot}, {Collins}, {Comoretto}, {Cross}, {Crosta}, {de Felice}, {Fienga},
  {Figueras}, {Fraile}, {Geyer}, {Hernandez}, {Hobbs}, {Hofmann}, {Liao},
  {Licata}, {Martino}, {McMillan}, {Michalik}, {Morbidelli}, {Parsons},
  {Pecoraro}, {Ramos-Lerate}, {Sarasso}, {Siddiqui}, {Steele},
  {Steidelm{\"u}ller}, {Taris}, {Vecchiato}, {Abreu}, {Anglada}, {Boudreault},
  {Cropper}, {Holl}, {Cheek}, {Crowley}, {Fleitas}, {Hutton}, {Osinde},
  {Rowell}, {Salguero}, {Utrilla}, {Blagorodnova}, {Soffel}, {Osorio},
  {Vicente}, {Cambras}, \& {Bernstein}}]{Fabricius2016}
{Fabricius}, C., {Bastian}, U., {Portell}, J., {et~al.} 2016, \aap, 595, A3

\bibitem[{{Gaia Collaboration} \& et~al.(2022)}]{DR3-top-level}
{Gaia Collaboration} \& et~al., V. 2022, \aap\ in prep.

\bibitem[{{Gaia Collaboration} {et~al.}(2016){Gaia Collaboration}, {Prusti},
  {de Bruijne}, {Brown}, {Vallenari}, {Babusiaux}, {Bailer-Jones}, {Bastian},
  {Biermann}, {Evans}, {Eyer}, {Jansen}, {Jordi}, {Klioner}, {Lammers},
  {Lindegren}, {Luri}, {Mignard}, {Milligan}, {Panem}, {Poinsignon},
  {Pourbaix}, {Randich}, {Sarri}, {Sartoretti}, {Siddiqui}, {Soubiran},
  {Valette}, {van Leeuwen}, {Walton}, {Aerts}, {Arenou}, {Cropper}, {Drimmel},
  {H{\o}g}, {Katz}, {Lattanzi}, {O'Mullane}, {Grebel}, {Holland}, {Huc},
  {Passot}, {Bramante}, {Cacciari}, {Casta{\~n}eda}, {Chaoul}, {Cheek}, {De
  Angeli}, {Fabricius}, {Guerra}, {Hern{\'a}ndez}, {Jean-Antoine-Piccolo},
  {Masana}, {Messineo}, {Mowlavi}, {Nienartowicz}, {Ord{\'o}{\~n}ez-Blanco},
  {Panuzzo}, {Portell}, {Richards}, {Riello}, {Seabroke}, {Tanga},
  {Th{\'e}venin}, {Torra}, {Els}, {Gracia-Abril}, {Comoretto},
  {Garcia-Reinaldos}, {Lock}, {Mercier}, {Altmann}, {Andrae}, {Astraatmadja},
  {Bellas-Velidis}, {Benson}, {Berthier}, {Blomme}, {Busso}, {Carry},
  {Cellino}, {Clementini}, {Cowell}, {Creevey}, {Cuypers}, {Davidson}, {De
  Ridder}, {de Torres}, {Delchambre}, {Dell'Oro}, {Ducourant}, {Fr{\'e}mat},
  {Garc{\'\i}a-Torres}, {Gosset}, {Halbwachs}, {Hambly}, {Harrison}, {Hauser},
  {Hestroffer}, {Hodgkin}, {Huckle}, {Hutton}, {Jasniewicz}, {Jordan},
  {Kontizas}, {Korn}, {Lanzafame}, {Manteiga}, {Moitinho}, {Muinonen},
  {Osinde}, {Pancino}, {Pauwels}, {Petit}, {Recio-Blanco}, {Robin}, {Sarro},
  {Siopis}, {Smith}, {Smith}, {Sozzetti}, {Thuillot}, {van Reeven}, {Viala},
  {Abbas}, {Abreu Aramburu}, {Accart}, {Aguado}, {Allan}, {Allasia},
  {Altavilla}, {{\'A}lvarez}, {Alves}, {Anderson}, {Andrei}, {Anglada Varela},
  {Antiche}, {Antoja}, {Ant{\'o}n}, {Arcay}, {Atzei}, {Ayache}, {Bach},
  {Baker}, {Balaguer-N{\'u}{\~n}ez}, {Barache}, {Barata}, {Barbier}, {Barblan},
  {Baroni}, {Barrado y Navascu{\'e}s}, {Barros}, {Barstow}, {Becciani},
  {Bellazzini}, {Bellei}, {Bello Garc{\'\i}a}, {Belokurov}, {Bendjoya},
  {Berihuete}, {Bianchi}, {Bienaym{\'e}}, {Billebaud}, {Blagorodnova},
  {Blanco-Cuaresma}, {Boch}, {Bombrun}, {Borrachero}, {Bouquillon}, {Bourda},
  {Bouy}, {Bragaglia}, {Breddels}, {Brouillet}, {Br{\"u}semeister},
  {Bucciarelli}, {Budnik}, {Burgess}, {Burgon}, {Burlacu}, {Busonero}, {Buzzi},
  {Caffau}, {Cambras}, {Campbell}, {Cancelliere}, {Cantat-Gaudin}, {Carlucci},
  {Carrasco}, {Castellani}, {Charlot}, {Charnas}, {Charvet}, {Chassat},
  {Chiavassa}, {Clotet}, {Cocozza}, {Collins}, {Collins}, {Costigan}, {Crifo},
  {Cross}, {Crosta}, {Crowley}, {Dafonte}, {Damerdji}, {Dapergolas}, {David},
  {David}, {De Cat}, {de Felice}, {de Laverny}, {De Luise}, {De March}, {de
  Martino}, {de Souza}, {Debosscher}, {del Pozo}, {Delbo}, {Delgado},
  {Delgado}, {di Marco}, {Di Matteo}, {Diakite}, {Distefano}, {Dolding}, {Dos
  Anjos}, {Drazinos}, {Dur{\'a}n}, {Dzigan}, {Ecale}, {Edvardsson}, {Enke},
  {Erdmann}, {Escolar}, {Espina}, {Evans}, {Eynard Bontemps}, {Fabre},
  {Fabrizio}, {Faigler}, {Falc{\~a}o}, {Farr{\`a}s Casas}, {Faye}, {Federici},
  {Fedorets}, {Fern{\'a}ndez-Hern{\'a}ndez}, {Fernique}, {Fienga}, {Figueras},
  {Filippi}, {Findeisen}, {Fonti}, {Fouesneau}, {Fraile}, {Fraser}, {Fuchs},
  {Furnell}, {Gai}, {Galleti}, {Galluccio}, {Garabato}, {Garc{\'\i}a-Sedano},
  {Gar{\'e}}, {Garofalo}, {Garralda}, {Gavras}, {Gerssen}, {Geyer}, {Gilmore},
  {Girona}, {Giuffrida}, {Gomes}, {Gonz{\'a}lez-Marcos},
  {Gonz{\'a}lez-N{\'u}{\~n}ez}, {Gonz{\'a}lez-Vidal}, {Granvik}, {Guerrier},
  {Guillout}, {Guiraud}, {G{\'u}rpide}, {Guti{\'e}rrez-S{\'a}nchez}, {Guy},
  {Haigron}, {Hatzidimitriou}, {Haywood}, {Heiter}, {Helmi}, {Hobbs},
  {Hofmann}, {Holl}, {Holland}, {Hunt}, {Hypki}, {Icardi}, {Irwin}, {Jevardat
  de Fombelle}, {Jofr{\'e}}, {Jonker}, {Jorissen}, {Julbe}, {Karampelas},
  {Kochoska}, {Kohley}, {Kolenberg}, {Kontizas}, {Koposov}, {Kordopatis},
  {Koubsky}, {Kowalczyk}, {Krone-Martins}, {Kudryashova}, {Kull}, {Bachchan},
  {Lacoste-Seris}, {Lanza}, {Lavigne}, {Le Poncin-Lafitte}, {Lebreton},
  {Lebzelter}, {Leccia}, {Leclerc}, {Lecoeur-Taibi}, {Lemaitre}, {Lenhardt},
  {Leroux}, {Liao}, {Licata}, {Lindstr{\o}m}, {Lister}, {Livanou}, {Lobel},
  {L{\"o}ffler}, {L{\'o}pez}, {Lopez-Lozano}, {Lorenz}, {Loureiro},
  {MacDonald}, {Magalh{\~a}es Fernandes}, {Managau}, {Mann}, {Mantelet},
  {Marchal}, {Marchant}, {Marconi}, {Marie}, {Marinoni}, {Marrese},
  {Marschalk{\'o}}, {Marshall}, {Mart{\'\i}n-Fleitas}, {Martino}, {Mary},
  {Matijevi{\v{c}}}, {Mazeh}, {McMillan}, {Messina}, {Mestre}, {Michalik},
  {Millar}, {Miranda}, {Molina}, {Molinaro}, {Molinaro}, {Moln{\'a}r},
  {Moniez}, {Montegriffo}, {Monteiro}, {Mor}, {Mora}, {Morbidelli}, {Morel},
  {Morgenthaler}, {Morley}, {Morris}, {Mulone}, {Muraveva}, {Musella},
  {Narbonne}, {Nelemans}, {Nicastro}, {Noval}, {Ord{\'e}novic},
  {Ordieres-Mer{\'e}}, {Osborne}, {Pagani}, {Pagano}, {Pailler}, {Palacin},
  {Palaversa}, {Parsons}, {Paulsen}, {Pecoraro}, {Pedrosa}, {Pentik{\"a}inen},
  {Pereira}, {Pichon}, {Piersimoni}, {Pineau}, {Plachy}, {Plum}, {Poujoulet},
  {Pr{\v{s}}a}, {Pulone}, {Ragaini}, {Rago}, {Rambaux}, {Ramos-Lerate},
  {Ranalli}, {Rauw}, {Read}, {Regibo}, {Renk}, {Reyl{\'e}}, {Ribeiro},
  {Rimoldini}, {Ripepi}, {Riva}, {Rixon}, {Roelens}, {Romero-G{\'o}mez},
  {Rowell}, {Royer}, {Rudolph}, {Ruiz-Dern}, {Sadowski}, {Sagrist{\`a}
  Sell{\'e}s}, {Sahlmann}, {Salgado}, {Salguero}, {Sarasso}, {Savietto},
  {Schnorhk}, {Schultheis}, {Sciacca}, {Segol}, {Segovia}, {Segransan},
  {Serpell}, {Shih}, {Smareglia}, {Smart}, {Smith}, {Solano}, {Solitro},
  {Sordo}, {Soria Nieto}, {Souchay}, {Spagna}, {Spoto}, {Stampa}, {Steele},
  {Steidelm{\"u}ller}, {Stephenson}, {Stoev}, {Suess}, {S{\"u}veges}, {Surdej},
  {Szabados}, {Szegedi-Elek}, {Tapiador}, {Taris}, {Tauran}, {Taylor},
  {Teixeira}, {Terrett}, {Tingley}, {Trager}, {Turon}, {Ulla}, {Utrilla},
  {Valentini}, {van Elteren}, {Van Hemelryck}, {van Leeuwen}, {Varadi},
  {Vecchiato}, {Veljanoski}, {Via}, {Vicente}, {Vogt}, {Voss}, {Votruba},
  {Voutsinas}, {Walmsley}, {Weiler}, {Weingrill}, {Werner}, {Wevers},
  {Whitehead}, {Wyrzykowski}, {Yoldas}, {{\v{Z}}erjal}, {Zucker}, {Zurbach},
  {Zwitter}, {Alecu}, {Allen}, {Allende Prieto}, {Amorim},
  {Anglada-Escud{\'e}}, {Arsenijevic}, {Azaz}, {Balm}, {Beck}, {Bernstein},
  {Bigot}, {Bijaoui}, {Blasco}, {Bonfigli}, {Bono}, {Boudreault}, {Bressan},
  {Brown}, {Brunet}, {Bunclark}, {Buonanno}, {Butkevich}, {Carret}, {Carrion},
  {Chemin}, {Ch{\'e}reau}, {Corcione}, {Darmigny}, {de Boer}, {de Teodoro}, {de
  Zeeuw}, {Delle Luche}, {Domingues}, {Dubath}, {Fodor}, {Fr{\'e}zouls},
  {Fries}, {Fustes}, {Fyfe}, {Gallardo}, {Gallegos}, {Gardiol}, {Gebran},
  {Gomboc}, {G{\'o}mez}, {Grux}, {Gueguen}, {Heyrovsky}, {Hoar}, {Iannicola},
  {Isasi Parache}, {Janotto}, {Joliet}, {Jonckheere}, {Keil}, {Kim},
  {Klagyivik}, {Klar}, {Knude}, {Kochukhov}, {Kolka}, {Kos}, {Kutka}, {Lainey},
  {LeBouquin}, {Liu}, {Loreggia}, {Makarov}, {Marseille}, {Martayan},
  {Martinez-Rubi}, {Massart}, {Meynadier}, {Mignot}, {Munari}, {Nguyen},
  {Nordlander}, {Ocvirk}, {O'Flaherty}, {Olias Sanz}, {Ortiz}, {Osorio},
  {Oszkiewicz}, {Ouzounis}, {Palmer}, {Park}, {Pasquato}, {Peltzer}, {Peralta},
  {P{\'e}turaud}, {Pieniluoma}, {Pigozzi}, {Poels}, {Prat}, {Prod'homme},
  {Raison}, {Rebordao}, {Risquez}, {Rocca-Volmerange}, {Rosen}, {Ruiz-Fuertes},
  {Russo}, {Sembay}, {Serraller Vizcaino}, {Short}, {Siebert}, {Silva},
  {Sinachopoulos}, {Slezak}, {Soffel}, {Sosnowska}, {Strai{\v{z}}ys}, {ter
  Linden}, {Terrell}, {Theil}, {Tiede}, {Troisi}, {Tsalmantza}, {Tur},
  {Vaccari}, {Vachier}, {Valles}, {Van Hamme}, {Veltz}, {Virtanen}, {Wallut},
  {Wichmann}, {Wilkinson}, {Ziaeepour}, \& {Zschocke}}]{Prusti2016}
{Gaia Collaboration}, {Prusti}, T., {de Bruijne}, J.~H.~J., {et~al.} 2016,
  \aap, 595, A1

\bibitem[{{Galluccio, L. et al.}(2022)}]{DR3-DPACP-89}
{Galluccio, L. et al.} 2022, \aap\ in prep.

\bibitem[{Harris {et~al.}(2020)Harris, Millman, van~der Walt, Gommers,
  Virtanen, Cournapeau, Wieser, Taylor, Berg, Smith, Kern, Picus, Hoyer, van
  Kerkwijk, Brett, Haldane, del R{'{\i}}o, Wiebe, Peterson,
  G{'{e}}rard-Marchant, Sheppard, Reddy, Weckesser, Abbasi, Gohlke, \&
  Oliphant}]{harris2020array}
Harris, C.~R., Millman, K.~J., van~der Walt, S.~J., {et~al.} 2020, Nature, 585,
  357

\bibitem[{Hunter(2007)}]{Hunter:2007}
Hunter, J.~D. 2007, Computing In Science \& Engineering, 9, 90

\bibitem[{{Jordi} {et~al.}(2006){Jordi}, {H{\o}g}, {Brown}, {Lindegren},
  {Bailer-Jones}, {Carrasco}, {Knude}, {Strai{\v{z}}ys}, {de Bruijne},
  {Claeskens}, {Drimmel}, {Figueras}, {Grenon}, {Kolka}, {Perryman},
  {Tautvai{\v{s}}ien{\.{e}}}, {Vansevi{\v{c}}ius}, {Willemsen},
  {Brid{\v{z}}ius}, {Evans}, {Fabricius}, {Fiorucci}, {Heiter}, {Kaempf},
  {Kazlauskas}, {Ku{\v{c}}inskas}, {Malyuto}, {Munari}, {Reyl{\'e}}, {Torra},
  {Vallenari}, {Zdanavi{\v{c}}ius}, {Korakitis}, {Malkov}, \&
  {Smette}}]{Jordi2006}
{Jordi}, C., {H{\o}g}, E., {Brown}, A.~G.~A., {et~al.} 2006, \mnras, 367, 290

\bibitem[{{Marinoni} {et~al.}(2016){Marinoni}, {Pancino}, {Altavilla},
  {Bellazzini}, {Galleti}, {Tessicini}, {Valentini}, {Cocozza}, {Ragaini},
  {Braga}, {Bragaglia}, {Federici}, {Schuster}, {Carrasco}, {Castro},
  {Figueras}, \& {Jordi}}]{Marinoni2016}
{Marinoni}, S., {Pancino}, E., {Altavilla}, G., {et~al.} 2016, \mnras, 462,
  3616

\bibitem[{{Montegriffo, P. et al.}(2022{\natexlab{a}})}]{Montegriffo2022}
{Montegriffo, P. et al.} 2022{\natexlab{a}}, \aap\ in prep.

\bibitem[{{Montegriffo, P. et al.}(2022{\natexlab{b}})}]{DR3-DPACP-93}
{Montegriffo, P. et al.} 2022{\natexlab{b}}, \aap\ in prep.

\bibitem[{Pancino {et~al.}(2021)Pancino, Sanna, Altavilla, Marinoni, Rainer,
  Cocozza, Ragaini, Galleti, Bellazzini, Bragaglia, Tessicini, Voss, Carrasco,
  Jordi, Harrison, De Angeli, Evans, \& Fanari}]{Pancino2021}
Pancino, E., Sanna, N., Altavilla, G., {et~al.} 2021, Monthly Notices of the
  Royal Astronomical Society, 503, 3660

\bibitem[{P\'erez \& Granger(2007)}]{PER-GRA:2007}
P\'erez, F. \& Granger, B.~E. 2007, Computing in Science and Engineering, 9, 21

\bibitem[{{Pickering}(1890)}]{pickering}
{Pickering}, E.~C. 1890, Annals of Harvard College Observatory, 27, 1

\bibitem[{{Riello} {et~al.}(2018){Riello}, {De Angeli}, {Evans}, {Busso},
  {Hambly}, {Davidson}, {Burgess}, {Montegriffo}, {Osborne}, {Kewley},
  {Carrasco}, {Fabricius}, {Jordi}, {Cacciari}, {van Leeuwen}, \&
  {Holland}}]{Riello2018}
{Riello}, M., {De Angeli}, F., {Evans}, D.~W., {et~al.} 2018, \aap, 616, A3

\bibitem[{{Riello} {et~al.}(2021){Riello}, {De Angeli}, {Evans}, {Montegriffo},
  {Carrasco}, {Busso}, {Palaversa}, {Burgess}, {Diener}, {Davidson}, {Rowell},
  {Fabricius}, {Jordi}, {Bellazzini}, {Pancino}, {Harrison}, {Cacciari}, {van
  Leeuwen}, {Hambly}, {Hodgkin}, {Osborne}, {Altavilla}, {Barstow}, {Brown},
  {Castellani}, {Cowell}, {De Luise}, {Gilmore}, {Giuffrida}, {Hidalgo},
  {Holland}, {Marinoni}, {Pagani}, {Piersimoni}, {Pulone}, {Ragaini}, {Rainer},
  {Richards}, {Sanna}, {Walton}, {Weiler}, \& {Yoldas}}]{Riello2021}
{Riello}, M., {De Angeli}, F., {Evans}, D.~W., {et~al.} 2021, \aap, 649, A3

\bibitem[{{Taylor}(2005)}]{Taylor2005}
{Taylor}, M.~B. 2005, in Astronomical Society of the Pacific Conference Series,
  Vol. 347, Astronomical Data Analysis Software and Systems XIV, ed.
  P.~{Shopbell}, M.~{Britton}, \& R.~{Ebert}, 29

\bibitem[{{van Leeuwen}(2007)}]{fvl2007}
{van Leeuwen}, F. 2007, {Hipparcos, the New Reduction of the Raw Data}, Vol.
  350

\end{thebibliography}

\begin{appendix} 

\section{Downloading BP/RP data from the \textit{Gaia} DR3 archive}\label{sec:archive}

Not all sources included in \gdr{3} will have \xp spectra available. 
The main \texttt{gaia\_source} table in the archive contains a field \texttt{has\_xp\_continuous} that is \texttt{true} if a \xp spectrum is available for that source. Users can therefore query the \texttt{gaia\_source} table to select sources with their favourite combination of parameters and use the additional criterion \texttt{has\_xp\_continuous='true'} to restrict their selection to sources that have \xp spectra available from the archive.

The support of the Datalink feature in the archive includes an independent service for the serialization of the \xp spectra. Other types of data such as photometric light curves are served using similar services. A dedicated tutorial is available \href{https://www.cosmos.esa.int/web/gaia-users/archive/ancillary-data#tutorial_datalinklc}{here}.

In this section we provide an example of how to download \xp spectra using the Python programming language. By splitting the list of sources identifiers (\texttt{ids['source\_id']} in the following code snippet), users can overcome the Datalink limitation on the number of sources. A bulk download option will also be implemented for users interested in getting all the \xp spectra in \gdr{3}.
\begin{lstlisting}
from astroquery.gaia import GaiaClass

# Connect to Gaia archive
gaia = GaiaClass(gaia_tap_server='https://gea.esac.esa.int/', gaia_data_server='https://gea.esac.esa.int/')
gaia.login()

# Run your ADQL query to get a list of source_ids
example_query = "select TOP 1000 source_id from gaiadr3.gaia_source where has_xp_continuous = 'True'"
job = gaia.launch_job_async(example_query, dump_to_file=False) 
ids = job.get_results()

# Now retrieve the BP/RP mean spectra in the continuous representation
result = gaia.load_data(ids=ids['source_id'], format='csv', data_release='Gaia DR3', data_structure='raw', retrieval_type='XP_CONTINUOUS', avoid_datatype_check=True)

# Result will be a dictionary, so you can check the available keys by running result.keys()
# In this example we are looking in particular for the XP_CONTINUOUS_RAW key
continuous_key = [key for key in result.keys() if 'continuous' in key.lower()][0] 
# The first element is the result we want as an Astropy table
data = result[continuous_key][0] 
# Astropy has a 'write' method for tables
# Write the table to CSV
data.write('filename.csv', format='csv') 
\end{lstlisting}

The data can be downloaded in different file formats. For a complete list of the available formats and for instructions on alternative download procedures, please refer to the archive pages and tutorials.

Once downloaded, the files can be given in input to GaiaXPy utilities to obtain sampled spectra or synthetic photometry. GaiaXPy also offers the possibility of providing a list of source IDs. In this case the download of the spectra from the archive is done within the GaiaXPy utility (users will be prompted for credentials). 

\section{Data format details}\label{sec:data_format}

This Section provides more detailed information on the structure of the data representing \xp mean spectra in the archive.
For completeness, all fields are described here, even though some have been mentioned and explained in the main part of the paper. Detailed descriptions are also available from the \gdr{3} documentation and from the archive documentation.

We first describe the fields available via DataLink when retrieving \texttt{XP\_CONTINUOUS} data:
\begin{itemize}
\item\texttt{source\_id} Source identifier. Among other information, this encodes the approximate position of the source in the equatorial system (ICRS) using the nested HEALPix scheme at level 12 ($Nside = 4096$), which divides the sky into $\simeq 200$~million pixels of about 0.7~arcmin$^2$.
\item\texttt{bp/rp\_basis\_function\_id} Identifier of the set of bases functions used in the Source Update process (see Sect. \ref{sec:processing:intcal}). Different sets were used during trial runs and validation but all the released spectra were created using the same set of bases. This implies that the identifier in \gdr{3} is different for BP and RP spectra, but the same for all sources in each band. When sampling the spectra in the internal reference system, care must be taken to ensure that the right basis configuration is used. 
\item\texttt{bp/rp\_degrees\_of\_freedom} Number of degrees of freedom in the Source Update least squares solution.
\item\texttt{bp/rp\_n\_parameters} Number of parameters in the Source Update least squares solution. This will be always 55 for the \gdr{3} \xp spectra.
\item\texttt{bp/rp\_n\_measurements} Number of measurements contributing to the Source Update least squares solution. This counts the single samples contributing rather than full epoch spectra.
\item\texttt{bp/rp\_n\_rejected\_measurements} Number of samples rejected in the Source Update least squares solution. This is based on a $k$-sigma rejection algorithm.
\item\texttt{bp/rp\_standard\_deviation} The final standard deviation of the Source Update least squares solution for this \xp and source.
\item\texttt{bp/rp\_chi\_squared} The $\chi^2$ of the Source Update least squares solution for this \xp and source.
\item\texttt{bp/rp\_coefficients} The array of coefficients of the mean spectrum representation as a superposition of basis functions. These are the $b_{s,n}$ in Eq. \ref{eq:mean:spec:bases}. This array will have length equal to \texttt{bp/rp\_n\_parameters}.
\item\texttt{bp/rp\_coefficient\_errors} The errors on the coefficients, one error per coefficient. This array will have length equal to \texttt{bp/rp\_n\_parameters}. The errors in this array are computed multiplying the formal errors (as obtained from the covariance matrix of the source update least square solution) by the standard deviation of the solution. This is a standard methodology and can also account for when the modelling of the data introduces a systematic that adds a pseudo-random error to the individual input data not accounted for in quoted errors.
\item\texttt{bp/rp\_coefficient\_correlations} The matrix containing the information on correlations between coefficients. Only the elements located in the upper triangular section of the matrix, excluding the diagonal where all elements are equal to 1.0 by definition, are stored as an array of constant size $n\,(n-1)/2$ where $n$ is equal to \texttt{bp/rp\_n\_parameters}. The order of the elements in the linear array follows a column-major scheme, i.e. for $n=55$
$${\bf M} = \begin{bmatrix} 
1   & C[0] & C[1] & C[3] & C[6] & \cdots & C[1431]\\
    &  1   & C[2] & C[4] & C[7] & \cdots & C[1432]\\
    &      &  1   & C[5] & C[8] & \cdots & C[1433]\\
    &      &      &  1   & C[9] & \cdots & C[1434]\\
    &      &      &      &  1   & \cdots & \vdots \\
    &      &      &      &      &   1    & C[1484]\\
    &      &      &      &      &        &   1    \\
\end{bmatrix}$$
\item\texttt{bp/rp\_n\_relevant\_bases} Number of coefficients that were considered above the noise according to the criterion described in Sect. \ref{sec:truncation}.
\item\texttt{bp/rp\_n\_relative\_shrinking} Ratio between the L2-norm of the truncated and full \xp spectrum. 
\end{itemize}

In the following, we also describe the additional fields available in the \texttt{xp\_summary} table (fields that duplicate information given in the above data structure are not repeated here): 
\begin{itemize}
\item\texttt{bp/rp\_n\_transits} Number of epoch spectra contributing to the mean spectrum.
\item\texttt{bp/rp\_n\_contaminated\_transits} Number of transits assessed as contaminated among those that contributed to the mean spectrum. A transits is considered contaminated when some of the flux within the window is estimated to come from a nearby (on the focal plane) source located outside the acquired window. Crowding assessment for \gdr{3} was based on the \gdr{2} source catalogue. The contaminating flux was estimated as detailed in Sect. 3.1 in \cite{Riello2021}.
\item\texttt{bp/rp\_n\_blended\_transits} Number of transits assessed as blended among those that contributed to the mean spectrum. A transit is considered blended when more than one source is within the acquires window. A transit is flagged as blended also when the non-target source is just outside the window (within 5 TDI periods in the AL direction and 2 pixel in the AC direction).
\end{itemize}

\section{Bases configuration and spectrum sampling}

The optimised bases finally adopted to represent the \gdr{3} mean spectra are defined as an orthogonal transformation of the first $N$ Hermite functions. The orthogonal transformations are different for BP and RP, and the $N \times N$ transformation matrices are denoted ${\bf V}_{BP}$ and ${\bf V}_{RP}$, respectively, where $N=55$ for both. The two transformation matrices are embedded in the Python package GaiaXPy, that uses them when computing sampled mean spectra in the internal reference system. The same \texttt{xml} configuration file used in GaiaXPy is also available via Zenodo.

Users that prefer to use this file directly rather than relying on GaiaXPy will have to pay attention to the following: 
\begin{itemize}
    \item The file contains a \texttt{bpConfig} and an \texttt{rpConfig} element. Each configuration element is identified with an unique id (\texttt{uniqueId}) which must agree with the \texttt{bp/rp\_basis\_function\_id} parameter in the \gdr{3} \xp spectral data.
    \item The ranges \texttt{range} and \texttt{normalizedRange} give the conversion rule from the pseudo-wavelength system to the argument of the Hermite functions. With reference to Eq. \ref{eq:mean:spec:bases}, the scaling factor $\Theta$ will be given by $\Theta = (r_{+}-r_{-}) / (n_{+}-n_{-})$ while the offset $\Delta \theta$ will be given by $\Delta \theta = r_{-} - n_{-} \cdot\Theta$ where $r_{\pm}$ and $n_{\pm}$ are used to indicate the higher ($+$) and lower ($-$) boundaries of the ranges \texttt{range} and \texttt{normalizedRange} respectively. 
    \item The element \texttt{transformationMatrix} lists all matrix elements for ${\bf V}_{BP}$ and ${\bf V}_{RP}$, stored in a row-major scheme. 
\end{itemize}

The sampled spectrum on a discrete grid of $n$ pseudo-wavelengths $u = [u_i]_{i=1,\ldots,n}$ is computed easily in a matrix formalism. First, the values of the first $N$ Hermite functions are computed on the pseudo-wavelength grid and arranged into an $n \times N$ matrix ${\bf D}$. The elements of this matrix are
\begin{equation}
    D_{i,j} = \varphi_{j-1}\left(\frac{u_i-\Delta \theta}{\Theta}\right)\; .
\end{equation}
Multiplying this matrix with ${\bf V}_{BP/RP}^{\mathsf T}$ from the right transforms from Hermite functions to the optimised Hermite basis. The sampled spectrum $f(u)$ is thus obtained as
\begin{equation}
    f(u) = {\bf D} \, {\bf V}_{BP/RP}^{\mathsf T} \, {\bf c}_{BP/RP}\; .
\end{equation}
The covariance matrix for $f(u)$, ${\bf C}^u$ is
\begin{equation}
    {\bf C}^u = {\bf D}\, {\bf V}_{BP/RP}^{\mathsf T} \, {\bf C}^{BP/RP} \, {\bf V}_{BP/RP} \, {\bf D}^{\mathsf T}\; ,
\end{equation}
with ${\bf C}^{BP/RP}$ the covariance matrix for the coefficient vector ${\bf c}_{BP/RP}$. Correlations might not be negligible in ${\bf C}^u$. In particular if $n > N$, ${\bf C}^u$ is singular.

If users desire to apply the suggested truncation, they will simply have to drop coefficient, coefficient error and associated row/column in the correlation matrix with index larger than \texttt{bp/rp\_n\_relevant\_bases}. Only the first \texttt{bp/rp\_n\_relevant\_bases} columns of the \texttt{transformationMatrix} will be required.

\section{The BP/RP split-epoch validation dataset}\label{sec:split:epoch}

During the validation activities leading to \gdr{3} (see Sects. \ref{sec:validation:errors} and \ref{sec:recommendations:noise}) and in the preparation of \cite{GSPPhot} and \cite{DR3-DPACP-93}, we have found very useful a dataset containing about 43 thousand sources for which two mean spectra per source were generated using only about half of the available epoch spectra (randomly chosen to avoid possible problems due to the distribution in time of their observations). 
This dataset, referred to as BP/RP split-epoch validation dataset, is made available via Zenodo, in the same format used in the archive for mean \xp spectra (with the exception of the truncation-related parameters \texttt{bp/rp\_n\_relevant\_bases} and \texttt{bp/rp\_n\_relative\_shrinking} that will not be available. 
We hope the wider community will find this useful to assess the uncertainties of their particular science cases.

The source list for this dataset was initially defined as a selection of the flux and LSF calibrators but was later augmented to include more bright sources and to increase the number of sources in the magnitude range [11, 12], i.e. around the boundary between 1D and 2D \xp configurations. The dataset covers the magnitude range $4.2 \le G \le 20.7$~mag and the colour range $-0.6 \le \bprp \le 7.1$~mag. 
While the initial selection came from the set of calibrators that were selected to have at least 10 usable FoV transits (thus leading to at least 5 transits when these are split in two groups, although the random generation of the two groups could in fact lead to smaller numbers), the following additions included also sources with fewer transits.
Moreover, the criterium based on the number of FoV transits for the selection of the calibrators was assessed on the number of usable observations and these were then subject to availability of calibrations and outlier rejection which could have the effect of decreasing the number of transits contributing to the mean spectrum below the quoted limit. This implies that this dataset contains mean spectra that have been generated from a number of transits that is lower to the limit adopted for the release. 
About 6 thousand of these sources will not have \xp spectra in \gdr{3}, mostly due to their magnitude being fainter than 17.65 (see Sect. \ref{sec:outputs}. Nevertheless they were not excluded from this dataset as they provide an opportunity to probe uncertainties at fainter magnitudes where some \xp spectra are still released.

Users are strongly discouraged from trying to look for consistency in number of transits and measurements between this dataset and the \gdr{3} catalogue of \xp spectra: rejection and filtering at epoch and sample level will act differently depending on the list of transits available to the software.

\section{\textit{Gaia}-related acronyms}\label{sec:acronyms}

\begin{table}[hp]
    \caption{\gaia-related acronyms used in the paper. Each acronym is also defined at its first occurrence in the paper.}
    \label{tab:acronyms}
    \centering
    \begin{tabular}{l|p{0.5\linewidth}|l}\hline\hline
Acronym & Description & See \\\hline
AC & ACross scan direction & \secref{inputs} \\
AF & Astrometric Field & \secref{inputs} \\
AL & ALong scan direction & \secref{inputs} \\
BP & Blue Photometer & \secref{introduction} \\
CCD(s) & Charge Coupled Device(s) & \secref{inputs} \\
DPAC & Data Processing and Analysis Consortium & \secref{introduction}\\
DR & Data Release & \secref{introduction} \\
ESA & European Space Agency & \secref{introduction} \\
FoV(s) & Field(s) of View & \secref{inputs} \\
GAPS & \gaia Andromeda Photometric Survey & \secref{introduction} \\
LSF & Line Spread Function & \secref{processing:calibrators} \\
OBMT & On-Board Mission Time & \secref{inputs} \\
\obmtrev & On-Board Mission Time in units of satellite revolutions & \secref{inputs} \\
RP & Red Photometer & \secref{introduction} \\
RVS & Radial Velocity Spectrometer & \secref{introduction} \\ 
SED & Spectral Energy Distribution & \secref{outputs} \\
SSC & Spectrum Shape Coefficient & \secref{processing:initial} \\
SSO & Solar System Objects & \secref{processing:intcal} \\
TCB & Barycentric Coordinate Time & \secref{processing:intcal} \\
TDI & Time Delayed Integration & \secref{inputs} \\
WC(s) & Window Class or strategy & \secref{inputs} \\
\end{tabular}
\end{table}

\section{Funding Agency Acknowledgements}\label{app:acknowledgements}

This work presents results from the European Space Agency (ESA) space mission \gaia. \gaia\ data are being processed by the \gaia\ Data Processing and Analysis Consortium (DPAC). Funding for the DPAC is provided by national institutions, in particular the institutions participating in the \gaia\ MultiLateral Agreement (MLA). The \gaia\ mission website is \url{https://www.cosmos.esa.int/gaia}. The \gaia\ archive website is \url{https://archives.esac.esa.int/gaia}.

The \gaia\ mission and data processing have financially been supported by, in alphabetical order by country:
\begin{itemize}
\item the Algerian Centre de Recherche en Astronomie, Astrophysique et G\'{e}ophysique of Bouzareah Observatory;
\item the Austrian Fonds zur F\"{o}rderung der wissenschaftlichen Forschung (FWF) Hertha Firnberg Programme through grants T359, P20046, and P23737;
\item the BELgian federal Science Policy Office (BELSPO) through various PROgramme de D\'{e}veloppement d'Exp\'{e}riences scientifiques (PRODEX) grants and the Polish Academy of Sciences - Fonds Wetenschappelijk Onderzoek through grant VS.091.16N, and the Fonds de la Recherche Scientifique (FNRS), and the Research Council of Katholieke Universiteit (KU) Leuven through grant C16/18/005 (Pushing AsteRoseismology to the next level with TESS, GaiA, and the Sloan DIgital Sky SurvEy -- PARADISE);  
\item the Brazil-France exchange programmes Funda\c{c}\~{a}o de Amparo \`{a} Pesquisa do Estado de S\~{a}o Paulo (FAPESP) and Coordena\c{c}\~{a}o de Aperfeicoamento de Pessoal de N\'{\i}vel Superior (CAPES) - Comit\'{e} Fran\c{c}ais d'Evaluation de la Coop\'{e}ration Universitaire et Scientifique avec le Br\'{e}sil (COFECUB);
\item the Chilean Agencia Nacional de Investigaci\'{o}n y Desarrollo (ANID) through Fondo Nacional de Desarrollo Cient\'{\i}fico y Tecnol\'{o}gico (FONDECYT) Regular Project 1210992 (L.~Chemin);
\item the National Natural Science Foundation of China (NSFC) through grants 11573054, 11703065, and 12173069, the China Scholarship Council through grant 201806040200, and the Natural Science Foundation of Shanghai through grant 21ZR1474100;  
\item the Tenure Track Pilot Programme of the Croatian Science Foundation and the \'{E}cole Polytechnique F\'{e}d\'{e}rale de Lausanne and the project TTP-2018-07-1171 `Mining the Variable Sky', with the funds of the Croatian-Swiss Research Programme;
\item the Czech-Republic Ministry of Education, Youth, and Sports through grant LG 15010 and INTER-EXCELLENCE grant LTAUSA18093, and the Czech Space Office through ESA PECS contract 98058;
\item the Danish Ministry of Science;
\item the Estonian Ministry of Education and Research through grant IUT40-1;
\item the European Commission’s Sixth Framework Programme through the European Leadership in Space Astrometry (\href{https://www.cosmos.esa.int/web/gaia/elsa-rtn-programme}{ELSA}) Marie Curie Research Training Network (MRTN-CT-2006-033481), through Marie Curie project PIOF-GA-2009-255267 (Space AsteroSeismology \& RR Lyrae stars, SAS-RRL), and through a Marie Curie Transfer-of-Knowledge (ToK) fellowship (MTKD-CT-2004-014188); the European Commission's Seventh Framework Programme through grant FP7-606740 (FP7-SPACE-2013-1) for the \gaia\ European Network for Improved data User Services (\href{https://gaia.ub.edu/twiki/do/view/GENIUS/}{GENIUS}) and through grant 264895 for the \gaia\ Research for European Astronomy Training (\href{https://www.cosmos.esa.int/web/gaia/great-programme}{GREAT-ITN}) network;
\item the European Cooperation in Science and Technology (COST) through COST Action CA18104 `Revealing the Milky Way with \gaia (MW-Gaia)';
\item the European Research Council (ERC) through grants 320360, 647208, and 834148 and through the European Union’s Horizon 2020 research and innovation and excellent science programmes through Marie Sk{\l}odowska-Curie grant 745617 (Our Galaxy at full HD -- Gal-HD) and 895174 (The build-up and fate of self-gravitating systems in the Universe) as well as grants 687378 (Small Bodies: Near and Far), 682115 (Using the Magellanic Clouds to Understand the Interaction of Galaxies), 695099 (A sub-percent distance scale from binaries and Cepheids -- CepBin), 716155 (Structured ACCREtion Disks -- SACCRED), 951549 (Sub-percent calibration of the extragalactic distance scale in the era of big surveys -- UniverScale), and 101004214 (Innovative Scientific Data Exploration and Exploitation Applications for Space Sciences -- EXPLORE);
\item the European Science Foundation (ESF), in the framework of the \gaia\ Research for European Astronomy Training Research Network Programme (\href{https://www.cosmos.esa.int/web/gaia/great-programme}{GREAT-ESF});
\item the European Space Agency (ESA) in the framework of the \gaia\ project, through the Plan for European Cooperating States (PECS) programme through contracts C98090 and 4000106398/12/NL/KML for Hungary, through contract 4000115263/15/NL/IB for Germany, and through PROgramme de D\'{e}veloppement d'Exp\'{e}riences scientifiques (PRODEX) grant 4000127986 for Slovenia;  
\item the Academy of Finland through grants 299543, 307157, 325805, 328654, 336546, and 345115 and the Magnus Ehrnrooth Foundation;
\item the French Centre National d’\'{E}tudes Spatiales (CNES), the Agence Nationale de la Recherche (ANR) through grant ANR-10-IDEX-0001-02 for the `Investissements d'avenir' programme, through grant ANR-15-CE31-0007 for project `Modelling the Milky Way in the \gaia era’ (MOD4Gaia), through grant ANR-14-CE33-0014-01 for project `The Milky Way disc formation in the \gaia era’ (ARCHEOGAL), through grant ANR-15-CE31-0012-01 for project `Unlocking the potential of Cepheids as primary distance calibrators’ (UnlockCepheids), through grant ANR-19-CE31-0017 for project `Secular evolution of galxies' (SEGAL), and through grant ANR-18-CE31-0006 for project `Galactic Dark Matter' (GaDaMa), the Centre National de la Recherche Scientifique (CNRS) and its SNO \gaia of the Institut des Sciences de l’Univers (INSU), its Programmes Nationaux: Cosmologie et Galaxies (PNCG), Gravitation R\'{e}f\'{e}rences Astronomie M\'{e}trologie (PNGRAM), Plan\'{e}tologie (PNP), Physique et Chimie du Milieu Interstellaire (PCMI), and Physique Stellaire (PNPS), the `Action F\'{e}d\'{e}ratrice \gaia' of the Observatoire de Paris, the R\'{e}gion de Franche-Comt\'{e}, the Institut National Polytechnique (INP) and the Institut National de Physique nucl\'{e}aire et de Physique des Particules (IN2P3) co-funded by CNES;
\item the German Aerospace Agency (Deutsches Zentrum f\"{u}r Luft- und Raumfahrt e.V., DLR) through grants 50QG0501, 50QG0601, 50QG0602, 50QG0701, 50QG0901, 50QG1001, 50QG1101, 50\-QG1401, 50QG1402, 50QG1403, 50QG1404, 50QG1904, 50QG2101, 50QG2102, and 50QG2202, and the Centre for Information Services and High Performance Computing (ZIH) at the Technische Universit\"{a}t Dresden for generous allocations of computer time;
\item the Hungarian Academy of Sciences through the Lend\"{u}let Programme grants LP2014-17 and LP2018-7 and the Hungarian National Research, Development, and Innovation Office (NKFIH) through grant KKP-137523 (`SeismoLab');
\item the Science Foundation Ireland (SFI) through a Royal Society - SFI University Research Fellowship (M.~Fraser);
\item the Israel Ministry of Science and Technology through grant 3-18143 and the Tel Aviv University Center for Artificial Intelligence and Data Science (TAD) through a grant;
\item the Agenzia Spaziale Italiana (ASI) through contracts I/037/08/0, I/058/10/0, 2014-025-R.0, 2014-025-R.1.2015, and 2018-24-HH.0 to the Italian Istituto Nazionale di Astrofisica (INAF), contract 2014-049-R.0/1/2 to INAF for the Space Science Data Centre (SSDC, formerly known as the ASI Science Data Center, ASDC), contracts I/008/10/0, 2013/030/I.0, 2013-030-I.0.1-2015, and 2016-17-I.0 to the Aerospace Logistics Technology Engineering Company (ALTEC S.p.A.), INAF, and the Italian Ministry of Education, University, and Research (Ministero dell'Istruzione, dell'Universit\`{a} e della Ricerca) through the Premiale project `MIning The Cosmos Big Data and Innovative Italian Technology for Frontier Astrophysics and Cosmology' (MITiC);
\item the Netherlands Organisation for Scientific Research (NWO) through grant NWO-M-614.061.414, through a VICI grant (A.~Helmi), and through a Spinoza prize (A.~Helmi), and the Netherlands Research School for Astronomy (NOVA);
\item the Polish National Science Centre through HARMONIA grant 2018/30/M/ST9/00311 and DAINA grant 2017/27/L/ST9/03221 and the Ministry of Science and Higher Education (MNiSW) through grant DIR/WK/2018/12;
\item the Portuguese Funda\c{c}\~{a}o para a Ci\^{e}ncia e a Tecnologia (FCT) through national funds, grants SFRH/\-BD/128840/2017 and PTDC/FIS-AST/30389/2017, and work contract DL 57/2016/CP1364/CT0006, the Fundo Europeu de Desenvolvimento Regional (FEDER) through grant POCI-01-0145-FEDER-030389 and its Programa Operacional Competitividade e Internacionaliza\c{c}\~{a}o (COMPETE2020) through grants UIDB/04434/2020 and UIDP/04434/2020, and the Strategic Programme UIDB/\-00099/2020 for the Centro de Astrof\'{\i}sica e Gravita\c{c}\~{a}o (CENTRA);  
\item the Slovenian Research Agency through grant P1-0188;
\item the Spanish Ministry of Economy (MINECO/FEDER, UE), the Spanish Ministry of Science and Innovation (MICIN), the Spanish Ministry of Education, Culture, and Sports, and the Spanish Government through grants BES-2016-078499, BES-2017-083126, BES-C-2017-0085, ESP2016-80079-C2-1-R, ESP2016-80079-C2-2-R, FPU16/03827, PDC2021-121059-C22, RTI2018-095076-B-C22, and TIN2015-65316-P (`Computaci\'{o}n de Altas Prestaciones VII'), the Juan de la Cierva Incorporaci\'{o}n Programme (FJCI-2015-2671 and IJC2019-04862-I for F.~Anders), the Severo Ochoa Centre of Excellence Programme (SEV2015-0493), and MICIN/AEI/10.13039/501100011033 (and the European Union through European Regional Development Fund `A way of making Europe') through grant RTI2018-095076-B-C21, the Institute of Cosmos Sciences University of Barcelona (ICCUB, Unidad de Excelencia `Mar\'{\i}a de Maeztu’) through grant CEX2019-000918-M, the University of Barcelona's official doctoral programme for the development of an R+D+i project through an Ajuts de Personal Investigador en Formaci\'{o} (APIF) grant, the Spanish Virtual Observatory through project AyA2017-84089, the Galician Regional Government, Xunta de Galicia, through grants ED431B-2021/36, ED481A-2019/155, and ED481A-2021/296, the Centro de Investigaci\'{o}n en Tecnolog\'{\i}as de la Informaci\'{o}n y las Comunicaciones (CITIC), funded by the Xunta de Galicia and the European Union (European Regional Development Fund -- Galicia 2014-2020 Programme), through grant ED431G-2019/01, the Red Espa\~{n}ola de Supercomputaci\'{o}n (RES) computer resources at MareNostrum, the Barcelona Supercomputing Centre - Centro Nacional de Supercomputaci\'{o}n (BSC-CNS) through activities AECT-2017-2-0002, AECT-2017-3-0006, AECT-2018-1-0017, AECT-2018-2-0013, AECT-2018-3-0011, AECT-2019-1-0010, AECT-2019-2-0014, AECT-2019-3-0003, AECT-2020-1-0004, and DATA-2020-1-0010, the Departament d'Innovaci\'{o}, Universitats i Empresa de la Generalitat de Catalunya through grant 2014-SGR-1051 for project `Models de Programaci\'{o} i Entorns d'Execuci\'{o} Parallels' (MPEXPAR), and Ramon y Cajal Fellowship RYC2018-025968-I funded by MICIN/AEI/10.13039/501100011033 and the European Science Foundation (`Investing in your future');
\item the Swedish National Space Agency (SNSA/Rymdstyrelsen);
\item the Swiss State Secretariat for Education, Research, and Innovation through the Swiss Activit\'{e}s Nationales Compl\'{e}mentaires and the Swiss National Science Foundation through an Eccellenza Professorial Fellowship (award PCEFP2\_194638 for R.~Anderson);
\item the United Kingdom Particle Physics and Astronomy Research Council (PPARC), the United Kingdom Science and Technology Facilities Council (STFC), and the United Kingdom Space Agency (UKSA) through the following grants to the University of Bristol, the University of Cambridge, the University of Edinburgh, the University of Leicester, the Mullard Space Sciences Laboratory of University College London, and the United Kingdom Rutherford Appleton Laboratory (RAL): PP/D006511/1, PP/D006546/1, PP/D006570/1, ST/I000852/1, ST/J005045/1, ST/K00056X/1, ST/\-K000209/1, ST/K000756/1, ST/L006561/1, ST/N000595/1, ST/N000641/1, ST/N000978/1, ST/\-N001117/1, ST/S000089/1, ST/S000976/1, ST/S000984/1, ST/S001123/1, ST/S001948/1, ST/\-S001980/1, ST/S002103/1, ST/V000969/1, ST/W002469/1, ST/W002493/1, ST/W002671/1, ST/W002809/1, and EP/V520342/1.
\end{itemize}

The GBOT programme  uses observations collected at (i) the European Organisation for Astronomical Research in the Southern Hemisphere (ESO) with the VLT Survey Telescope (VST), under ESO programmes
092.B-0165,
093.B-0236,
094.B-0181,
095.B-0046,
096.B-0162,
097.B-0304,
098.B-0030,
099.B-0034,
0100.B-0131,
0101.B-0156,
0102.B-0174, and
0103.B-0165;
%
%
and (ii) the Liverpool Telescope, which is operated on the island of La Palma by Liverpool John Moores University in the Spanish Observatorio del Roque de los Muchachos of the Instituto de Astrof\'{\i}sica de Canarias with financial support from the United Kingdom Science and Technology Facilities Council, and (iii) telescopes of the Las Cumbres Observatory Global Telescope Network.

\end{appendix}

\end{document}